\documentclass[twoside,12pt]{article}

\usepackage{epsfig,latexsym}
\usepackage{amsmath}

\usepackage[latin1]{inputenc}

\def\xslash#1{{\rlap{$#1$}/}}
\newcommand{\bfm}[1]{\mbox{\boldmath$#1$}}
\newcommand{\bff}[1]{\mbox{\scriptsize\boldmath${#1}$}}
\newcommand{\Eqn}[1]{Eq.~(\ref{#1})}

\def\bfnabla{\mbox{\boldmath $\nabla$}}

\def\bfsigma{\mbox{\boldmath $\sigma$}}
\def\lQ{\Lambda_{QCD}}
\newcommand{\nn}{\nonumber}
\newcommand{\be}{\begin{equation}}
\newcommand{\ee}{\end{equation}}
\newcommand{\bea}{\begin{eqnarray}}
\newcommand{\eea}{\end{eqnarray}}
\def\al{\alpha}

\def\siml{{\
    \lower-1.2pt\vbox{\hbox{\rlap{$<$}\lower6pt\vbox{\hbox{$\sim$}}}}\ }} 
\def\simg{{\ \lower-1.2pt\vbox{\hbox{\rlap{$>$}\lower6pt\vbox{\hbox{$\sim$}}}}\ }}

\newcommand{\MS}{\overline{\rm MS}}
\newcommand{\RS}{\rm RS}

\newcommand{\braQM}[1]{\ensuremath{\langle#1|}}
\newcommand{\ketQM}[1]{\ensuremath{|#1\rangle}}
\def\em{{\rm em}}
\def\dsl{\,\raise.15ex\hbox{/}\mkern-13.5mu D}

\def \bk {\mathbf{k}}
\def \br {\mathbf{r}}
\def \mbk {\vert\bk\vert}

\def\lQ{\Lambda_{\rm QCD}}

\newcommand{\beqa}{\begin{eqnarray}}
\newcommand{\eeqa}{\end{eqnarray}}

\newcommand{\arccot}{\mathop{\mbox{arccot}}\nolimits}

\def\eul{\gamma_{\rm E}}
\def\bdm{\begin{displaymath}}
\def\edm{\end{displaymath}}

\def\Vt{\widetilde{V}}

\def\eps{\epsilon}

\begin{document}


\title{ \vspace{1cm} Review of Heavy Quarkonium at weak coupling}
\author{Antonio Pineda\\
\\
Grup de F\'\i sica Te\`orica, Universitat\\
Aut\`onoma de Barcelona, E-08193 Bellaterra, Barcelona, Spain}
\maketitle
\begin{abstract} 
We review weakly-bound heavy quarkonium systems 
using effective field theories of QCD. We concentrate on potential Non-Relativistic QCD, which provides with a well founded connection between QCD and descriptions of the 
heavy quarkonium dynamics in terms of Schrödinger-like equations. This connection is obtained using standard quantum field theory techniques 
such as dimensional regularization, which is used throughout, and renormalization. Renormalization group equations naturally follow. Certain effort is made to illustrate how computations are performed 
and the necessary techniques, providing some examples. 
Finally, we briefly review a selected 
set of applications, which include spectroscopy, radiative transitions, non-relativistic sum rules, inclusive decays, and electromagnetic threshold 
production.
\end{abstract}

\eject
\tableofcontents
\vfill
\newpage

\section{Introduction}

Quark--antiquark systems near threshold and with very large masses (or Heavy Quarkonium for short) are  extremely appealing. 
The reason is both theoretical and experimental. On the theoretical side, the large mass of 
its heavy constituents makes plausible the description of its dynamical properties by solving a proper non-relativistic (NR)
Schr\"odinger equation. In this respect the Heavy Quarkonium is often thought as "the Hydrogen atom" problem of QCD. On the experimental side, the appeal comes from the existence of several candidates in nature for Heavy Quarkonium, such as  charmonium, bottomonium, or $t$-$\bar t$ systems near threshold.
Therefore, it is not a surprise that many studies have been dwell on Heavy Quarkonium over the years, basically since the 
birth of QCD (see \cite{Yndurain} for older references). In particular, in the last years, we have witnessed the
development of effective field theories (EFTs) directly derived from
QCD like NRQCD \cite{Caswell:1985ui} or potential NRQCD (pNRQCD) \cite{Pineda:1997bj}
(for a comprehensive review see Ref.~\cite{Brambilla:2004jw}) aiming to describe heavy quarkonium systems. This has opened the door to
a systematic study and model independent determination of their properties. 
Instrumental in this development is the fact that, for large enough
masses, the NR nature of the Heavy Quarkonium characterizes its dynamics by, at least, three widely separated scales: 
\begin{itemize}
\item
{\bf hard}: $m$. The mass of the heavy quarks; 
\item
{\bf soft}: $|{\bf p}| \sim mv$, $ v \ll 1$. The relative momentum of the
heavy-quark--antiquark pair in the center of mass frame, and 
\item
{\bf ultrasoft}: $E \sim mv^2$. The
typical kinetic energy of the heavy quark in the bound state
system. \end{itemize}

In 1986, NRQED \cite{Caswell:1985ui}, an EFT for
NR leptons, was presented, providing the first and decisive link 
in a chain of developments that is still growing. NRQED is obtained from QED by
integrating out the hard scale $m$.  NRQCD \cite{NRQCD1} was
born soon afterwards. NRQCD has proved to be extremely successful in studying
$Q$-${\bar Q}$ systems near threshold. The Lagrangian of NRQCD can be
organized in powers of $1/m$, thus making explicit the NR
nature of the physical system. Yet, it still does not provide with a direct connection 
to a NR Schr\"odinger-like formulation of the problem. For instance, 
in a first approximation, the dynamics of the Hydrogen atom can be described 
by the solution of the Schr\"odinger equation with a Coulomb potential.
However, the derivation of this equation 
from the more fundamental quantum field theory, QED, or from NRQED is cumbersome (to say the least). 
The complications quickly increase when corrections are considered, to the point that is very difficult to incorporate them 
in a systematic way. This problem becomes even more acute for 
Heavy Quarkonium, since the weak-coupling computations are much more complicated due to the non-abelian nature of the interactions (and on top of that non-perturbative effects, due to $\lQ$, also show up).
One efficient solution to this problem comes from the use of 
EFTs and in particular of pNRQCD to which we focus in this review. 
This EFT takes full advantage of the hierarchy of scales that appear 
in the system:
\be
\label{hierarchy}
m \gg mv \gg mv^2 \cdots,
\ee 
and makes systematic and natural the connection of the Quantum Field Theory 
with the Schrödinger equation. Roughly speaking the EFT turns out to be 
something like:
\begin{eqnarray*}
\,\left.
\begin{array}{ll}
&
\displaystyle{\left(i\partial_0-{{\bf p}^2 \over m}-V_s^{(0)}(r)\right)\phi({\bf r})=0}
\\
&
\displaystyle{{\rm \ + \ corrections\; to\; the\; potential}}
\\
&
\displaystyle{+{\rm interaction \;with\; other\; low}-{\rm energy\; degrees \;of\; freedom}}
\end{array} \right\} 
{\rm pNRQCD}
\end{eqnarray*}

For very large masses one has that $mv \gg \lQ$. The static potential $V_s^{(0)}(r)$ can then be determined within perturbation theory and would 
be equal to the Coulomb potential $V_s^{(0)}(r)\simeq V_s^C(r)\equiv -C_f\al(\nu)/r$ at leading order (LO), whereas $\phi({\bf r})$ is the
${\bar Q}$--$Q$ wave-function. 

The construction of any EFT is determined  by  
the kinematic situation aimed to describe.  This fixes the (energy of the) degrees of freedom that appear as 
physical states (and not only as loop fluctuations). In our case the degrees of freedom in pNRQCD have 
{\bf $E \sim mv^2$}. In order to derive pNRQCD we sequentially integrate out larger scales. 
This is the path we follow for the construction of pNRQCD: 
\begin{figure}[htb]
\hspace{-0.3in}
\put(250,2){{\bf\Large $E \sim mv^2$}.}
\epsfxsize=3.3in
\centerline{\epsffile{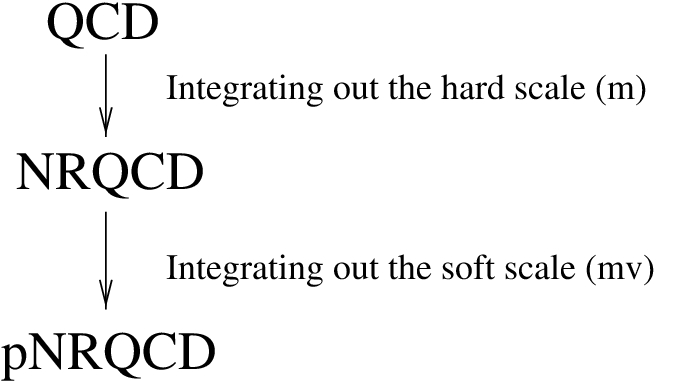}}
\end{figure}

The specific construction details of pNRQCD are slightly different depending upon the relative size between the soft and the 
$\lQ$ scale. Two main situations are distinguished, named by the weak \cite{Pineda:1997bj,Brambilla:1999xf} ($mv \gg \lQ$)  and the strong \cite{Brambilla:2000gk} ($mv \simeq \lQ$) coupling version of pNRQCD.
One major difference between them is that in the former the potential can be computed in perturbation theory unlike in the latter. A general overview 
of the general formalism of both versions of pNRQCD was given in Ref. \cite{Brambilla:2004jw}, with which we will overlap in some aspects. Here we focus on the strict weak coupling version of pNRQCD, as it provides us with a closed body of research. One major aim is pedagogical, with special emphasis in showing, in a comprehensive way, how
computations are performed in practice in dimensional regularization/renormalization from the beginning to the end: starting from QCD and ending up
in bound state systems\footnote{Yet some previous knowledge in Heavy Quark Effective Theory (HQET) is advisable. See \cite{HQET} for reviews.}. All steps can be computed using Feynman diagrams, 
to which we focus in this review. Therefore, 
we give some few examples of explicit computations in detail.  Due to lack of space, we will not dwell on Wilson loop analysis for which we refer to \cite{Brambilla:2004jw}.
We profit to display several results scattered in different papers 
in a more unified and coherent notation. We show some details of how the renormalization of the effective theory is carried out, which will naturally lead us to 
obtain the associated Renormalization Group (RG) equations, and to obtain the resummation of logarithms.
Overall we hope that this review can serve as a self-contained summary of results and techniques on which one 
can rely for future computations, as well as a good starting point for beginners. We also expect so for those aiming to study pNRQCD at 
strong coupling or in more complicated setups, as the weak coupling limit is theoretically the cleanest.

The derivation of the theory is performed at pure weak coupling. Therefore, we will neither incorporate renormalons
nor non-perturbative effects, which still exist in a weak coupling analysis.
Nevertheless, they are incorporated in the phenomenological analysis reviewed here when necessary, particularly renormalon effects, which are crucial to get 
convergent series for some physical observables. 

In this review we focus on QCD, yet the same techniques can and has been applied to QED. For instance, the application of pNRQED and, in general, of factorization with dimensional regularization, 
has also led to a plethora of results for the spectra of positronium 
\cite{Pineda:1998kn,Czarnecki:1998zv,Melnikov:1999uf,Melnikov:2000zz,Kniehl:2000cx}.
Therefore, we also expect this review to be useful in this context.

\section{From QCD to NRQCD}
More details of NRQCD can be found in Refs. \cite{Bodwin:1994jh,Brambilla:2004jw,Mannel:1994xh}. 
In particular, in the last reference expressions for different operators when the heavy quarks have been boosted to a general 
frame can be found. Here we only take the needed results and fix the notation. 
The only thing we will explain in some detail is the matching procedure and the derivation of the soft RG equations.

\subsection{NRQCD. How to build it? General procedure}
In order to write the effective theory Lagrangian we need the following:
\begin{itemize}
\item
{\bf The degrees of freedom/cut-off}. The degrees of freedom of NRQCD are a quark-antiquark pair,
gluons and light quarks with the cut-off $\nu_{\rm NR}=\{\nu_p,\nu_s\}$ satisfying $E, |{\bf p}|, \lQ \ll \nu_{\rm NR} \ll m$;
$\nu_p$ is the UV cut-off of the relative three-momentum of the heavy quark
and antiquark; $\nu_s$ is the UV cut-off of the energy of the heavy quark and
the heavy antiquark, and of the four-momentum of the gluons and light quarks.
\item
{\bf Symmetries}. NRQCD should be invariant under rotations, gauge transformations, C, P and T. The Poincare symmetry is realized nonlinearly (see \cite{Luke:1992cs,Brambilla:2003nt}). 
\item
{\bf Power counting (the scales of the problem)}. The NRQCD Lagrangian can be
organized as a power series in $1/m$.  Since several scales ($E$, $|{\bf p}|$, $\lQ$)  remain dynamical, it is not possible to unambiguously 
assign a size to each operator without extra
assumptions: no homogeneous power counting exists. As we will see below, the introduction of pNRQCD facilitates this
task.  The original power  counting introduced by \cite{Bodwin:1994jh} assumes
$\lQ\sim E \sim mv^2$, and  hence $|{\bf p}| \sim mv\gg \lQ$, $v\sim \al (mv)\ll 1$. 
\item
{\bf Matching}. This item is only used if the coefficients can be determined from the underlying theory, as it is the case. Otherwise they are fixed by experiment.
The Wilson coefficients of NRQCD are determined imposing suitable QCD and NRQCD Green's functions to be equal.  The Wilson coefficients of each operator depend logarithmically on
$m$, $\nu_{\rm NR}$ and can be calculated
in perturbation theory in $\al (\nu_{\rm NR})$. Hence the importance of a given
operator for a practical calculation not only depends on its size (power
counting), but also on the leading power
of $\al$ that its Wilson coefficient has. 
\end{itemize}

\subsection{NRQCD Lagrangian}
\label{sec:NRLag}
The allowed
operators in the Lagrangian are constrained by the symmetries of
QCD. We restrict ourselves to the equal mass case (for the non-equal mass case see \cite{Brambilla:2004jw}). 
It reads \cite{Caswell:1985ui,Bodwin:1994jh,Manohar:1997qy,Bauer:1997gs} (we do not write the $1/m^2$ operators 
in the pure light fermion sector as they are suppressed compared with present day analysis):
\bea
&& 
{\cal L}_{\rm NRQCD}={\cal L}_g+{\cal L}_l+{\cal L}_{\psi}+{\cal L}_{\chi}+{\cal L}_{\psi\chi},
\label{LagNRQCD}
\\
\nn
\\
&&
{\cal L}_g=-\frac{1}{4}G^{\mu\nu \, a}G_{\mu \nu}^a +
{1\over 4}{c_1^{g} \over m^2}  
g f_{abc} G_{\mu\nu}^a G^{\mu \, b}{}_\alpha G^{\nu\alpha\, c},
\label{Lg}
\\
\nn
\\
&&
{\cal L}_l = \sum_{i=1}^{n_f} \bar q_i \, i \dsl \, q_i 
+ {\cal O}\left(\frac{1}{m^2}\right)
\label{Ll}
,
\\
\nn
\\
&&
{\cal L}_{\psi}=
\psi^{\dagger} \Biggl\{ i D_0 + {c_k\over 2 m} {\bf D}^2 + {c_4 \over 8 m^3} {\bf D}^4 
+ {c_F \over 2 m} {\bfsigma \cdot g{\bf B}} 
\nn
\\
&& \qquad\qquad
+ { c_D \over 8 m^2} \left({\bf D} \cdot g{\bf E} - g{\bf E} \cdot {\bf D} \right) 
+ i \, { c_S \over 8 m^2} 
{\bfsigma \cdot \left({\bf D} \times g{\bf E} -g{\bf E} \times {\bf D}\right) }
\Biggr\} \psi
\nn
\\
&& \qquad\qquad
+{c_1^{hl} \over 8m^2}\, g^2 \,\sum_{i=1}^{n_f}\psi^{\dagger} T^a \psi \ \bar{q}_i\gamma_0 T^a q_i 
+{c_2^{hl} \over 8m^2}\, g^2 \,\sum_{i=1}^{n_f}\psi^{\dagger}\gamma^\mu\gamma_5
T^a \psi \ \bar{q}_i\gamma_\mu\gamma_5 T^a q_i 
\nn
\\
&& \qquad\qquad
+{c_3^{hl}\over 8m^2}\, g^2 \,\sum_{i=1}^{n_f}\psi^{\dagger} \psi \ \bar{q}_i\gamma_0 q_i
+{c_4^{hl}\over 8m^2}\, g^2 \,\sum_{i=1}^{n_f}\psi^{\dagger}\gamma^\mu\gamma_5
\psi \ \bar{q}_i\gamma_\mu\gamma_5 q_i,
\label{Lhl}
\\
\nn
\\
\label{Lchic}
&& {\cal L}_{\chi_c} = {\cal L}_{\psi}(\psi \rightarrow \chi_c, g \rightarrow -g, T^a \rightarrow (T^a)^T),
\\
\nn
\\
&&
{\cal L}_{\psi\chi_c} =
 - {d_{ss} \over m^2} \psi^{\dag} \psi \chi_c^{\dag} \chi_c
+
  {d_{sv} \over m^2} \psi^{\dag} {\bfsigma} \psi
                         \chi_c^{\dag} {\bfsigma} \chi_c
-
  {d_{vs} \over m^2} \psi^{\dag} {\rm T}^a \psi
                         \chi_c^{\dag} ({\rm T}^a)^T \chi_c
+
  {d_{vv} \over m^2} \psi^{\dag} {\rm T}^a {\bfsigma} \psi
                         \chi_c^{\dag} ({\rm T}^a)^T {\bfsigma} \chi_c
\,.
\label{Lhh}
\eea
$\psi$ stands for a NR fermion, represented by a Pauli spinor, $\chi_c=-i\sigma^2\chi^\ast$, its antiparticle, is also represented by a Pauli spinor.
$(T^a)^T$ stands for the transpose matrix of $T^a$, and $T^a \rightarrow (T^a)^T$ in Eq. (\ref{Lchic}) only applies to the matrices contracted
 to the heavy quark color indexes. 
$\bfsigma$ are the Pauli matrices,
$i D^0=i\partial^0 -gA^0$, $i{\bf D}=i\bfnabla+g{\bf A}$,
${\bf E}^i = G^{i0}$, ${\bf B}^i = -\epsilon_{ijk}G^{jk}/2$, $\epsilon_{ijk}$ being 
the usual three-dimensional antisymmetric tensor\footnote{
In dimensional regularization several prescriptions are possible for 
the $\epsilon_{ijk}$ tensors and $\bfsigma$.
Therefore, if dimensional regularization is used, one has to make sure that 
one uses the same prescription as the one used to calculate the Wilson coefficients.}
($({\bf a} \times {\bf b})^i \equiv \epsilon_{ijk} {\bf a}^j {\bf b}^k$)
with $\epsilon_{123}=1$.

The NRQCD Lagrangian is defined up to field redefinitions. In the expression 
adopted here, we have used of this freedom. Powers larger than one of $iD_0$ 
applied to the quark fields have been eliminated. We have also redefined the 
gluon fields in such a way that the coefficient in front of $- G^{\mu\nu \, a}G_{\mu \nu}^a/4$ 
in  ${\cal L}_g$ is one. This turns out to be equivalent to redefining the coupling constant 
in such a way that it runs with $n_f$ light flavors (heavy quarks do not contribute to the running). 
A possible term $D^\mu\, G_{\mu \alpha}^a\, D_\nu\, G^{\nu \alpha \, a}$ has been 
eliminated through the identity \cite{Manohar:1997qy}: 
\be
\displaystyle \int d^4x \, \left(
2\, D^\mu\, G_{\mu \alpha}^a\, D_\nu\, G^{\nu \alpha \, a} 
+ 2\ g \, f_{abc} \, G_{\mu\nu}^a G^{\mu \, b}{}_\alpha G^{\nu\alpha\, c}
+ G_{\mu \nu}^a\, D^2\, G^{\mu \nu \, a}\right) = 0 .
\ee 
Finally, a possible term like $c\, G_{\mu\nu}^a D^2 G^{\mu\nu \, a}$
has been eliminated through the field redefinition 
$A_\mu \to A_\mu + 2 \, c \, [D^\alpha,G_{\alpha\mu}]$ \cite{Pineda:2000sz}. 

Expressions for 
the Feynman rules associated to the first two lines of
Eq.~(\ref{Lhl}), and Eq. (\ref{Lhh}), can be found in the Appendix.

\medskip

\noindent
{\bf Coupling to hard photons}.\\
Heavy Quarkonium can be produced or annihilated through hard-photons mediated processes, 
which can be described in NRQCD in terms of NR currents. Similarly to the Lagrangian, they can 
be written as an expansion in $1/m$ times some hard Wilson coefficients 
times some NR (local) operators composed of the NR two-component spinor fields $\psi^{\dagger}$ 
and $\chi$. 

The one-photon mediated processes are induced by the electromagnetic
current $j^\mu$. Its space components have the following decomposition:
\begin{equation}
\bfm{j}=b_1\psi^\dagger{\bfm\sigma}\chi+{d_1\over6m^2}
\psi^\dagger\bfsigma{\bf D}^2\chi
+\cdots\,.
\label{vcurr}
\end{equation}
By using the equations of motion, \Eqn{NRQCDcurrent}
can also be written in the following way \cite{Gremm:1997dq}
\begin{equation}
 \bfm{j}= b_1\chi^{\dagger}\bfsigma^i\psi
- {d_1 \over 6m}i \partial_0 
\left( \chi^{\dagger}\bfsigma^i\psi \right) + \cdots
\,.
\label{nrcurrent}
\end{equation}

The operator responsible for the two-photon $S$-wave  processes in the
NR limit is generated by the expansion of the product of
two electromagnetic currents and has the following representation
\begin{equation}
O_{\gamma\gamma}=b_0\psi^\dagger\chi
+{d_0\over6m^2}
\psi^\dagger{\bf D}^2\chi
+\cdots,
\label{gcurr}
\end{equation}
which, up to the Wilson coefficient, reduces to the NR limit of the 
pseudoscalar current.

The determination of the Wilson coefficients is discussed in the following section.

\subsection{Matching QCD to NRQCD}
\label{secmatchingNRQCD}

\begin{figure}
\epsfxsize=6cm
\hfil\epsfbox{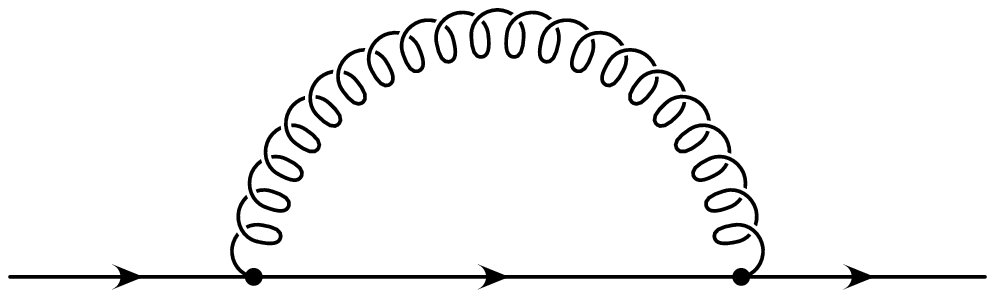}\hfill
\caption{\it One-loop heavy quark self-energy.
\label{fig:wave}}
\end{figure}

The procedure we follow is mainly based on Refs. \cite{Manohar:1997qy,Pineda:1998kj}. 
The Wilson coefficients of NRQCD are fixed by imposing
QCD and NRQCD Green's functions (or matrix elements) with different number of heavy quarks (0,1, ...) and gluons to be equal for scales below $\nu_{\rm NR}$ 
to a given order in an expansion in $\al$, $E/m$ and $|{\bf p}|/m$, where $E$ and $|{\bf p}|$ are the 
external energy and three-momenta.  It extraordinarily simplifies calculations if
these expansions are done {\it before} the loop integrals are performed, particularly if dimensional regularization is used 
as the regulator in QCD and NRQCD for the infrared and ultraviolet divergences. This is so because all loop integrals in the NRQCD
calculations will be scaleless and can be set to zero. Moreover by using the same regulator for the infrared divergences in QCD and NRQCD ensures 
that they will cancel in the difference. Schematically \cite{Manohar:1997qy},
one has
\be
A_{eff}\left( {1 \over \epsilon_{UV}} - {1 \over \epsilon_{IR}} \right)
\ee
in the EFT, which is zero if $\epsilon_{UV}=\epsilon_{IR}$
in dimensional regularization. For instance ($D=4+2\epsilon$),
\begin{equation}
\label{11}
\int {d^D\,k\over \left(2\pi\right)^D} {1\over k^4} = 
\int {d^D\,k\over \left(2\pi\right)^D} \left[{1\over k^2\left(k^2+m^2\right)}
+{m^2\over k^4\left(k^2 + m^2\right)} \right] = {1\over 16\pi^2}\left[{1\over
\epsilon_{UV}}-{1\over \epsilon_{IR}}\right] =0.
\end{equation}
Therefore, we only have to calculate loop integrals in QCD that depend on a
single scale ($m$). Typical integrals one has to compute are the following 
(note that these expressions correspond to the hard contribution in the threshold expansion of QCD diagrams \cite{Beneke:1997zp}):
\bea
I_{n,s}&=&\int {d^{D}q\over (2\pi)^{D}}{1\over \left(q^2 +i\eta\right)^{s}}{1\over (q^2+2mq^0 +i\eta)^n}
\\
\nn
&=&
{i\over (4\pi)^{2}}\left(- m^2\right)^{2-n-s}\left({m^2 \over 4\pi}\right)^{\epsilon}\frac{\Gamma[n+s-2-\epsilon]\Gamma[4-n-2s+2\epsilon]}{\Gamma[n]\Gamma[4+2\epsilon-n-s]}
\eea
in the case 
that a single heavy quark appear in the diagram (see Figs. \ref{fig:wave}, \ref{fig:NRQCDmatching} for illustration),
and 
\be
I_{n}=\int {d^{D}q\over (2\pi)^{D}}{1\over \left(q^2 +i\eta\right)^{n}}{1\over q^2+2mq^0 +i\eta}
{1\over q^2-2mq^0 +i\eta}
={i\over (4\pi)^{2}}
\left({-1 \over m^2}\right)^{n}
\left({m^2 \over 4\pi}\right)^{\epsilon}
{\Gamma(n-\epsilon)\Gamma(2\epsilon-2n+1) \over \Gamma(2\epsilon+2-n)}
\ee
in the case where a heavy quark and a heavy antiquark appear (see Figs. \ref{1loopsoftspin}, \ref{fig:NRQCDmatching} for illustration)\footnote{Note that in the process to reduce the Feynman diagrams to these master integrals one may introduce spurious ultraviolet and infrared divergences. This is not a problem, since it is not necessary to distinguish the origin of the divergence in order to obtain the renormalized Wilson coefficient.}.
After expanding in $1/\epsilon$ one would have
\be
A {1 \over \epsilon_{UV}} + B {1 \over \epsilon_{IR}}  + \left( A+B
\right) \ln {\nu_{\rm NR} \over m} + D
\,.
\ee
Since the full and the effective theory share the same infrared behavior
$B=-A_{eff}$. Moreover the ultraviolet divergences are absorbed in the
coefficients of the full and effective theory. In this way the difference
between the full and the effective theory reads
\be
\left( A+B \right) \ln{\nu_{\rm NR}\over m}+D
\,,
\ee
which provides the renormalized one-loop contribution to the Wilson coefficients for the effective theory:
\be
 c_i \sim 1 + \al\left( A_{c_i} \log {m \over \nu_{\rm NR}} + B_{c_i} \right)
\qquad \qquad
d_i \sim \al \left(1 + \al\left( A_{d_i} \log {m \over \nu_{\rm NR}} + B_{d_i} \right)\right)
\,.
\ee
In this procedure the same renormalization scheme is used
for  both ultraviolet and infrared divergences in NRQCD. In the QCD calculation both the ultraviolet and infrared divergences can also be
renormalized in the same way, for instance using the  $\MS$ scheme, which
is the standard one for QCD calculations. This fixes the ultraviolet
renormalization scheme in which the NRQCD Wilson coefficients have
been calculated. This means that for these Wilson coefficients to be
consistently used in a NRQCD  calculation, this calculation must be
carried out in the same scheme, for instance in dimensional regularization and in the $\MS$ scheme.
We could also obtain the bare expression 
for the Wilson coefficients if the ultraviolet divergences of the QCD calculation are known. Those can be absorbed by 
$\al$ and the masses. Therefore, it is relatively easy to obtain the bare expressions for the NRQCD Wilson coefficients in case of need.

The matching calculation can be carried out in any gauge. However, since one is usually matching
gauge-dependent Green functions, the same gauge must be chosen in  QCD
and NRQCD. Using different gauges or, in general, different ways to
carry out the matching procedure, may lead to apparently different
results for the Wilson coefficients (within the same regularization
and renormalization scheme). These results must be related by local
field redefinitions, or, in other words, if both matching calculations
had agreed to use the same {\it minimal} basis  of operators
beforehand, the results would have coincided. Irrespectively of this, 
some Wilson coefficients are gauge independent and can be computed independently in any gauge, as they can be directly related with observables. One example would be $c_F$. In general one has to be careful and compute all the matching coefficients in the same gauge, specially if one is not working in a minimal basis. For instance the computation of $c_D$ and $d_{vs}$ in different gauges would lead to wrong results, as only an specific combination of them is free of ambiguities. If the matching is
carried out as described above, it is more convenient  to choose a
covariant gauge (i.e. Feynman gauge), since only QCD calculations,
which are manifestly covariant, are to be carried out.  

In order to address the matching calculation, 
we also need the relation between the QCD and NRQCD quark (antiquark) fields:
\be
Q(x) \rightarrow Z^{1\over 2}{1+\gamma_0 \over 2} e^{-im t}\psi (x) + Z^{1\over 2} {1- \gamma_0 \over 2} e^{im t}\chi (x)
\,.
\ee
Actually the computation of $Z$ provides us with the simplest possible example on which to illustrate the above matching discussion. One only needs the computation in QCD of the self-energy shown in Fig. \ref{fig:wave} \cite{Pokorski:1987ed}:
\begin{eqnarray}\label{6}
-i \Sigma\left(p\right) &=& -i C_f{\alpha\over 4 \pi}
\left(A\left(p^2\right)m + B\left(p^2\right) 
\xslash{p}\right) \\
A\left(p^2\right) &=& \int_0^1 dx\ \Gamma\left(-\epsilon\right) 
\left(4+2\epsilon\right) \left[ m^2 x - p^2 x (1-x) \right]^{\epsilon}\\
B\left(p^2\right) &=& -\int_0^1 dx\ \Gamma\left(-\epsilon\right) 
\left(2+2\epsilon\right)\left(1-x\right) \left[ m^2 x - p^2 x (1-x) 
\right]^{\epsilon}
\,.
\end{eqnarray}
At one loop, the bare $Z$ then reads
\begin{eqnarray}\label{7}
Z_B &=&1  + C_f{\alpha_s\over 4 \pi} \left[B\left(m^2\right)
+ 2m^2 \left(
{\partial A \over \partial p^2} + {\partial B \over \partial
p^2}\right)_{p^2=m^2}\right] \nonumber \\
&=& 1-C_f {\alpha\over \pi}\left[ -{3\over 2 \epsilon}
+ 1 - {3\over 2} \log {m\over \nu_{\rm NR} }\right].
\end{eqnarray}
and the renormalized one,
\be
Z = 1 + C_f {\al \over \pi} \left( {3 \over 4}\ln{m^2\over \nu_{\rm NR}^2} -1
\right) + {\cal O}\left(\left({\al \over \pi}\right)^2 \right) 
\,.
\ee
Note that we did not have to distinguish the ultraviolet or infrared origin of the $1/\epsilon=1/3(1/\epsilon_{UV}+2/\epsilon_{IR})$. 
Note also that the states are differently normalized in relativistic 
($\left< {\bf p}\vert {\bf p}'\right>=(2\pi)^3 2 \sqrt{{\bf p}^2 + m^2} 
\delta^3({\bf p}-{\bf p}')$) or
NR ($\left< {\bf p}\vert {\bf p}'\right>=(2\pi)^3 \delta^3({\bf p}-{\bf p}')$) theories. 
Hence, in order to compare the S-matrix elements between the
two theories, a factor $(2\sqrt{{\bf p}^2 + m^2})^{1/2}$ has to
be introduced for each external fermion.

The NRQCD Wilson coefficients have been computed over the years. In 
Refs. \cite{Eichten:1990vp,Manohar:1997qy} one can find them for the 
one heavy quark and pure gluonic sector at NLO, whereas in Ref. \cite{Pineda:1998kj} one can find them 
for the two heavy quark sector both in the equal and non-equal mass case (although with a difference of scheme for $d_{vv}$ with respect the one presented here). 
 All of them have been computed at 
${\cal O}(1/m^2)$ and are displayed in the Appendix.

In the matching computation a single factorization scale appears: $\nu_{\rm NR}$, and it is not possible to distinguish between $\nu_s$ and 
$\nu_p$. This is not a problem because, in strict finite order computations it is not necessary to distinguish between factorization scales as they all cancel to the required accuracy, whereas for RG analysis the obtained result is the initial condition of the RG equation. 

Finally, we also would like to emphasize that 
performing the matching as explained above also produces an extremely welcome side effect: Coulomb pole 
singularities exactly vanish in the matching computation.  This is an very important simplification, as the 
Coulomb poles are infrared effects that cancel in the matching, so their appearance in the intermediate computation 
is unnecessary and would make the computation much more cumbersome.

\medskip

\noindent
{\bf NR current Wilson coefficients}.\\
Their determination analogously follows  the matching procedure of the 
four-fermion operators. For instance for the vector current we have
\begin{equation}
{\bar Q} \gamma^{i} Q(0) \bigg|_{\rm QCD}= 
b_1\chi^{\dagger}\sigma^i\psi(0) -
\frac{d_1}{6m^2}\chi^{\dagger}\sigma^i\left(i {\bf D} \right)^2 \psi(0) +
\ldots\bigg|_{\rm NRQCD}\ .
\label{NRQCDcurrent}
\end{equation}
The Wilson coefficients $b_s$
and $d_s$ represent the contributions from the hard modes and may be
evaluated as a series in $\al$ in
full QCD for free on-shell on-threshold external (anti)quark fields.
We define it through
\begin{eqnarray}
  b_s(\nu_{\rm NR}) &=& \sum_{i=0}^\infty\left(\al(\nu_{\rm NR})\over
  \pi\right)^i b_s^{(i)}(\nu_{\rm NR})\
  \,, \qquad b_s^{(0)}=1\,,
\end{eqnarray}
and similarly for other  coefficients. $b_s^{(1)}$ has been known for quite a
long time \cite{KalSar,HarBro}. $b_1^{(2)}$ has been computed in Ref. \cite{Hoang:1997ui} for QED and 
in Refs. \cite{Czarnecki:1997vz,Beneke:1997jm} for QCD. There are also 
some partial results at ${\cal O}(\al^3)$ \cite{Marquard:2006qi}. $b_0^{(2)}$ was determined in
semi-numerical form \cite{CzaMel2}, where the $gg\to\gamma\gamma$ contribution induced by a light-fermion
box was estimated to be small and not included in
the result. Note that the $b_0^{(2)}$ result depends on the definition of the nonrelativistic
axial current. We show the explicit expressions in the $\overline{\rm MS}$ up to two loops in the Appendix. 

\subsection{RG: Soft running}
\label{sec:NRRG}
\begin{figure}[htb]
\makebox[0.0cm]{\phantom b}
\put(20,50){$\displaystyle{\delta d_{vv}\dot=}$}
\put(80,15){\epsfxsize=6truecm \epsfbox{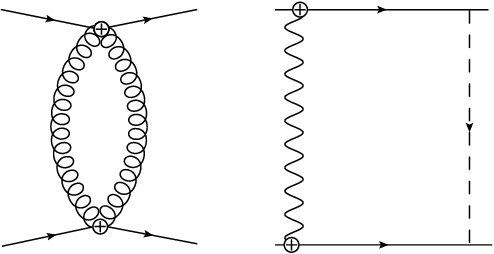}}
\put(155,50){$\displaystyle{+}$}
\put(270,50){$\displaystyle{+\cdots \simeq-\frac{C_A}{2}\al^2c_F^2\ln \nu_s+\cdots}$}
\caption{\it One loop ultraviolet soft contribution to $d_{vv}$.}
\label{1loopsoftspin}
\end{figure}
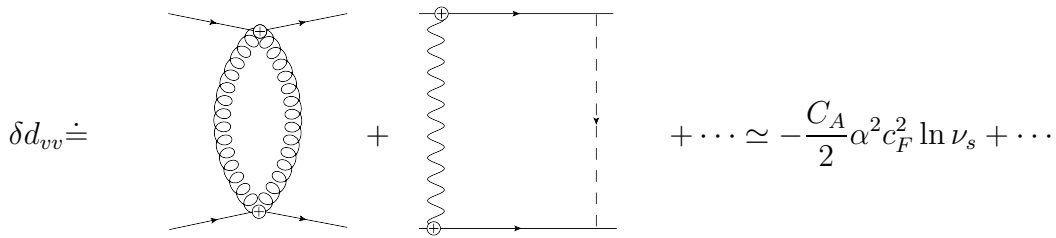

The Wilson coefficients of the zero and two heavy-quark operators, $\{c\}$, only depend on $\nu_s$: $c(\nu_s)$, as 
they are insensitive to the the other heavy quark. This is not so for the Wilson coefficients of the four heavy-quark operators, 
$\{d\}$, which depend on both $\nu_s$ and $\nu_p$: $d(\nu_p,\nu_s)$. 
Irrespectively, the factorization scale dependence on $\nu_{\rm NR}$ is eventually traded for a lower 
scale ($|{\bf p}|$, $E$, $\lQ$), introducing large logarithms, which can be summed 
up. This resummation is obtained by solving the appropriate RG equations of the Wilson coefficients, which in the case 
of NRQCD are not known, nor it is expected one can get them in the near future. 
The reason is the entanglement of the soft and ultrasoft scale. This handicap is solved in pNRQCD, 
where a complete set of RG equations will be given. 
Yet, the NRQCD Lagrangian is useful in order to obtain the pure soft running (associated to $\nu_s$), which can be obtained considering the 
quasi-static limit (HQET version) of NRQCD. In this approximation, 
one can always perform the computation with static propagators for the 
heavy quarks and order by order in $1/m$.
Formally, we can write the NRQCD Lagrangian as an expansion in $1/m$:
\be
{\cal L}_{\rm NRQCD} =\sum_{n=0}^{\infty}{1\over m^n}\lambda_nO_n
, 
\ee
where the RG equations of the Wilson coefficients read
\be
\nu_s {d \over d \nu_s}\lambda=B_{\lambda}(\lambda)
.
\ee
The RG equations have a triangular structure (the standard
structure one can see, for instance, in HQET RG equations):
\bea
\nu_s {d \over d \nu_s}\lambda_0&=&B_0(\lambda_0)
\,,
\nn
\\
\nu_s {d \over d \nu_s}\lambda_1&=&B_1(\lambda_0)\lambda_1
\,,
\nn
\\
\nu_s {d \over d
\nu_s}\lambda_2&=&B_{2(2,1)}(\lambda_0)\lambda_2+B_{2(1,2)}(\lambda_0)\lambda_1^2
\,,
\eea
$$
\cdots\,,
$$
where the different $B$'s can be expanded into a power series in $\lambda_0$
($\lambda_0$ corresponds to the marginal operators (renormalizable
interactions)). For NRQCD we have $\lambda_0=\al$ and
$\lambda_{1}=\{c_k, c_F\}$, $\lambda_{2}=\{c_1^g, c_D, c_S,
\{c^{ll}\},\{c^{hl}\}, \{d\} \}$. Note that in order to properly identify the ultraviolet soft divergences we can not work as we did for the matching computation, where infrared and ultraviolet divergences are kept equal. We now need a method to discern them, usually by regulating the infrared using some off-shell small $E$ energies. In the case of the spin-dependent coefficient, 
there are not infrared divergences at LO and one could actually work as in the matching computation. On the other hand it is quite easy to obtain the logarithmic divergent contribution directly, without even considering dimensional regularization. We  
 illustrate the computation in Fig. \ref{1loopsoftspin}, which produces the following RG equation for $d_{vv}$ at LO:
 \be
 \nu_s {d \over d \nu_s} d_{vv}=-\frac{C_A}{2}\al^2c_F^2\,.
 \ee
 For the rest of the Wilson coefficients, the LL running
for the $\{c\}$ can be read off from the 
results of \cite{Bauer:1997gs}, and the LL running of the four-fermion operators $\{d\}$ 
can be found in \cite{Pineda:2001ra}. Both of them are displayed in the Appendix.
 At this order there is no dependence on $\nu_p$, which appears at the next order.
On the other hand the leading factorization scale dependence of $b_s$ starts at two loops and is logarithmically proportional to 
$\nu_p$.

\section{From NRQCD to pNRQCD}

\subsection{Motivation and Physical picture}
\label{sec:mot}

NRQCD simplifies NR bound-state problems by making explicit the NR nature of the system.
Nevertheless, it is not optimal yet.  The main problem stems from the fact that degrees of freedom with soft 
energy are still dynamical in the EFT. This has effects on the power counting rules, which are not
homogeneous (the power counting by \cite{Lepage:1992tx}, which assumed that
$\lQ \siml mv^2$, catches the LO contribution of the matrix
elements but there are also subleading contributions in $v$); and on the
perturbative calculations, which were dependent on two scales and, therefore, still difficult to compute. 
Initially, it was tried to disentangle the soft and ultrasoft scales in NRQCD by classifying the different momentum
regions existing in a purely perturbative version of NRQCD and/or reformulating
NRQCD in such a way that some of these regions were explicitly displayed by
introducing new fields in the NRQCD Lagrangian.  
In particular, we mention \cite{Labelle:1996en} where a diagrammatic approach to NRQED was used and the
subsequent work by \cite{Luke:1996hj,Grinstein:1997gv,Luke:1997ys} in
NRQCD. All these early attempts turned out to be missing some relevant
intermediate degrees of freedom.

The first complete solution came in \cite{Pineda:1997bj}. It tried to answer the
question: How would we like the effective theory for ${\bar Q}$--$Q$ systems near
threshold to be? The main observation was that some degrees of freedom included in NRQCD
 only appear as virtual fluctuations and never as asymptotic states, unlike those with ultrasoft energy. 
Therefore, the unwanted degrees of freedom should be integrated out, keeping only those with ultrasoft energy.
Moreover, we wanted to get a closer connection with a Schr\"odinger-like
formulation for these systems (see also \cite{Lepage:1997cs}). The idea was to connect NRQCD with potential
models also, eventually, in the non-perturbative regime, profiting, as much as possible, from dimensional regularization. 
The resulting EFT was called potential NRQCD (pNRQCD).

Let us now partially illustrate the physical picture behind the previous discussion. We consider a four-fermion Green function near threshold in NRQCD. 
At tree level in the Feynman gauge the leading contribution from longitudinal gluons reads (${\bf k}={\bf p}-{\bf p}'$)
\begin{figure}[htb]
\makebox[0.0cm]{\phantom b}
\put(40,-5){\epsfxsize=4.5truecm \epsfbox{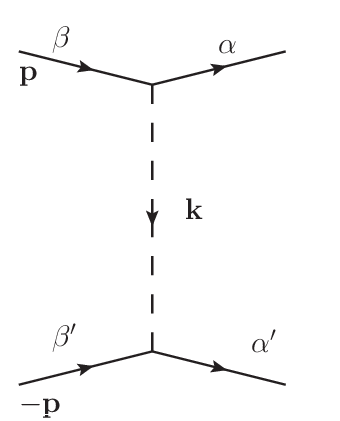}}
\put(143,75){$\displaystyle{
=-T^a_{\alpha\beta}T^a_{\beta'\alpha'}g^2\frac{1}{k^2}=\tilde V^C+{\cal O}(\frac{k_0^2}{{\bf k}^4})}\equiv$}
\put(325,-5){\epsfxsize=4.5truecm \epsfbox{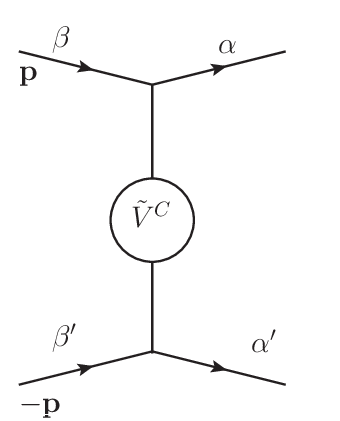}}
\put(435,75){$\displaystyle{+{\cal O}(\frac{k_0^2}{{\bf k}^4})}$.}
\label{figtreelevelA0}
\end{figure}
\\
Note that the LO term gives the Coulomb potential (actually, in the Coulomb gauge, there are no subleading corrections):
\be
\label{VCtree}
\tilde V^C\equiv-T^a_{\alpha\beta}T^a_{\beta'\alpha'}g^2\frac{1}{{\bf k}^2}
\,.
\ee
We can also easily project the potential to the singlet/octet sector($|s\rangle=\frac{1}{\sqrt{N_c}}\delta_{\alpha\alpha'}$ and $|o\rangle=\frac{1}{\sqrt{T_F}}T^a_{\alpha\alpha'}$) so
\be
\tilde V^C_s\equiv-C_fg^2\frac{1}{{\bf k}^2}\;,\qquad \tilde V^C_o\equiv\frac{1}{2N_c}g^2\frac{1}{{\bf k}^2}
\,.
\ee

Let us now single out the following one-loop contribution for illustration
\begin{figure}[htb]
\makebox[0.0cm]{\phantom b}
\put(40,-5){\epsfxsize=6truecm \epsfbox{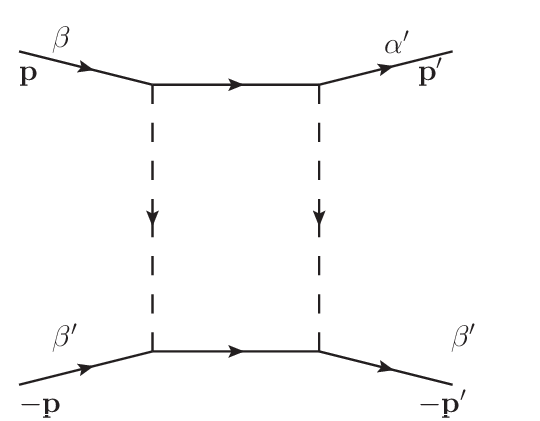}}
\put(203,50){$\displaystyle{
=I_A+I_B}=$}
\label{1loop}
\end{figure}
\be
=ig^4(T^aT^b)_{\alpha\beta}(T^bT^a)_{\beta'\alpha'}\int \frac{d^Dq}{(2\pi)^D}\frac{1}{(q-p)^2+i\eta}\frac{1}{(q-p')^2+i\eta}
\frac{i}{q^0+E/2-\frac{{\bf q}^2}{2m}+i\eta}\frac{i}{-q^0+E/2-\frac{{\bf q}^2}{2m}+i\eta}
.
\ee
To simplify the discussion we project to the singlet sector from now on: $(T^aT^b)_{\alpha\beta}(T^bT^a)_{\beta'\alpha'} \rightarrow C_f^2$. 
We have split the integral into two contributions, $I_A$ and $I_B$, 
depending on the positions of the poles of $q^0$ in the complex plane (note that the closer the pole to the origin the bigger the contribution will be):

\medskip

\noindent
{\bf A)} The leading contribution will come from the following kinematical configuration: 
$E \sim p^0 \sim p^{'0} \sim q^0 \sim mv^2, \quad|{\bf p}| \sim {\bf q} \sim 
{\bf p}' \sim mv$. We can then neglect the energy dependence in the gluon propagator and only integrate the $q^0$ poles coming from the quark propagators: 
\medskip
\begin{figure}[h]
\hspace{-0.1in}
\epsfxsize=3.8in
\put(50,50){$\displaystyle{I_A\sim \int {d^3\, {\bf q} \over (2\pi)^3} \tilde V^C_s ({\bf p},{\bf q})
{1 \over E -{\bf q}^2/m+i\eta }
\tilde V^C_s ({\bf q},{\bf p}^{\prime} )}\sim$}
\put(307,-5){\epsfxsize=5truecm \epsfbox{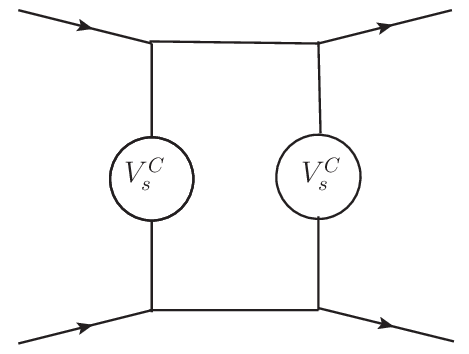}}
\end{figure}

\noindent
Note that this object is non-local in energy. Therefore, it can not be associated to a (local in time) potential (nor it should be integrated out if looking for a new EFT). Actually this term is nothing but the usual iteration of potentials in NR Quantum Mechanics. $I_A$ can then be written as
$$
I_A \sim \langle {\bf p}| {\hat V}{1 \over E- {\bf p}^2/m +
i\eta} {\hat V} | {\bf p^{\prime}} \rangle
\,.
$$
Actually it is possible to singlet out the analogous leading contribution from any ladder diagram obtaining
$$
i{\cal A} = -i \langle {\bf p}| \left({\hat V} + {\hat V} {1 \over
E- {\bf \hat p}^2/m + i\eta}  {\hat V} + \ldots \right) | {\bf
p^{\prime}} \rangle
=-
i \langle {\bf p}| {\hat V}\left( 1+  {1 \over
E- {\bf \hat p}^2/m + {\hat V}+ i\eta}   \right){\hat V}  | {\bf
p^{\prime}} \rangle
\,.
$$
For the bound state $V \sim E \sim mv^2$, 
$p \sim mv$ and one has to sum up the whole series. This is nothing but solving the Lippmann-Schwinger/Schrödinger equation. A similar discussion in the context of nuclear physics can be found in Ref. \cite{Kaplan:1996xu}.

\medskip

{\bf B)} We now consider the other possible momentum configuration:
$E \sim mv^2 \sim p^0 \sim p^{'0}$, $ |{\bf p}| \sim q^0  \sim {\bf q} \sim 
{\bf p}' \sim mv$. In this situation the poles of the quarks do not contribute to the integral (one then says that the heavy quarks are {\it off-shell}) and 
their propagators can be approximated by the static ones: ${i \over q^0+i\eta}$. Therefore,
\bea
\label{IB}
I_B&\dot=& ig^4C_f^2\int \frac{d^Dq}{(2\pi)^D}\frac{1}{(-q+p)^2+i\eta}\frac{1}{(-q+p')^2+i\eta}
\frac{i}{q^0+i\eta}\frac{i}{-q^0+i\eta}+{\cal O}\left(\frac{1}{m}\right)
\\
\nn
&\equiv&
\delta \tilde V(k)+{\cal O}(1/m)
\,.
\eea
where in the above expression it has to be understood that the poles of the quark propagators at $q_0=\pm i\eta$ do not have to be taken into account, as the corresponding contribution has already been included in $I_A$. Within Eq. (\ref{IB}) they produce the pinch singularity, which is nothing but the iteration of the Coulomb potential in the static limit, and has to be subtracted to get $\delta \tilde V(k)$. Therefore, strictly speaking,
\bea
\delta {\tilde V}&=& ig^4C_f^2\int \frac{d^Dq}{(2\pi)^D}
\left(
\frac{1}{(-q+p)^2+i\eta}\frac{1}{(-q+p')^2+i\eta}
-\frac{1}{(-{\bf q}+{\bf p})^2}\frac{1}{(-{\bf q}+{\bf p}')^2}
\right)
\frac{i}{q^0+i\eta}\frac{i}{-q^0+i\eta}
\nn
\\
&=&4C^2_Fg^2\frac{1}{{\bf k}^2}g^2{\bf k}^{2\epsilon}
\frac{\Gamma[1-\epsilon]\Gamma^2[\epsilon]}
{2^{3+2\epsilon}\pi^{2+\epsilon}\Gamma[2\epsilon]}
\,.
\label{deltaVexp}
\eea
The last equality in Eq. (\ref{deltaVexp}) can be obtained in several ways. One is using formulas like 
\be
\label{eqlambda}
\frac{1}{(q^2)^n(q\cdot v)^m}=\frac{(n+m-1)!}{(n-1)!(m-1)!}\int_0^{\infty}d\lambda\frac{2^m\lambda^{m-1}}{(q^2+2\lambda q\cdot v)^{n+m}}
\,,
\ee
familiar in the context of HQET. Other possibility is to perform the $q^0$ integration first (eliminating the quark poles as we mentioned). The resulting expression 
becomes a $D-1 \equiv d$ dimensional integral in Euclidean space that can be computed using standard one-loop formulas, as those one can find in Ref. \cite{Pascual:1984zb}. Even another possibility is performing the three-dimensional integration first. Either way, the outcome is the same. We have also seen that pinch singularities have to be subtracted to 
avoid double counting, so that Eq. (\ref{deltaVexp}) is always pinch-singularity free, yet some regularizations may set the pinch singularity to zero simplifying the computation. 

Eq. (\ref{deltaVexp}) is energy independent. Therefore, it corresponds to a correction to the potential that is ${\cal O}(\al)$ 
suppressed compared with the Coulomb one, Eq. (\ref{VCtree}).

The idea now is to quantify this discussion in the most efficient possible way (and in dimensional regularization).

\subsection{pNRQCD. How to build it? General procedure}
\label{sec:pproc}
In order to write the effective theory Lagrangian we need the following:
\begin{itemize}
\item
{\bf The degrees of freedom/cut-off}. The degrees of freedom of pNRQCD are a quark-antiquark pair,
gluons and light quarks with the cut-offs
$\nu_{\rm pNR}=\{\nu_p,\nu_{us}\}$. $\nu_p$ is the cut-off of the relative
three-momentum of the heavy quarks and $\nu_{us}$ is the cut-off of 
the energy of the heavy quark-antiquark pair and of the four momentum 
of the gluons and light quarks. They satisfy the following
inequalities: $|{\bf p}| \ll \nu_p \ll m$ and ${\bf p}^2/m \ll
\nu_{us} \ll |{\bf p}|$. 
The degrees of freedom of pNRQCD can be arranged in several ways  by using different fields. 
Two main possibilities can be considered:

\noindent
{\bf A)} Using a field for the heavy quark, $\psi(t,{\bf x})$, and another for the heavy antiquark, $\chi_c(t,{\bf x})$.
This allows a smooth connection with the NRQCD chapter.\\

\medskip

\noindent
{\bf B)} Using a single field (and then time) for the heavy quark-antiquark pair: 
\be
 \Psi ({\bf x}_1, {\bf x}_2, t)_{\alpha\beta}
 \sim
 \psi_{\alpha} ({\bf x}_1, t) \chi_{\beta}^\dagger ({\bf x}_2, t)
 \sim
 {1 \over N_c}\delta_{\alpha\beta}\psi_{\sigma} ({\bf x}_1, t)
 \chi_{\sigma}^\dagger ({\bf x}_2, t)
 +
 {1 \over T_F} T^a_{\alpha\beta}T^a_{\rho\sigma}\psi_{\sigma} ({\bf x}_1, t)
 \chi_{\rho}^\dagger ({\bf x}_2, t)
 \,.
 \label{Psipsichi}
 \ee
This can be rigorously achieved in a NR system, since particle and antiparticle numbers
are separately conserved. If we are interested in the one-heavy-quark 
one-heavy-antiquark sector, there is no loss of generality if we project our
 theory to that subspace of the Fock space, which is described by the wave
function field $\Psi ({\bf x}_1, {\bf x}_2, t)$. Furthermore, this wave
function field can be uniquely decomposed into singlet and octet field components 
(where $P$ stands for path ordered,  
${\bf R} \equiv \frac{m_1}{m_1+m_2}{\bf x}_1+\frac{m_2}{m_1+m_2}{\bf x}_2$ 
is the center position of the system, and ${\bf r}={\bf x}_1-{\bf x}_2$): 
\be
\Psi ({\bf x}_1 ,{\bf x}_2 , t)= 
P\bigl[e^{ig\int_{{\bf x}_2}^{{\bf x}_1} {\bf A} \cdot 
                   d{\bf x}} \bigr]\;{\rm S}({{\bf r}}, {{\bf R}}, t)
+P\bigl[e^{ig\int_{{\bf R}}^{{\bf x}_1} {\bf A} \cdot  d{\bf x}}
\bigr]\; {\rm O} ({\bf r} ,{\bf R} , t) \;
P\bigl[e^{ig\int^{{\bf R}}_{{\bf x}_2} {\bf A} \cdot d{\bf x}}
\bigr]
\,,
\label{SinOct}
\ee
with
homogeneous (ultrasoft) gauge transformations ($g({\bf R} ,t)$) with respect 
to the centre-of-mass coordinate:
\be
{\rm S} ({\bf r} ,{\bf R} , t)\rightarrow {\rm S} ({\bf r} ,{\bf R} , t)
\,, \qquad
O ({\bf r} ,{\bf R} , t)\rightarrow g({\bf R} ,t)O ({\bf r} ,{\bf R} ,
t)g^{-1}({\bf R} ,t)
\label{Ogaugetrans}
\,.
\ee
Using these fields has the advantage that the relative coordinate ${\bf r}$
is explicit, and hence the fact that ${\bf r}$ is much smaller than the
typical  length of the light degrees of freedom
 can be easily implemented via a multipole expansion. This implies that the gluon
fields will always appear evaluated at the centre-of-mass coordinate. Note
that this is nothing but translating to real space the constraint $\nu_p \gg
\nu_{us}$. In addition, if we restrict ourselves to the singlet field only, we
are left with a theory which is totally equivalent to a NR quantum-mechanical
Hamiltonian. Yet, the complete theory also contains the singlet-to-octet
transitions mediated by the emission of an ultrasoft gluon, which could not be
encoded in a pure NR quantum-mechanical Hamiltonian.
\item
{\bf Symmetries}. 
pNRQCD should be invariant under rotations, gauge transformations, C, P and T. The Poincare symmetry is realized nonlinearly (see \cite{Brambilla:2003nt}).
\item
{\bf Power counting (the scales of the problem)}.
 The power counting is easier to establish when the pNRQCD
Lagrangian is written in terms of singlet and
octet fields. Since quark and antiquark particle numbers are separately
conserved, the Lagrangian will be bilinear in these fields and, hence, we only
have to estimate the size of the terms multiplying those bilinears. $m$ and
$\al (m)$, inherited from the hard Wilson coefficients, have well-known
values. Derivatives with respect to the relative coordinate $i\bfnabla_{\bf
r}$ and $1/r \sim k$ (the transfer momentum) must be assigned the soft scale 
$\sim |{\bf p}|$. 
Time derivatives
$i\partial_0$, centre-of-mass derivatives $i\bfnabla_{\bf R}$, and the fields
of the light degrees of freedom must be assigned the ultrasoft scale $E\sim
{\bf p}^2/m$. The $\al$ arising in the matching calculation from NRQCD, namely
those in the potentials, must be assigned the size $\al (1/r)$ and those
associated with the light degrees of freedom (gluons, at lower orders) the size $\al
(E)$, though smaller scales could appear.
\item
{\bf Matching}. The Wilson coefficients of pNRQCD are determined imposing suitable NRQCD and pNRQCD Green's functions to be equal. The Wilson coefficients of each operator depend logarithmically on
$1/r$, $\nu_{\rm pNR}$ and can be calculated
in perturbation theory in $\al (\nu_{\rm pNR})$. 
\end{itemize}

\subsection{pNRQCD Lagrangian. Different formulations}

\noindent
{\bf pNRQCD Lagrangian with quark and antiquark fields}.\\
As we have seen, the degrees of freedom of pNRQCD can be arranged in several ways and accordingly the pNRQCD Lagrangian. We first write it in terms of quarks and gluons. It reads
\be
L_{\rm pNRQCD}=L_{\rm NRQCD}^{\rm US}+L_{\rm pot}
\,,
\label{lagpnrqcda}
\ee
where $L_{\rm NRQCD}^{\rm US}$ has the form of the NRQCD Lagrangian, 
but all the gluons must be understood as ultrasoft.
$L_{\rm pot}$ reflects one of the 
most distinct features of the pNRQCD Lagrangian: the appearance of non-local in $r$ (space) Wilson coefficients for the 
4-fermion operators after integrating out 
the mv scale (note though that the Lagrangian is still local in time as it should)\footnote{We do not consider six-fermion operators and so on. They are not 
necessary if we restrict to situations with a single Heavy Quarkonium, but they could be needed in other processes like 
charmonium-charmonium scattering. If the center-of-mass velocities were different one 
should then proceed as in HQET, adding a velocity label for each heavy quark with different velocity.}:
\be
\label{lpot}
L_{\rm pot} = -\int d^3{\bf x}_1 d^3{\bf x}_2 \;
\psi^{\dagger} (t,{\bf  x}_1)  \chi(t,{\bf  x}_2)
\;
V({\bf r}, {\bf p}_1, {\bf p}_2, {\bf S}_1,{\bf S}_2)
(\hbox{ultrasoft gluon fields}) 
\;
\chi^\dagger(t,{\bf  x}_2) \psi(t,{\bf  x}_1)
\,,
\ee
where ${\bf p}_j = -i \bfnabla_{{\bf x}_j}$ and ${\bf S}_j= {\bfsigma}_j/2$, for $j = 1,2$,  
act on the fermion and antifermion, respectively 
(the fermion and antifermion spin indices are contracted with the 
indices of $V$, which are not explicitly displayed). Typically, ultrasoft 
gluon fields show up at higher order, making possible to create gauge-independent structures. 
For instance, the inclusion of ultrasoft gluons would change the Coulomb potential term in the following way:
\be
\label{lpotcoulomb}
\delta L_{pot}=
-{g^2\over2}
\int d^3{\bf  x}
\int d^3{\bf  y}
\bar Q\gamma^0T^{r}Q({\bf x},t) \left(  {1\over {\bf D}^{2}} 
\right)^{rs}({\bf x}, {\bf y},t)
\bar Q\gamma^0T^{s}Q({\bf y},t)
.
\ee

\medskip

\noindent
{\bf Projecting to  the one-heavy-quark one-heavy-antiquark sector}.\\
We have seen that the degrees of freedom of pNRQCD can be represented by the same fields as in NRQCD.
This representation is suitable (in some cases) for explicit
perturbative matching calculations using standard 
Feynman diagram techniques. On the other hand, for the study of the Heavy Quarkonium 
it is convenient, before calculating physical quantities,  
to project the Lagrangian in Eq. (\ref{lagpnrqcda}) onto the quark-antiquark sector 
of the Fock space (actually these projections could also be done in NRQCD). This 
makes easier to establish the power counting rules and, in particular, to make explicit the 
multipole expansion at the Lagrangian level. It is also useful 
when the matching is made via Wilson loops.
The projection onto the quark-antiquark sector is easily  done at the
Hamiltonian level by projecting onto the subspace spanned by
\be
\int d^3{\bf x}_1 d^3{\bf x}_2 \, \Psi ({\bf x}_1,{\bf x}_2 ) \, \psi^{\dagger}({\bf x}_1)
\chi({\bf x}_2) \vert \hbox{ultrasoft gluons} \rangle
\,,
\ee
where $\vert \hbox{ultrasoft gluons}\rangle$ is a generic state belonging to 
the Fock subspace with no quarks and antiquarks but an arbitrary number of
ultrasoft gluons. The pNRQCD Lagrangian then has the form:
\bea
\label{lagpnrqcdb} 
L_{\rm pNRQCD} &=&
\int d^3{\bf x}_1 \, d^3{\bf x}_2 \; 
{\rm Tr}\, \left\{\Psi^{\dagger} (t,{\bf
  x}_1 ,{\bf x}_2 ) \left(
iD_0  +{{\bf D}_{{\bf x}_1 }^2\over 2\, m_1}+{{\bf D}_{{\bf x}_2 }^2\over 2\,
  m_2} + \cdots \right)\Psi (t,{\bf x}_1 ,{\bf x}_2 )\right\}
\\
&-&
\int d^3 x \; {1\over 4} G_{\mu \nu}^{a}(x) \,G^{\mu \nu \, a}(x) + 
\int d^3 x \;
\sum_{i=1}^{n_f} \bar q_i(x) \, i \dsl \,q_i(x)
+ \cdots
\nn
\\
\nn
&+& 
\int d^3{\bf x}_1 \, d^3{\bf x}_2 \; 
{\rm Tr} \left\{ \Psi^{\dagger} (t,  {\bf x}_1,{\bf x}_2)\,
V( {\bf r}, {\bf p}_1, {\bf p}_2, {\bf S}_1,{\bf S}_2)
(\hbox{ultrasoft gluon fields}) \,
\Psi(t, {\bf x}_1, {\bf x}_2 )
\right\},
\eea
where the first two lines stand for the NRQCD Lagrangian  
projected onto the quark-antiquark sector and 
\be
iD_0 \Psi (t,{\bf x}_1 ,{\bf x}_2)
= i\partial_0\Psi (t,{\bf x}_1 ,{\bf x}_2 ) -g A_0(t,{\bf x}_1)\,  
\Psi (t,{\bf x}_1 ,{\bf x}_2) + \Psi (t,{\bf x}_1 ,{\bf x}_2)\, g A_0(t,{\bf
  x}_2).
\ee 
The dots in Eq.~(\ref{lagpnrqcdb}) stand for higher terms in the $1/m$ expansion.
The last line contain the 4-fermion terms specific of pNRQCD. The expression for $V$ is 
equal to the one in Eqs. (\ref{lpot}), (\ref{lpotcoulomb}).

Since the gluons are ultrasoft we can multipole expand them in ${\bf r}$: 
\be
\label{multipoleA}
A_{\mu}({\bf x}_{1(2)},t)=A_{\mu}({\bf X})+(-)
\frac{m_{2(1)}}{m_1+m_2}{\bf x}\cdot\bfnabla A_{\mu}({\bf X},t)+\cdots \,.
\ee  
Eq. (\ref{lpotcoulomb}) then becomes the Coulomb potential at leading order in the multipole expansion:
\be
{\al \over |{\bf x}_1 - {\bf x}_2|}\, {\rm Tr} \, \biggl( 
T^{a} \, \Psi^{\dagger}  (t,{\bf x}_1 ,{\bf x}_2 ) \, T^{a}\Psi(t,{\bf x}_1 ,
{\bf x}_2 )\biggr),
\label{coulcol}
\ee
and, according to the power counting rules given in Sec. \ref{sec:pproc}, the multipole
expansion makes explicit the size of each term in the Lagrangian. 
On the other hand applying Eq. (\ref{multipoleA}) spoils the manifest gauge invariance of the Lagrangian. 
This may be restored by introducing singlet and octet fields as in 
Eq.~(\ref{SinOct}), which we do next\footnote{Obviously the restoration of the gauge symmetry happens in any gauge, yet it is 
worth mentioning the ${\bf x}\cdot {\bf A}({\bf X}+{\bf x},t)=0$ gauge, where the gluon links in Eq. (\ref{SinOct}) vanish and 
the multipole expansion in Eq. (\ref{multipoleA}) can be written in a gauge invariant manner:
\be
A^0({\bf X}+{\bf x},t)=A^0({\bf X},t)+\sum_{n=0}^{\infty}\frac{x^rx^{r_1}\ldots x^{r_n}}{(n+1)!}
[{\bf D}_{\bf X}^{r_1},[{\bf D}_{\bf X}^{r_2}\ldots[{\bf D}_{\bf X}^{r_n},G_{r 0}({\bf X},t)]]\ldots]
=A^0({\bf X},t)+\int_0^1du{\bf r}\cdot{\bf E}({\bf X}+u{\bf x},t)
\,,
\ee
\be
{\bf A}^i({\bf X}+{\bf x},t)=-\sum_{n=0}^{\infty}\frac{x^rx^{r_1}\ldots x^{r_n}}{n!(n+2)}
[{\bf D}_{\bf X}^{r_1},[{\bf D}_{\bf X}^{r_2}\ldots[{\bf D}_{\bf X}^{r_n},G^{r i}({\bf X},t)]]\ldots]
=
-\int_0^1du{\bf r}^jG^{ji}({\bf X}+u{\bf x},t)
\,.
\ee
}.

\medskip

\noindent
{\bf pNRQCD Lagrangian with Singlet and Octet fields}.\\
We choose the following normalization with respect to color:
\begin{equation}
{\rm S} \equiv { 1\!\!{\rm l}_c \over \sqrt{N_c}} S\,, \qquad {\rm O} \equiv  { T^a \over \sqrt{T_F}}O^a, 
\label{norm}
\end{equation}
to have the proper free field normalization in color space.
We will not always explicitly display their dependence on ${\bf R}$, {\bf r} and $t$
in the following.
After multipole expansion, the pNRQCD Lagrangian (density) may be organized as an
expansion in $1/m$ and $r$ (and $\al$). 
Up to NLO in the multipole expansion
it reads~\cite{Pineda:1997bj,Brambilla:1999xf}:
\begin{eqnarray}
& & {\cal L}_{\rm us} =
{\rm Tr} \Biggl\{ {\rm S}^\dagger \left( i\partial_0  - h_s(r)   \right) {\rm S} 
 + {\rm O}^\dagger \left( iD_0 - h_o(r)  \right) {\rm O} \Biggr\} 
\nonumber\\
& &\qquad
+ g V_A ( r) {\rm Tr} \left\{  {\rm O}^\dagger {\bf r} \cdot {\bf E} \,{\rm S}
+ {\rm S}^\dagger {\bf r} \cdot {\bf E} \,{\rm O} \right\} 
  + g {V_B (r) \over 2} {\rm Tr} \left\{  {\rm O}^\dagger \left\{{\bf r} \cdot {\bf E} , {\rm O}\right\}\right\} ,
\label{pnrqcd0}
\end{eqnarray}
plus the gluon and light fermion terms, which do not change with respect to Eq. (\ref{lagpnrqcdb}). 
All gluon and scalar fields in Eq. (\ref {pnrqcd0}) are evaluated 
in ${\bf R}$ and the time $t$, in particular the chromoelectric field ${\bf E} \equiv {\bf E}({\bf R},t)$ and the ultrasoft covariant derivative
$iD_0 {\rm O} \equiv i \partial_0 {\rm O} - g [A_0({\bf R},t),{\rm O}]$. We will not always display this dependence.

 $h_s$ can be split in the kinetic term and the potential: 
\bea
& &
h_s({\bf r}, {\bf p}, {\bf S}_1,{\bf S}_2) = 
{{\bf p}^2 \over 2\, m_r}+ 
V_s({\bf r}, {\bf p}, {\bf S}_1,{\bf S}_2)\equiv h^C_s+\delta h_s 
,
\\
& & 
h_o({\bf r}, {\bf p}, {\bf S}_1,{\bf S}_2) = 
 {{\bf p}^2 \over 2\, m_r}  
 + 
V_o({\bf r}, {\bf p}, {\bf S}_1,{\bf S}_2)\equiv h^C_o+\delta h_o ,
\eea
where $h^C_{s/o}={{\bf p}^2 \over 2\, m_r}+V^C_{s/o}$, $m_r=m_1m_2/(m_1+m_2)$ and ${\bf p}=-i\bfnabla_{\bf r}$. 
$h^C$ represents the LO Hamiltonian, as its power counting is $h^C \sim {\cal O}(mv^2)$ (LO) and $\delta h \sim mv^3$ (NLO) at least. In general $\delta h \sim 1/m^{n-s-1} p^nr^s\al^q \sim {\cal O}(m\al^{n+q-s})$ (at least). If the resummation 
of logarithms is incorporated in the potentials the counting changes to LO $\rightarrow$ LL, NLO $\rightarrow$ NLL, and so on. 

For the equal mass case: $m_1=m_2=m$, the potential has the following structure at ${\cal O}(1/m^2)$
(except for the static potential we drop the labels $s$ and $o$ for the singlet and octet, which have to be understood implicitly)\footnote{
The ${\cal O}(1/m^3)$ contribution
\be
\label{deltahm3}
\delta h=-\frac{c_4}{4m^3}{\bf p}^4
\ee
has to be introduced for nowadays precision, as $c_4$ is not suppressed by any power of $\al$.}:
\bea
V_{s/o}(r)&=&V_{s/o}^{(0)}(r) +{V^{(1)}(r) \over m}+{V^{(2)} \over m^2}+\cdots,  
\label{ppot}                                                                      
\\
V^{(2)}&=&V^{(2)}_{SD}+V^{(2)}_{SI},\nn\\
V^{(2)}_{SI}
&=&                                   
{1 \over 2}\left\{{\bf p}^2,V_{{\bf p}^2}^{(2)}(r)\right\}
+{V_{{\bf L}^2}^{(2)}(r)\over r^2}{\bf L}^2 + V_r^{(2)}(r),
\\
\label{VSD}
V^{(2)}_{SD} &=&
V_{LS}^{(2)}(r){\bf L}\cdot{\bf S} + V_{S^2}^{(2)}(r){\bf S}^2
 + V_{{\bf S}_{12}}^{(2)}(r){\bf S}_{12}({\hat {\bf r}}), 
\eea
where we split the $1/m^2$ potential into the spin-dependent (SD) and spin-independent (SI) terms, ${\bf S}={\bf S}_1+{\bf S}_2$, ${\bf L}={\bf r}\times {\bf p}$, and ${\bf S}_{12}({\hat {\bf r}}) \equiv 
3 {\hat {\bf r}}\cdot \bfsigma_1 \,{\hat {\bf r}}\cdot \bfsigma_2 - \bfsigma_1\cdot \bfsigma_2$.

Other forms of the potential can be brought to the one above by using 
unitary transformations, or the relation
\be
- \left\{ {1 \over r},{\bf p}^2 \right\} + 
{1 \over r^3}{\bf L}^2 + 4\pi\delta^{(3)}({\bf r})
= - {1 \over r} \left( {\bf p}^2 + 
{ 1 \over r^2} {\bf r} \cdot ({\bf r} \cdot {\bf p}){\bf p} \right).
\ee
Note that this equality is true in four dimensions. For the equivalent one in D dimensions is better to work in momentum space, where the angular
momentum operator is generalized to
\begin{equation}
\frac{{\bf L}^2}{2\pi r^3} \to 
\left(\frac{{\bf p}^2-{\bf p}^{\prime\,2}}{{\bf q}^2}\right)^2 - 1
\label{lddef}
\end{equation}
to be compatible with $D$ dimensional calculations in momentum
space. 

Schematically, we can also write the pNRQCD Lagrangian as an expansion in 
$r$ (which we use as a shorthand notation of $r$ and  $1/p$. Note though that only positive powers of $p$ appear, and with maximum power ${\bf p}^{n+1}/m^n$, coming from the expansion of the kinetic term) and $1/m$ in the following way
\be
\label{lagalpha}
{\cal L}_{\rm pNRQCD} =\sum_{n=-1}^{\infty}r^n{\al^{(0,n)}_V} O_n
+{1 \over m}\sum_{n=-2}^{\infty}r^n{\al^{(1,n)}_V}O_n^{(1)}
+{\cal O}\left({1 \over m^2}\right),
\ee
where $\al^{(s,n)}_V$ are dimensionless (in four dimensions) constants.
Since they absorb the divergences of the EFT, they will depend on $\nu_{\rm pNR}$.
This form of ${\cal L}_{\rm pNRQCD}$ will be used to discuss the general structure of the RG in secs. \ref{sec:RGUS}, \ref{sec:RGsoftp}, 
\ref{sec:RGpot}.

Equations (\ref{lagpnrqcda}), (\ref{lagpnrqcdb}) and (\ref{pnrqcd0}) provide
three different ways to write the pNRQCD Lagrangian. We have also shown how to
derive one from the other. This works (and is useful) at tree level.  In
general, each form of the pNRQCD Lagrangian may be constructed independently of the others by
identifying the degrees of freedom, using symmetry arguments and matching directly to NRQCD, where we 
turn next.

\subsection{Matching NRQCD to pNRQCD}
\label{pNRmatchingI}

The pNRQCD Wilson coefficients, mainly the potentials, are obtained by  
matching NRQCD with pNRQCD. This matching can be done order by order in $\al$, and in several ways. 
The main difference comes from the operators used for the 
matching. Using Wilson loops is specially suitable if we want to match directly to Eq. (\ref{pnrqcd0}). This we will not
discuss in this review (see Ref. \cite{Brambilla:2004jw}). If we are interested in a diagrammatic approach, specially convenient for perturbative computations, it is quite suitable 
to match NRQCD and pNRQCD in the formulation of Eq.~(\ref{lagpnrqcda}). We can still devise two ways 
to achieve our goal:

\medskip

\noindent
{\bf A) Off-shell matching}. It goes along the
lines of \cite{Pineda:1997bj,Pineda:1997ie,Pineda:1998kn}. The philosophy is pretty much similar 
to the one of Sec. \ref{secmatchingNRQCD} but demanding off-shell Green functions in NRQCD and pNRQCD to be
equal. The reason is that we are eventually
interested in bound states, 
and the LO equation of motion of pNRQCD 
includes the Coulomb potential. Therefore, quarks and gluon fields in a bound state do 
not satisfy the equations of motion of free particles. We then obtain the Wilson coefficients of pNRQCD by enforcing
2- and 4-fermion Green functions with arbitrary ultrasoft external gluons to
be equal to those of NRQCD at any desired order in $E/k$, where $E$ generically
denotes the external momentum or the kinetic term ${\bf p}^2/m$. Note 
that this implies the use of static (HQET) propagators for the fermions.
By expanding the energy of the external quark and the energy and momenta
of the ultrasoft gluons around zero {\it before} carrying out the loop integrals 
the integrals become homogeneous in the soft scale and hence easier to
evaluate. This may produce infrared divergences, which are most conveniently (but
not necessarily) regulated in dimensional regularization, in the same way as the ultraviolet divergences
are. Since the infrared behavior of NRQCD and pNRQCD is the same, these infrared
divergences will cancel out in the matching, provided the same infrared regulator is
used in both theories. The ultraviolet divergences of NRQCD must be renormalized in the
$\MS$ scheme if we want to use the Wilson coefficients of the NRQCD Lagrangian
computed themselves in the $\MS$ scheme. We still have a choice in the
renormalization scheme of pNRQCD. However, it is most advantageous to use
again the $\MS$ scheme. Indeed, with this choice we can blindly subtract any
divergence in the matching calculation regardless of whether it is ultraviolet or infrared. For the ultraviolet
divergences of NRQCD and pNRQCD, this just corresponds to our choice of scheme, and for the
infrared divergences this is possible since, as long as we use the same treatment in both
theories, their infrared behavior is the same. This allows to set integrals with
no scale equal to zero.

In short, loops in NRQCD will have the structure 
\be
\int d^Dq \, f(q,k,E)= \int d^Dq \, f(q,k,0)+ {\cal O}\left({E \over k} \right)
\,,
\ee
whereas in pNRQCD:
\be
\int d^Dq \, f(q,E)= \int d^Dq \, f(q,0)+{\cal O} \left({E \over k} \right)= 0\,,
\ee
and we can directly identify the (renormalized) potentials from a
 calculation in NRQCD. We would like to stress again the similarity in 
the procedure
 with the matching between QCD and NRQCD as carried out in Sec. \ref{secmatchingNRQCD}.
 The potentials in pNRQCD play the role of Wilson
coefficients in the matching procedure. We also note that one can obtain the bare expressions for the 
potential if the ultraviolet divergences are known (which in many cases can be obtained with little effort).

As a summary for the practitioner, the final set of rules are the following:

\begin{itemize}
\item
Compute (off-shell) NRQCD Feynman diagrams within an expansion in
$\al$, $1/m$ and $E$. In case loops appear, they have to be computed using
static propagators for the heavy quark and antiquark, which makes the
integrals depend on $k$ only.
\item
Match the resulting expression to the {\it tree level} expression in pNRQCD
(i.e. the potentials that appear in the pNRQCD Lagrangian) to the required
order in $\al$, $1/m$ and $E$.
\item
In case pinch singularities appear, one must isolate them in expressions which are identical 
to those which appear in the pNRQCD computation and set them to zero. Alternatively, one may just
subtract the pNRQCD diagrams with the same pinch singularity, as illustrated in Sec. \ref{sec:mot}.
\end{itemize}

Let us mention here, that when this procedure is used to match local NRQCD
4-fermion operators or local currents, these do not get any loop correction.  Indeed, due to the
use of HQET propagators, all NRQCD integrals become scaleless and hence
vanish. We often say that they are {\it inherited} in pNRQCD.

\medskip

\noindent
{\bf B) On-shell matching}. One can also perform the matching using (free) on-shell quarks (see for instance  \cite{Kniehl:2001ju,Kniehl:2002br}). 
We can not then use the set of rules 
described for the off-shell matching. In particular loops in pNRQCD do not vanish (since the energy is not
left as a free parameter in which one can expand) and have to be
subtracted accordingly. In addition, the on-shell condition may set 
to zero some terms in the (off-shell) potential. When these terms 
enter in a NRQCD subdiagram of a higher-loop matching calculation, 
they may give rise to new contributions to the potential due to 
quark potential loops. 

\medskip

Different matching procedures may lead to different potentials. This is not necessarily wrong as 
far as they can be related by unitary transformations. In particular these transformations can be 
used to remove time derivatives in higher order terms and to write the
pNRQCD Lagrangian in a standard form. 

\subsection{The potential}

Out of the previous matching computation one gets the potentials with $\nu_p=\nu_{us}\equiv \nu$. 
This is not a limitation. In fixed order computations the factorization scale dependence (no matter its origin) vanishes 
once all contributions to an observable are added. If one wants to resum logarithms, the expressions obtained are the initial condition of the RG equations.

We now present the Wilson coefficients with N$^3$LO precision and in a way suitable for a RG analysis.
We only consider the equal mass case, since the ${\cal O}(1/m^2)$ Wilson coefficients at one loop are not fully known in the non-equal mass case (except for QED, see for instance \cite{Pineda:1998kn}), though many partial results exist 
\cite{Gupta:1981pd,Gupta:1982qc,Buchmuller:1981aj,Pantaleone:1985uf,
Titard:1993nn,Pineda:1998kn,Brambilla:1999xj,Manohar:2000hj,Kniehl:2002br}.
We now study each potential separately. Note that the result will depend on the basis 
of potentials used and on field redefinitions except for the singlet static potential, which is 
unambiguous\footnote{This observation also applies to the RG corrections to the potentials obtained in 
Sec. \ref{Renormalization}.}.

The initial matching conditions for the static potential in our $\MS$ scheme reads 
\begin{eqnarray}
V^{(0)}_{s,\MS}(r;\nu)
&=&
 -\frac{C_f\,\al(\nu)}{r}\,
\bigg\{1+\sum_{n=1}^{3}\left(\frac{\al(\nu)}{4\pi}\right)^n a_n(r;\nu)\bigg\}
\end{eqnarray}
with coefficients
\begin{eqnarray}
a_1(r;\nu)
&=&
a_1+2\beta_0\,\ln\left(\nu e^{\gamma_E} r\right)
\,,
\nonumber\\
a_2(r;\nu)
&=&
a_2 + \frac{\pi^2}{3}\beta_0^{\,2}
+\left(\,4a_1\beta_0+2\beta_1 \right)\,\ln\left(\nu e^{\gamma_E} r\right)\,
+4\beta_0^{\,2}\,\ln^2\left(\nu e^{\gamma_E} r\right)\,
\,,
\nonumber \\
a_3(r;\nu)
&=&
a_3+ a_1\beta_0^{\,2} \pi^2+\frac{5\pi^2}{6}\beta_0\beta_1 +16\zeta_3\beta_0^{\,3}
\nonumber \\
&+&\bigg(2\pi^2\beta_0^{\,3} + 6a_2\beta_0+4a_1\beta_1+2\beta_2+\frac{16}{3}C_A^{\,3}\pi^2\bigg)\,
  \ln\left(\nu e^{\gamma_E} r\right)\,
\nonumber \\
&+&\bigg(12a_1\beta_0^{\,2}+10\beta_0\beta_1\bigg)\,
  \ln^2\left(\nu e^{\gamma_E} r\right)\,
+8\beta_0^{\,3}  \ln^3\left(\nu e^{\gamma_E} r\right)\,
\,.
\label{eq:Vr}
\end{eqnarray}
The ${\cal O}(\al)$ term was computed in Ref. \cite{Fischler:1977yf}, the ${\cal O}(\al^2)$  in Ref. \cite{Schroder:1998vy}, 
the ${\cal O}(\al^3)$ logarithmic term in Refs. \cite{Brambilla:1999qa,Kniehl:1999ud}, the light-flavour finite piece in Ref. 
\cite{Smirnov:2008pn}, and the pure gluonic finite piece in Refs. \cite{Anzai:2009tm,Smirnov:2009fh}. 
We display the explicit expressions for $a_i$ in the Appendix. 

For the $1/m$ potential the initial matching condition reads
\bea
\label{V1MS}
V_{\MS}^{(1)}(r;\nu)
&=&
\frac{C_f\al^2(e^{-\gamma_E}/r)}{2r^2}
\left(
\frac{C_f}{2}-C_A+\frac{\al(1/r)}{\pi}\left[-\frac{2}{3}(C_A^2+2C_AC_f)\ln(\nu^2 r^2e^{2\gamma_E})
\right.
\right.
\nn
\\
&&
\left.
\left.
-\frac{89}{36}C_A^2+\frac{17}{18}C_AC_f+\frac{49}{36}C_AT_Fn_f-\frac{2}{9}C_fT_Fn_f
\right]
\right)
\,.
\eea
The ${\cal O}(\al^3)$ log-dependent term was computed in Refs. \cite{Kniehl:1999ud,Brambilla:1999xj}. 
The log-independent ${\cal O}(\al^3)$ result has been taken from Ref. \cite{Kniehl:2001ju} and changed accordingly to fit our renormalization scheme for the 
ultrasoft computation. This explains the difference between its expression in Eq. (\ref{V1MS}) and in Ref. \cite{Kniehl:2001ju}. Ours is the proper 
one to be combined with the ultrasoft correction obtained in Eq. (\ref{deltaEusMSnl}) in the next section.

For the momentum-dependent $1/m^2$ potential the Wilson coefficient reads at one loop 
\be
V_{p^2,\MS}^{(2)}(r;\nu) 
=
\frac{ C_f\al(e^{-\gamma_E}/r)}{4r}
\left(
-4+\frac{\al(1/r)}{\pi}\left(-\frac{31}{9}C_A+\frac{20}{9}T_Fn_f-\frac{8}{3}C_A\ln(\nu^2 r^2e^{2\gamma_E})
\right)
\right)
\,,
\ee
where the ${\cal O}(\al^2)$ log-dependent term was computed in Ref. \cite{Kniehl:1999ud,Brambilla:1999xj} and the log-independent ${\cal O}(\al^2)$ term in Ref. \cite{Kniehl:2002br}. On the other hand $V_{L^2}^{(2)}=0$ at this order for this choice 
of Wilson coefficients.

$V_r$ depends logarithmically on the mass of the heavy quark through the Wilson coefficients 
inherited from NRQCD. It is convenient to write it 
in terms of the potential in momentum space (otherwise ill-defined distributions appear)
\be
\label{Vrmatching}
V_{r,\MS}^{(2)}(r;\nu)
=
\int \frac{d^3q}{(2\pi)^3}e^{-i{\bf q}\cdot {\bf r}}
\tilde V_{r,\MS}^{(2)}(q;\nu)
\,,
\ee
where
\bea
\tilde V_{r,\MS}^{(2)}(q;\nu)
&=&\pi C_f
\Bigg[
\al(q)(1+c_D(\nu)-2c_F^2(\nu))
\\
\nn
&&
\left.
+\frac{1}{\pi}(d_{vs}(\nu,\nu)+3d_{vv}(\nu,\nu)+\frac{1}{C_f}(d_{ss}(\nu,\nu)+3d_{sv}(\nu,\nu)))
 +\delta \tilde V_{soft}(\nu,q)
\right]
\,,
\eea
\be
\delta \tilde V_{soft}=\frac{\al^2}{\pi}
\left[
\left(\frac{9}{4}+\frac{25}{6}\ln\frac{\nu^2}{q^2}\right)C_A
+
\left(\frac{1}{3}-\frac{7}{3}\ln\frac{\nu^2}{q^2}\right)C_f
\right]
\,.
\ee
The abelian term of $\delta \tilde V_{soft}$ was computed in Ref. \cite{Pineda:1998kn}, the logarithmic term in Refs. \cite{Kniehl:1999ud,Brambilla:1999xj} and its complete expression in Ref. \cite{Kniehl:2002br}. The rest encodes the hard part through the non-trivial dependence on $m$ of the NRQCD matching coefficients. 
Upon expanding the NRQCD Wilson coefficients in powers of 
$\al(\nu)$ the above expression agrees with the hard contribution of Ref.  \cite{Kniehl:2002br} at ${\cal O}(\al^2)$.
In order to fit with their scheme 
we had to change the scheme for the Pauli ${\bfsigma}$ matrices 
with respect the computation performed in Ref. \cite{Pineda:1998kj}
 for $d_{vv}$ (see also the discussion in Refs. \cite{Manohar:2000hj,Pineda:2000sz}). The new expression for $d_{vv}$ can be found in the 
Appendix, as well as the other NRQCD Wilson coefficients.

The spin-dependent potentials at ${\cal O}(1/m^2)$ and at one loop have been computed before \cite{Gupta:1981pd,Gupta:1982qc,Buchmuller:1981aj,Pantaleone:1985uf}. Within the pNRQCD framework they read 
\be
V_{LS,\MS}^{(2)}(r;\nu) = \frac{3C_f}{2}\frac{1}{r^3}
\al(e/r)
\left\{
\frac{c_S(\nu)+2c_F(\nu)}{3}
+
\frac{\al}{\pi}
\left[
\left(\frac{7}{36}-\frac{2}{3}\ln(\nu r/e^{4/3})\right)C_A-\frac{5}{9}T_Fn_f
\right]
\right\}
,
\ee
\be
V_{S_{12},\MS}^{(2)} (r;\nu)= \frac{C_f}{4}\frac{1}{r^3}
\al(e^{4/3}/r)
\left\{
c_F^2(\nu)
+
\frac{\al}{\pi}
\left[
\left(\frac{13}{36}-\ln(\nu r/e)\right)C_A-\frac{5}{9}T_Fn_f
\right]
\right\}
.
\ee
$V_{S^2,\MS}^{(2)}$ suffers from the same disease as $V_r$. It is convenient to write it first 
in terms of the potential in momentum space (otherwise ill-defined distributions appear)
\be
\label{VS2matching}
V_{S^2,\MS}^{(2)}(r;\nu)
=
\int \frac{d^3q}{(2\pi)^3}e^{-i{\bf q}\cdot {\bf r}}
\tilde V_{S^2,\MS}^{(2)}(q;\nu)
\,,
\ee
\bea
\tilde V_{S^2,\MS}^{(2)}(q;\nu)
&=&
\frac{4\pi}{3}C_f
\left\{
\al(q)c_F^2(\nu)
-\frac{3}{2\pi C_f}\left(d_{sv}(\nu,\nu)+C_f d_{vv}(\nu,\nu)\right)
\right.
\nn
\\
&&
\left.
-
\frac{\al^2}{\pi}
\left[
\left(\frac{7}{18}+\frac{7}{4}\ln(\nu/q)\right)C_A+\frac{5}{9}T_Fn_f
\right]
\right\}
\,.
\eea

For the off-shell matching the individual potentials can be 
related with gauge invariant Wilson loops, see Refs. \cite{Eichten:1980mw,Brambilla:2000gk,Pineda:2000sz}. 
In the case of the on-shell matching the result should also be 
gauge invariant, as one computes on-shell matrix elements, which are gauge invariant. Yet,
as we have already mentioned the $1/m$ and $1/m^2$ spin-independent potentials 
suffer from field redefinitions ambiguities. Therefore, in some circumstances it can be convenient to cast the
initial matching conditions of $V_s$ in the unified form:
\be
V_{s,\MS}(r;\nu)=V^{(0)}_{s,\MS}(r;\nu)
+
\frac{V^{(1)}_{\MS}(r;\nu)}{m}+
\frac{1}{m^2}
\left(
{1 \over 2}\left\{{\bf p}^2,V_{{\bf p}^2,\MS}^{(2)}(r;\nu)\right\}
 + V_{r,\MS}^{(2)}(r;\nu)+V^{(2)}_{SD,\MS}(r;\nu)
\right)
.
\ee
In this unified form the potential is explicitly gauge independent. For an explicit and illustrative 
check of the gauge independence of the hamiltonian see Appendix A of \cite{Pineda:1998kn}.

One also has to consider soft corrections to $V_A$. They have been studied  in Ref. \cite{Brambilla:2006wp} with NLO accuracy reaching to the conclusion that $V_A=1+{\cal O}(\al^2)$. So we can set
$V_A=1$.

We have written the potentials in position space. We can transform them to momentum space using the 
Fourier transform formulas displayed in the Appendix. 

\section{Renormalization (Group) in pNRQCD}
\label{Renormalization}

We have finally obtained the pNRQCD Lagrangian and the renormalized expressions for its Wilson coefficients, ie. for $\al$ and $V_{\{s,o,A,B\}}(r)$ in terms 
of the underlying theory, QCD. Their bare counterparts will be generically denoted by $V_B$, and $\al_B$ ($g_B$).
Those bare Wilson coefficients are in charge of absorbing the divergences produced by ultrasoft and/or potential loops in the EFT. 
From now on we will use the index ``$B$'' to explicitly denote bare quantities. Parameters without this index are understood to be renormalized. 
 As we have seen in the previous section 
their renormalized counterparts are fixed 
 at the scale $\nu$ where the matching has been performed.

In our convention 
$\al_B$ is dimensionless and related to $g_B$ by
\be
\alpha_B=\frac{g_B^2\nu_{us}^{2\epsilon}}{4\pi}\,.
\ee
It has a special status since 
it does not receive corrections from other Wilson coefficients of the 
effective theory. Its divergences are of pure ultrasoft origin and 
it can be renormalized multiplicatively:
\be
\al_B=Z_{\al}\al
\,,
\ee
where
\be
Z_{\al}
=1+
\sum_{s=1}^{\infty}Z^{(s)}_{\al}\frac{1}{\epsilon^s}\,,
\ee
and $Z_\al$ is equal to the standard one of QCD.  

The bare potentials $V_B$ in position space have integer mass dimensions (note that this is not so in 
momentum space) and, due to the structure of the theory, we do not renormalize them multiplicatively
(see the discussion in Ref.~\cite{Pineda:2000gza}). We define
\be
\label{VBsplitting}
V_B=V+\delta V\,.
\ee
$\delta V$ depends on the Wilson coefficients of the effective theory, i.e. on $\al$ and $V$, 
and on the number of space-time dimensions.
In $D$ dimensions, using the MS renormalization scheme, we define
\be
\delta V
=
\sum_{s=1}^{\infty}Z^{(s)}_{V}\frac{1}{\epsilon^s}\,,
\ee
where the divergences could be of both ultrasoft and potential origin. Nevertheless, whereas all potentials will eventually absorb 
ultrasoft divergences, some potentials, such as the static and $1/m$ potential, are free of potential divergences.

In order to get $\delta V$ (and illustrate how all the divergences can be absorbed in the potential), we compute 
the NR propagator (Green function) of a quark-antiquark pair and the gluonic vacuum. We will focus here on the singlet sector: 
\be
\label{Pirrp}
G(E,{\bf r},{\bf r}') \equiv i\int d t \, d^3{\bf R} \,e^{iEt}
\langle {\rm vac} |
T\{ S({\bf r}', {\bf 0}, 0)
S^{\dagger}({\bf r}, {\bf R}, t) \}
| {\rm vac}\rangle= \langle {\bf r}'|G(E)|{\bf r} \rangle\,,
\ee
\be
\label{Gsfull}
G(E)\equiv P_s \langle {\rm vac} |{1 \over H-E-i\eta} | {\rm vac} \rangle P_s=G_c(E)+\delta G
\,,
\ee
where $H$ is the pNRQCD Hamiltonian, $P_s$ the projector to the singlet sector, $G_c$ the Coulomb Green function, 
defined in Fig.~\ref{pnrqcdfig} of Appendix \ref{sec:pNRFR}, and $E$ the energy measured from the threshold $2m$. This will be a key quantity for this review.
If we work at NLO in the multipole expansion the singlet propagator can be approximated by the following expression:
\bea
\nn
G(E,{\bf r},{\bf r}') 
&\sim& 
\langle {\bf r}'|
\frac{1}{h_s^B+\Sigma_B(E)-E-i\eta}|{\bf r}\rangle 
\,,
\eea
where $\Sigma_B(E)$ accounts for the effects due to the ultrasoft scale. $\Sigma_B(E)$ and (the iteration of) $\delta h_s$ produce ultraviolet ultrasoft and potential divergences. 
After Taylor expanding them the divergences get absorbed by $\delta V_s$, which we also Taylor expand in the Green function: 
\begin{figure}[htb]
\makebox[0.0cm]{\phantom b}
\put(50,1){$\delta G^{\rm c.t.}=$}
\put(100,-5){\epsfxsize=4.5truecm \epsfbox{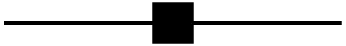}}
\put(153,-20){$\delta V_s$}
\put(231,1){$
\displaystyle{~~+~~\cdots ~~\sim ~~ -G_c\delta V_s G_c}~~+~~\cdots$.}
\label{figdeltaV}
\end{figure}

We discuss the renormalization details in Secs. \ref{USren} and \ref{QMPT}.

\subsection{Ultrasoft divergences}
\label{USren}

$\Sigma_B(E)$ can be expressed in a compact form at NLO in the multipole expansion (but exact to any order in $\al$) through the 
chromoelectric correlator. It reads (in the Euclidean)
\be
\Sigma_{B}(E)
  = V_A^2{T_F \over N_c\,d}  \int_0^\infty \!\! dt {\bf r} e^{-t(h^B_{o}-E)} {\bf r}
  \langle vac|
g_B{\bf E}_E^a(t) 
\phi_{\rm adj}^{ab}(t,0) g_B{\bf E}_E^b(0) |vac \rangle
\label{deltaVUS}
\,,
\end{equation}
where 
$$
\phi_{\rm adj}^{ab}(t,0) 
= P\, {\rm exp}\left\{ - ig_B \displaystyle \int_{0}^{t} \!\! dt' \, A_0(t') \right\}
$$
is evaluated in the adjoint representation. 

The pNRQCD one-loop computation gives the following contribution to the Green function: 
\begin{figure}[htb]
\makebox[0.0cm]{\phantom b}
\put(00,1){$\delta G^{\rm us}=$}
\put(50,-2){\epsfxsize=5truecm \epsfbox{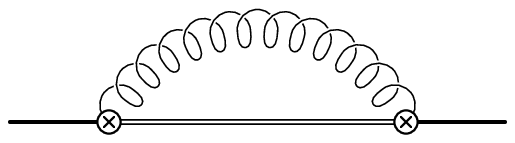}}
\put(201,1){$\displaystyle{\sim - G_c(E)\Sigma^{\rm 1-loop}_B(E)\,G_c(E)}$}
\end{figure}

\noindent
where \cite{Pineda:1997ie,Brambilla:1999qa,Kniehl:1999ud}
\bea
\Sigma_{B}^{\rm 1-loop}
&=&
-g_B^2C_fV_A^2{\bf r}\frac{d-1}{2\,d}\int \frac{d^dk}{(2\pi)^d}\frac{k}{h_o^B-E+k}{\bf r}
\\
\nn
&=&
-g_B^2C_fV_A^2(1+\epsilon)\frac{\Gamma(2+\epsilon)\Gamma(-3-2\epsilon)}{\pi^{2+\epsilon}}
{\bf r}\,(h^B_{o}-E)^{3+2\epsilon}{\bf r}
\nn
\\
&=&
-C_fV_A^2\frac{\al_B}{3\pi}{\bf r}\, (h_o^B-E)^3
\left\{
{1 \over \epsilon}+\gamma_E-\ln \pi +\ln{(h_o^B-E)^2 \over \nu_{us}^2}+\frac{5}{3}+{\cal O}(\epsilon)
\right\}\,
{\bf r}
\,,
\label{USbare1loop}
\eea

This result can be obtained in several ways.  Above we have chosen to do the integration over $k^0$ first. One could also use Eq. (\ref{eqlambda}). In time space one would have to integrate 
\be
\int_0^{\infty}dt_Ee^{-t_E(h_o^B-E)}(t_E^2)^{-2-\epsilon}=\frac{\Gamma(-3-2\epsilon)}{(h_o^B-E)^{-3-2\epsilon}}
\,.
\ee
Actually, in time space, the structure is exactly the same at any loop (it would only change the exponent of $t_E^2$).

The two-loop bare expression reads~\cite{Eidemuller:1997bb,Brambilla:2006wp,Pineda:2011db,Pineda:2011aw} 
\be
\label{USbare2loop}
\Sigma_{B}^{\rm 2-loops}
=
g_B^4C_fC_AV_A^2\Gamma(-3-4\epsilon)
\left[
{\cal D}^{(1)}(\epsilon)-(1+2\epsilon){\cal D}_1^{(1)}(\epsilon)
\right]
{\bf r}\,(h^B_{o}-E)^{3+4\epsilon}{\bf r}\,,
\ee
where
\be
{\cal D}^{(1)}(\epsilon)
=
\frac{1}{(2\pi)^2}\frac{1}{4\pi^{2+2\epsilon}}\Gamma^2(1+\epsilon)g(\epsilon)\,,
\ee
\be
{\cal D}_1^{(1)}(\epsilon)
=
\frac{1}{(2\pi)^2}\frac{1}{4\pi^{2+2\epsilon}}\Gamma^2(1+\epsilon)g_1(\epsilon)\,,
\ee
and
\be
g(\epsilon)=\frac{2 \epsilon^3+6 \epsilon^2+8 \epsilon+3}{\epsilon \left(2 \epsilon^2+5 \epsilon+3\right)}
-\frac{2 \epsilon \Gamma
   (-2 \epsilon-2) \Gamma (-2 \epsilon-1)}{(2 \epsilon+3) \Gamma (-4 \epsilon-3)}
\,,
\ee
\be
g_1(\epsilon)=\frac{6 \epsilon^3+17 \epsilon^2+18 \epsilon+6}{\epsilon^2 
\left(2 \epsilon^2+5 \epsilon+3\right)}+\frac{4 (\epsilon+1)
   n_f T_F}{\epsilon (2 \epsilon+3) N_c}+\frac{2
   \left(\epsilon^2+\epsilon+1\right) \Gamma (-2 \epsilon-2) \Gamma (-2 \epsilon-1)}{\epsilon (2 \epsilon+3)
   \Gamma (-4 \epsilon-3)}
\,.
\ee

We now discuss how to obtain  $\delta V_s$ from $\Sigma_B(E)$. $\delta G^{c.t.}+\delta G^{us}$ should be free of divergences. This determines $\delta V_s$ from $\Sigma_B(E)$. Let us see how. We first 
Taylor expand $\Sigma_B(E)$ in $\epsilon$ and $\al$:
\bea
\nn
\Sigma_{B}(E)
&=&
-\frac{1}{\epsilon}
{\bf r}(h_o-E)^3{\bf r} C_fV_A^2
\left[
\frac{\al(\nu_{us})}{3\pi}
-
\frac{\al^2(\nu_{us})}{36\pi^2}
(C_A(-\frac{47}{3}-2\pi^2)+\frac{10}{3}T_Fn_f)
\right]
\\
&&
\nn
-
\frac{1}{\epsilon^2}{\bf r}(h_o-E)^3{\bf r} C_fV_A^2
\frac{2}{3}\beta_0\frac{\al^2(\nu_{us})}{(4\pi)^2}
\\
&&
\nn
+C_f {\bf r}(h_o-E)^3 V_A^2
\left[
-\frac{\al(\nu_{us}) }{9 \pi } \left(6 \ln \left(\frac{h_o-E}{\nu_{us} }\right)+6\ln 2-5\right)
\right.
\\
&& 
\nn  
+\frac{\al^2(\nu_{us}) }{108 \pi ^2} 
\left(18 \beta_0 \ln ^2\left(\frac{h_o-E}{\nu_{us}
   }\right)-6 \left(C_A \left(13+4 \pi ^2\right)+2 \beta_0 (5-3\ln 2)\right) \ln \left(\frac{h_o-E}{\nu_{us}
   }\right)
   \right.
   \\
   &&
   \nn
-2 C_A \left(-84+39 \ln 2+4 \pi ^2 (-2+3\ln 2)+72 \zeta (3)\right)
\\
&&
\left.
+\beta_0 \left(67+3 \pi ^2-60 \ln 2+18\ln 2\right)\right)
   \Bigg]{\bf r}
+{\cal O}(\epsilon,\al^3)
\,.
\eea
Note that the $1/\epsilon^2$ term comes from two sources: the two-loop bare result and 
the $1/\epsilon$ of $\al_B$ in the one-loop result.
Quite remarkable, there is not log dependence of the $1/\epsilon$ and $1/\epsilon^2$ terms. This is of 
fundamental importance for renormalizability and a check of consistency (see Eq. (\ref{Z2})). 
This result has to be reexpressed in terms of the potentials of the 
singlet/octet Hamiltonian and $h_s-E$. For the renormalization of the potentials we can neglect 
positive powers of $h_s-E$. Therefore, we are rather interested in the identity (valid in $D$ dimensions)
\bea
\nn
&&
{\bf r}(h_o-E)^3{\bf r}=
{\bf r}^2(\Delta V)^3
-\frac{1}{2m_r^2}\left[{\bf p},\left[{\bf p},V^{(0)}_o\right]\right]
+\frac{1}{2m_r^2}\left\{{\bf p}^2,\Delta V\right\}
+\frac{2}{m_r}\Delta V  \left(r\frac{d}{dr}V^{(0)}_s\right)
\\
&&
\nn
\qquad\qquad
+\frac{1}{2m_r}\left[
(\Delta V)^2(3d-5)+4\Delta V\left( \left(r\frac{d}{dr}\Delta V\right)+\Delta V \right)
+\left( \left(r\frac{d}{dr}\Delta V\right)+\Delta V \right)^2
\right]
\\
&&
\qquad\qquad
+{\cal O}((h_s-E))
\,,
\eea
where we have approximated $h_o-h_s=V^{(0)}_o-V_s^{(0)}$, which is enough for our discussion, and
defined $\Delta V \equiv V^{(0)}_o-V^{(0)}_s$.
We have used the combination $\left( \left(r\frac{d}{dr}\Delta V\right)+\Delta V \right)$,
since it has a ${\cal O}(\epsilon)$ suppression with respect $\Delta V$. 
We can now obtain the counterterms of the singlet Hamiltonian due 
to the ultrasoft divergences up to NLO in the following compact expression
\bea
\nn
\delta V_s
&=&
\Biggl(
{\bf r}^2(\Delta V)^3
-\frac{1}{2m_r^2}\left[{\bf p},\left[{\bf p},V^{(0)}_o\right]\right]
+\frac{1}{2m_r^2}\left\{{\bf p}^2,\Delta V\right\}
+\frac{2}{m_r}\Delta V  \left(r\frac{d}{dr}V^{(0)}_s\right)
\\
&&
\nn
+\frac{1}{2m_r}\left[
(\Delta V)^2(3d-5)+4\Delta V\left( \left(r\frac{d}{dr}\Delta V\right)+\Delta V \right)
+\left( \left(r\frac{d}{dr}\Delta V\right)+\Delta V \right)^2
\right]
\Biggr)
\\
&&
\times
\Biggl[
\frac{1}{\epsilon}
C_fV_A^2
\left[
\frac{\al(\nu_{us})}{3\pi}
-
\frac{\al^2(\nu_{us})}{36\pi^2}
(C_A(-\frac{47}{3}-2\pi^2)+\frac{10}{3}T_Fn_f)
\right]
+
\frac{1}{\epsilon^2} C_fV_A^2
\frac{2}{3}\beta_0\frac{\al^2(\nu_{us})}{(4\pi)^2}
\Biggr]
\,.
\label{deltaVs}
\eea
In the above expressions we choose to keep the full $d$-dependence (also in the potentials). This 
is  potentially important once potential divergences are included, necessary for a 
complete NNNLL or N$^4$LO evaluation of the heavy quarkonium mass. As we have already mentioned,
 in this prescription we also take $V_s^{(0)}$ and $V^{(0)}_o$ in $d$ dimensions. At one loop, 
 the ultrasoft  divergences of $V^{(0)}_{s/o}$ do not show up yet and the renormalized and bare potentials (which one can find in Ref. \cite{Schroder:1999sg}) are equal: $V_{s/o}^{(0)}\simeq V_{s/o,B}^{(0)}$, and 
$V^{(0)}_{s/o,B}$ are finite when we take the $\epsilon \rightarrow 0$ limit. Moreover at one loop we also have the equality $V_o^{(0)}=-1/(N_c^2-1)V_s^{(0)}$.
We prefer the scheme used in Eq. (\ref{deltaVs}), since it allows to keep the ultrasoft counterterms in a very compact manner. One is always free 
to change to a more standard $\MS$ scheme. Note that $\MS$ in momentum and position space is not the same.

\subsection{RG: Ultrasoft running}
\label{sec:RGUS}

The RG equation of $\al$ is
\be
\nu_{us}\frac{d}{d\nu_{us}}\al\equiv \al\beta(\al;\epsilon)=2\epsilon\al+\al\beta(\al;0)\,.
\ee
In the limit $\epsilon \rightarrow 0$
\be
\nu_{us}\frac{d}{d\nu_{us}}\al\equiv \al\beta(\al;0)
\equiv \al\beta(\al)=-2\al \frac{d}{d\al}Z^{(1)}_{\al}\,,
\ee
where
\be
Z^{(1)}_{\al}=\frac{\al}{4\pi}\beta_0+\cdots 
\qquad 
\al\beta(\al)=-2\al\left(\beta_0\frac{\al}{4\pi}+\beta_1\frac{\al^2}{(4\pi)^2}
+\cdots 
\right) 
\ee
and expressions for $\beta_0$, $\beta_1$, etc, can be found in the Appendix. 

From the scale independence of the bare potentials
\be
\nu_{us} \frac{d}{d\nu_{us}}V_B=0
\,,
\ee
one obtains the RG equations of the different
renormalized potentials. They can schematically be written as one (vector-like) equation including all potentials:
\be
\nu_{us} \frac{d}{d\nu_{us}}V=B(V)\,, \label{VRGE}
\ee
\be
B(V)\equiv -\left(\nu_{us} \frac{d}{d\nu_{us}}\delta V\right)\,.
\ee

Note that Eq.~(\ref{VRGE}) implies that all the $1/\epsilon$ poles disappear once the derivative 
with respect to the renormalization scale is performed.
 This imposes some constraints on $\delta V$:
\bea
\label{Z1}
{\cal O}(1/\epsilon): \qquad
&&B(V)=-2\al \frac{\partial}{\partial\al}Z^{(1)}_{V}\,,\; \label{1loopBV}
\\
{\cal O}(1/\epsilon^2): \qquad
&&
\label{Z2}
B(V)\frac{\partial}{\partial V}Z^{(1)}_{V}
+
\al \beta (\al)\frac{\partial}{\partial\al}Z^{(1)}_{V}
+
2\al \frac{\partial}{\partial\al}Z^{(2)}_{V}=0\,,
\eea
and so on. 
At this respect note that the complete $1/\epsilon^2$ term in Eq. (\ref{deltaVs}) fulfills Eq. (\ref{Z2}) and hence it is a check of the computation.

Using Eq. (\ref{deltaVs}) and Eq. (\ref{Z1}) we obtain the following RG equation: 
\be
\nu_{us}\frac{d}{d\nu_{us}} V_{s,\MS}
=
B_{V_s}
\,,
\ee 
where
\bea
\label{BVs}
B_{V_s}&=&
C_fV_A^2
\left[
{\bf r}^2(\Delta V)^3
+\frac{2}{m_r}\Biggl(\Delta V  \left(r\frac{d}{dr}V^{(0)}_s\right)
+
(\Delta V)^2
\Biggr)
-\frac{1}{2m_r^2}\left[{\bf p},\left[{\bf p},V^{(0)}_o \right]\right]
\right.
\\
\nn
&&
\left.
+\frac{1}{2m_r^2}\left\{{\bf p}^2,\Delta V \right\}
\right]
\times
\left[
-\frac{2\al(\nu_{us})}{3\pi}
+
\frac{\al^2(\nu_{us})}{9\pi^2}
(
C_A(-\frac{47}{3}-2\pi^2)+\frac{10}{3}T_Fn_f
)
+
{\cal O}(\al^3)
\right]
\,,
\eea
and now one can take the four-dimensional expression for the potentials. This result holds true in both schemes, the MS and $\MS$ scheme (in a way this is due to the 
fact that the subdivergencies associated to $\al$ also change to make the result scheme independent). 
After solving the RG equation we find
\be
\label{VsRGnus}
V^{RG}_{s,\MS}(r;\nu_p=\nu,\nu_s=\nu,\nu_{us})=V_{s,\MS}(r;\nu)+\delta V_{s,RG}(r;\nu,\nu_{us})
\ee
where
\bea
\nn
\delta V_{s,RG}(r;\nu,\nu_{us})
&=&
\left[
\Biggl(
{\bf r}^2(\Delta V)^3
+\frac{2}{m_r}\left(\Delta V  \left(r\frac{d}{dr}V^{(0)}_s\right)+(\Delta V)^2\right)
\Biggr)F(\nu;\nu_{us})
\right.
\\
&&
\left.
-\frac{1}{2m_r^2}\left[{\bf p},\left[{\bf p},V^{(0)}_o(r) F(\nu;\nu_{us})\right]\right]
+\frac{1}{2m_r^2}\left\{{\bf p}^2,\Delta V(r) F(\nu;\nu_{us})\right\}
\right]
\label{VsRG}
\,,
\eea
and
\bea
F(\nu;\nu_{us})
&=&C_fV_A^2
\frac{2\pi}{\beta_0}
\left\{
\frac{2}{3\pi}\ln\frac{\al(\nu_{us})}{\al(\nu)}
\right.
\\
\nn
&&
\left.
-(\al(\nu_{us})-\al(\nu))
\left(
\frac{8}{3}\frac{\beta_1}{\beta_0}\frac{1}{(4\pi)^2}-\frac{1}{27\pi^2}\left(C_A\left(47+6\pi^2\right)-10T_Fn_f\right)
\right)
\right\}
\,.
\eea
This expression resums the LL \cite{Pineda:2000gza,Pineda:2001ra} and NLL \cite{Brambilla:2009bi,Pineda:2011aw} ultrasoft logs of the potential\footnote{In vNRQCD \cite{Luke:1999kz}, the LL expression was checked in Ref. \cite{Hoang:2002yy}, and the NLL one in Refs. \cite{Hoang:2006ht,Hoang:2011gy} except for the static potential.}. 

From Eq. (\ref{VsRG}) we can easily identify the RG contribution to each potential\footnote{And from 
Eqs. (\ref{deltaVs}), (\ref{BVs})  one could also 
easily obtain the counterterms and RG equations for each potential.} (now we work in the equal mass case). The total RG improved static potential then reads 
\be
\label{VRGIsMS}
V^{(0),RG}_{s,\MS}(r;\nu,\nu_{us})=V^{(0)}_{s,\MS}(r;\nu)+\delta V^{(0)}_{s,RG}(r;\nu,\nu_{us})
\equiv - C_f {\alpha_{V_s}(r;\nu,\nu_{us}) \over r}
\,,
\ee
where we have made explicit that $V_s^{(0)}$  does not depend on $\nu_p$ 
(and similarly for some of the following potentials) and
\be
\delta V^{(0)}_{s,RG}(r;\nu,\nu_{us})
=
{\bf r}^2(\Delta V)^3
F(\nu;\nu_{us})
\,.
\ee

The total RG improved $1/m$ potential reads
\be
\label{VRGI1MS}
V^{(1),RG}_{\MS}(r;\nu,\nu_{us})=V^{(1)}_{\MS}(r;\nu)+\delta V^{(1)}_{RG}(r;\nu,\nu_{us})
\equiv -{C_fC_A D^{(1)}(r;\nu,\nu_{us}) \over 2r^2}
\,,
\ee
where
\be
\delta V^{(1)}_{RG}(r;\nu,\nu_{us})
=4
\Biggl(
\Delta V  \left(r\frac{d}{dr}V^{(0)}_s\right)
+
(\Delta V)^2
\Biggr)F(\nu;\nu_{us})
\,.
\ee
The total RG improved momentum-dependent $1/m^2$ potential reads
\be
\label{VRGp21MS}
V^{(2),RG}_{p^2,\MS}(r;\nu,\nu_{us})=V^{(2)}_{p^2,\MS}(r;\nu)+\delta V^{(2)}_{p^2,RG}(r;\nu,\nu_{us})
\equiv
-  C_f D^{(2)}_{1} (r;\nu,\nu_{us})
\,,
\ee
where
\bea
\delta V^{(2)}_{p^2,RG}(r;\nu,\nu_{us})
=
4 \Delta V(r) F(\nu;\nu_{us})
\label{Vp2RG}
\,.
\eea

The RG correction of $V_r$ is first defined in momentum space. From Eq. (\ref{VsRG}) we have 
\be
\delta V^{(2)}_{r,RG}(r;\nu,\nu_{us})
=
-2\left[{\bf p},\left[{\bf p},V^{(0)}_o F(\nu;\nu_{us})\right]\right]
=\int \frac{d^3q}{(2\pi)^3}e^{i{\bf q}\cdot {\bf r}}\delta \tilde V^{(2)}_{r,RG}(q;\nu,\nu_{us})
\label{VsrRG}
\,,
\ee
where (note that $\delta V^{(2)}_{r,RG}$ vanishes in the large $N_c$ limit)

\be
\delta \tilde V^{(2)}_{r,RG}(q;\nu,\nu_{us})
=
-2 {\bf q}^2\tilde V^{(0)}_o(q)F(\nu;\nu_{us})
\ee
and $\tilde V^{(0)}_o$ is the Fourier transform of $V^{(0)}_o$. We can then give the following expression for the potential
\be
\label{VrRG}
V_{r,\MS}^{(2),RG}(r;\nu_p,\nu_s,\nu_{us})\Big|_{\nu_s=\nu_p=\nu}
\equiv
\int \frac{d^3q}{(2\pi)^3}e^{i{\bf q}\cdot {\bf r}} \tilde V^{(2)}_{r,RG}(q;\nu_p,\nu_s,\nu_{us})\Big|_{\nu_s=\nu_p=\nu}
=V_{r,\MS}^{(2)}(r;\nu)+\delta V^{(2)}_{r,RG}(r;\nu,\nu_{us})
\,.
\ee
This allows to obtain the associated dimensionless constant for $\nu=\nu_s=\nu_p$:
\be
\label{D2RGnu}
\pi C_f D^{(2)}_{d} (q;c(\nu),d(\nu,\nu);\nu,\nu_{us})\equiv  \tilde V^{(2)}_{r,RG}(q;\nu,\nu_{us})
\,.
\ee

For completeness, we also give the relation between the rest of the ${\cal O}(1/m^2)$ potentials and the associated dimensionless Wilson coefficients defined in Eq. (\ref{lagalpha}): $\al_V^{(s,n)}$. These constants depend logarithmically on $r$ (or 
$k$ in momentum space) and the
renormalization scale $\nu_{\rm pNR}$ (note that at the order discussed in this review they do not depend on $\nu_{us}$, yet we introduce its dependence for completeness):
\bea
V_{{\bf L}^2}^{(2)}(r;\nu,\nu_{us})&=& { C_f  \over 2 }{1 \over r}D^{(2)}_{2}(r;c(\nu);\nu,\nu_{us}),
\qquad
V_{LS}^{(2)}(r;\nu,\nu_{us})={ 3 C_f  \over 2 }{1 \over r^3}D^{(2)}_{LS}(r;c(\nu);\nu,\nu_{us}),\\
V_{{\bf S}_{12}}^{(2)}(r;\nu,\nu_{us})&=& { C_f  \over 4 }{1 \over r^3}D^{(2)}_{S_{12}}(r;c(\nu);\nu,\nu_{us}),
\label{V2}
\nn
\quad
\tilde V_{S^2}^{(2)}(q;\nu,\nu,\nu_{us})
={4\pi C_f  \over 3}D^{(2)}_{S^2}(q;c(\nu),d(\nu,\nu);\nu,\nu_{us})
 \,.
 \eea 
 
Note that $D_{r/S^2}^{(2)}$, the dimensionless Wilson coefficients associated to $V_{r}^{(2)}$ and $V_{S^2}^{(2)}$, are first defined in 
momentum space. A direct definition in position space is possible but more cumbersome, due to the fact that a naive definition produces $\delta^{(3)}({\bf r})\times \ln r$ distributions, which are ill-defined. We discuss further this issue below.

One can also organize the above RG equations
within an expansion (besides in $\al(\nu_{us})$) in $1/m$, as the ultrasoft computation does not mix different orders in $1/m$.
At ${\cal O}(1/m^0)$, the analysis corresponds to the study of the static
limit of pNRQCD, which has been carried out in Ref. \cite{Pineda:2000gza} at LL and in Ref. \cite{Brambilla:2006wp} at NLL 
(see also \cite{Pineda:2011db,Pineda:2011aw}).
Since $\al_V^{(0,-1)}\not= 0$, there are
relevant operators (super-renormalizable terms) in the Lagrangian
and the ultrasoft RG equations lose the triangular structure that we enjoyed
for the RG equations of $\nu_s$. Still, if $\al_V^{(0,-1)} \ll 1$, the RG
equations can be obtained as a double expansion in $\al_V^{(0,-1)}$ and
$\al_V^{(0,0)}$, where the latter corresponds to the marginal operators (renormalizable interactions).
At short distances ($1/r \gg \lQ$), the static limit of pNRQCD is
in this situation. Specifically, we have $\al_V^{(0,-1)}=\{\al_{V_s},\al_{V_o}\}$, 
that fulfills $\al_V^{(0,-1)} \sim \al(r) \ll 1$;
$\al_V^{(0,0)}=\al(\nu_{us})$  and $\al_V^{(0,1)}=\{V_A,V_B\} \sim 1$.
Therefore, we can calculate the anomalous dimensions order by
order in $\al(\nu_{us})$. In addition, we also have an expansion in
${\tilde V}_{-1}$. Moreover, the specific form of the pNRQCD
Lagrangian severely constrains the RG equations' general structure.
Therefore, for instance, the leading non-trivial RG equation for $\al_{V_s}$ reads
\be
\label{alVsrunning}
\nu_{us} {d\over d\nu_{us}}\al_{V_{s}}
=
{2 \over 3}{\al\over\pi}V_A^2\left( \left({C_A \over 2} -C_f\right)\al_{V_o}+C_f\al_{V_s}\right)^3
+{\cal O}([\al_V^{(0,-1)}]^4\al_V^{(0,0)},[\al_V^{(0,0)}]^2[\al_V^{(0,-1)}]^3)
\,.
\ee
At higher orders in $1/m$ the analysis has been carried out in Ref. \cite{Pineda:2001ra} at LL and in 
Ref. \cite{Pineda:2011aw} at NNL. The same considerations than for the static limit apply 
as far as the non-triangularity of the RG equations is
concerned. At $O(1/m,1/m^2)$, we have the following Wilson coefficients:
$\al_V^{(1,-2)}=\{D^{(1)},c_k\}$ and 
$\al_V^{(2,-3)}=\{D_{1}^{(2)},D_{2}^{(2)},D_{d}^{(2)},D_{S^2}^{(2)},D_{LS}^{(2)},D_{S_{12}}^{(2)}\}$.
In general, one has the structure
\be
\label{nusRG}
\nu_{us} {d\over d\nu_{us}}\al_V^{(\ell,n)}
\sim
\sum_{\{n_i\}\{\ell_i\}}\al_V^{(\ell_1,n_1)}\al_V^{(\ell_2,n_2)}\cdots\al_V^{(\ell_j,n_j)}\,, 
\quad {\rm with}\quad \sum_{i=1}^j \ell_i=\ell\;,\, \sum_{i=1}^j n_i=n\,,
\ee
and one has to pick up the leading contributions from all possible
terms. Finally, 
by solving Eq. (\ref{nusRG}) between $\nu$ and $\nu_{us}$, we
will have $\al_V^{(\ell,n)}(r;c(\nu),d(\nu,\nu);\nu,\nu_{us})$, where the
running with respect to $\nu_{us}$ is known. 

Finally, one should also consider the ultrasoft running of $V_A$. It has been studied in Ref. 
\cite{Pineda:2000gza} with LL accuracy,  and in Ref. \cite{Brambilla:2009bi} with NLL accuracy,
reaching to the conclusion that:
\be
\nu_{us}\frac{d}{d\nu_{us}}V_A=0+{\cal O}(\al^3)
\,.
\ee
In the previous section we got that the initial matching condition $V_A=1+{\cal O}(\al^2)$. So we can set
$V_A=1$.

\subsection{RG: Soft running}
\label{sec:RGsoftp}
In the previous section we have kept the dependence on $\nu_s$ explicit, yet this dependence should cancel in the sum in Eq. 
(\ref{VsRGnus}), as the Wilson coefficients are $\nu_s$ independent. 
Being more specific,
the potentials have the following structure (either in position or momentum space):
$$
\al^{(n,s)}_V(r; c(\nu_s), d(\nu_p,\nu_s);\nu_s,\nu_{us})
\,,
$$
and the independence on $\nu_s$ gets reflected in the following equation:
\bea
\label{nus}
&&\nu_s {d\over d\nu_s}\al^{(n,s)}_V(r; c(\nu_s), d(\nu_p,\nu_s);\nu_s,\nu_{us})
\\
\nn
&&
\qquad\qquad
=
\left[\nu_s {\partial \over \partial \nu_s}+\nu_s \left({d\over d\nu_s}d\right){\partial \over
    \partial d}+\nu_s \left({d\over d\nu_s}c\right){\partial \over
    \partial c}\right]\al^{(n,s)}_V(r; c(\nu_s), d(\nu_p,\nu_s);\nu_s,\nu_{us})=0
\,,
\eea
At the practical level, with the accuracy we are working, it
is equivalent to set $\nu_s= 1/r$ up to factors of order one:
\be
\al^{(n,s)}_V(r; c(\nu_s), d(\nu_p,\nu_s);\nu_s,\nu_{us}) \rightarrow \al^{(n,s)}_V(r; c(1/r), d(\nu_p,1/r);1/r,\nu_{us})
\,.
\ee
 Therefore, one can also deduce the (logarithmic)
dependence of $\al_V$ on $r$, and resum some associated logarithms of $r$. This works fine for most potentials, 
and, in particular, we can also set $\nu_s=1/r$ as far as $F(1/r,\nu_{us})$ is kept inside the 
(anti)commutators, i.e. in the way displayed in Eq. (\ref{VsRG}). 
One works similarly for $V_r$ and $V_{S^2}$ but in momentum space. Actually one could try to do the whole analysis in momentum space but then ultrasoft computations are more difficult. We then chose to 
set $\nu_s=q$ in the Fourier transform and define 
\be
\label{Vr1}
V_{r/S^2,\MS}^{(2),RG}(r;\nu_p,\nu_{us})
\equiv
\int \frac{d^3q}{(2\pi)^3}e^{i{\bf q}\cdot {\bf r}} \tilde V^{(2)}_{r/S^2,RG}(q;\nu_p,q,\nu_{us})
\,.
\ee
In other words this is equivalent to
\be
\label{D2RGq}
\pi C_f D^{(2)}_{d/S^2,s} (q;c(q),d(\nu_p,q);q,\nu_{us})\equiv  \tilde V^{(2)}_{r/S^2,RG}(q;\nu_p,q,\nu_{us})
\,.
\ee
Note that the complete determination of $D_{d/S^2}^{(2)}$ requires to distinguish $\nu_p$ from $\nu_s$. This can be done iteratively as, at LO, there is no 
dependence on $\nu_p$. 
We take Eq. (\ref{Vr1})  as our final expressions 
for $V_{r/S^2,\MS}^{(2),RG}(r;\nu_p,\nu_{us})$.

We have then obtained the RG improved potentials that 
compose $V_s$ to the order of interest. As we have already mentioned, the $1/m$ and $1/m^2$ potentials 
suffer from field redefinitions ambiguities. Therefore, in some circumstances, it can be convenient to cast the
total potential to be introduced in the Schrödinger equation in the following unified form:
\bea
\label{VsMS}
V^{RG}_{s,\MS}(r;\nu_p,\nu_{us})&=&V^{(0),RG}_{s,\MS}(r;1/r,\nu_{us})
+
\frac{V^{(1),RG}_{s,\MS}(r;1/r,\nu_{us})}{m}
\\
\nn
&&
+
\frac{1}{m^2}
\left(
{1 \over 2}\left\{{\bf p}^2,V_{{\bf p}^2,\MS}^{(2),RG}(r;1/r,\nu_{us})\right\}
 + V_{r,\MS}^{(2),RG}(r;\nu_p,\nu_{us})
\right)
\\
\nn
&&
+
\frac{1}{m^2}
\left(
V_{LS}^{(2)}(r;1/r,\nu_{us}){\bf L}\cdot{\bf S} + V_{S^2}^{(2)}(r;\nu_p,\nu_{us}){\bf S}^2
 + V_{{\bf S}_{12}}^{(2)}(r;1/r,\nu_{us}){\bf S}_{12}({\hat {\bf r}})
 \right)
\,.
\eea

\subsection{NR Quantum mechanics perturbation theory}
\label{QMPT}

Within pNRQCD, talking 
about potential loops is essentially talking about NR quantum-mechanics 
perturbation theory:
\begin{figure}[htb]
\makebox[0.0cm]{\phantom b}
\put(50,1){$\delta G^{\rm pot.}=$}
\put(100,-5){\epsfxsize=4.5truecm \epsfbox{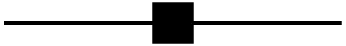}}
\put(153,-20){$\delta h_s$}
\put(231,1){$
\displaystyle{~~+~~\cdots ~~\sim ~~ G_c\delta h_s G_c}~~+~~\cdots$,}
\label{deltaVbis}
\end{figure}

\noindent
where the black square represents a generic $\delta h_s$ correction to the 
singlet Coulomb Hamiltonian. 

It is not possible to compute potential loops analytically, in particular when seeking for divergences, 
using $\al(1/r)$ (or $\al(k)$ in momentum space) as the expansion parameter in the Wilson coefficient/potential. 
One rather has to Taylor expand around a momentum-independent factorization scale, $\nu$. The expansion parameter in 
the Wilson coefficient is $\al(1/r)=\al(\nu)+{\cal O}({\al^2})$ and each Wilson coefficient 
generates an infinite tower of Wilson coefficients with different powers of logarithms of $r$. 
If one is also working at fixed order in perturbation theory (not aiming for the 
resummation of logarithms) all factorization scales are set equal and only a single factorization scale appears in the computation.

In Sec. \ref{USren} the leading and NLO ultrasoft contribution to $\delta V_s$ was given. 
The LO ultrasoft contribution is enough to completely renormalize the potentials at the order necessary to obtain the spectroscopy with 
${\cal O}(m\al^5)$ precision. For the ${\cal O}(m\al^6)$ precision, besides the NLO ultrasoft divergences, potential divergences also appear, and show up through the iteration of potentials\footnote{At higher orders, potential divergences can also show up if singular enough potentials appear such that the potential itself need regularization, or if its expectation value is singular.}:
\be
\label{QMperturbation}
G_c(E)\delta h_s G_c(E) \cdots \delta h_s G_c (E)
\,.
\ee
The complete expression for $\delta V_s$ at ${\cal O}(m\al^6)$ coming from potential divergences 
is at present unknown, though for the case of QED and the spin-dependent potential it could be deduced from the results of 
\cite{Czarnecki:1999mw} and \cite{Kniehl:2003ap,Penin:2004xi} respectively. Therefore, we do not aim here 
to give the complete expression. We will only singlet out a contribution and see how it is absorbed by the counterterms.

\begin{figure}[htb]
\hspace{-0.1in}
\epsfxsize=4.5in
\centerline{\epsffile{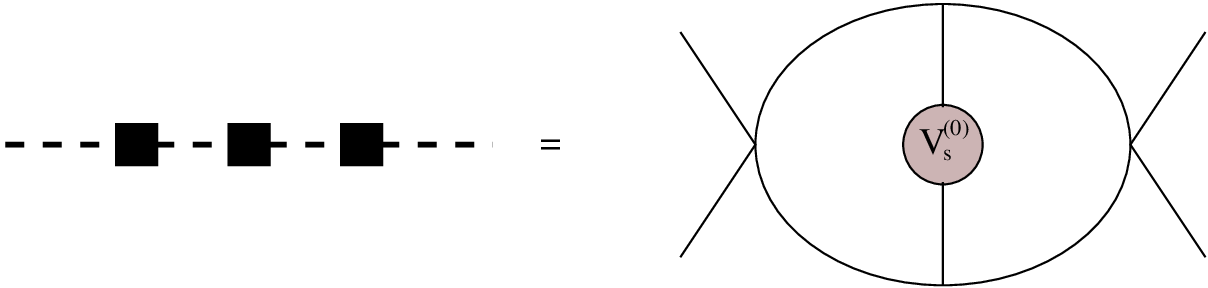}}
\bigskip
\put(-400,0){$D_{d}^{(2)}$}
\put(-365,0){$\alpha_{V_s}$}
\put(-335,0){$D_{d}^{(2)}$}
\put(-230,-15){$D_{d}^{(2)}$}
\put(-190,-15){$\alpha_{V_s}$}
\put(-150,-15){$D_{d}^{(2)}$}
\caption {{\it One possible contribution to the running of $D_{d}^{(2)}$ at
NLL. The first picture represents the calculation in 
terms of the free quark-antiquark propagator $G_c^{(0)}$ and the 
potentials (the small rectangles). The picture on the right 
is the representation within a more standard diagrammatic
interpretation in terms of quarks and antiquarks. The delta potentials
are displayed as local interactions and the Coulomb potential as an
extended object in space (but not in time).}}
\label{obs12}
\end{figure}
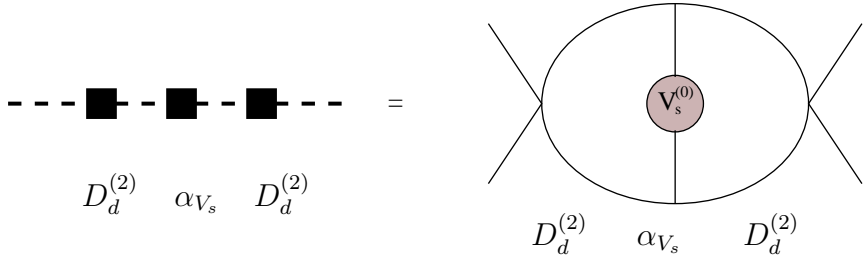

We are interested in the ultraviolet divergences produced by corrections of the type of Eq. (\ref{QMperturbation}), i.e. NR quantum mechanics 
perturbation theory. These divergences are absorbed by the
Wilson coefficients of the local potentials (those proportional 
to $\delta^{(3)}({\bf r}$) or its derivatives). 
Let us explain how this works in detail. Since the singular behavior of
the potential loops appears in the ultraviolet for $|{\bf p}| \gg \al/r$, a perturbative
expansion in $\al$ (or in other words in $V_s^C$) is allowed in $G_c(E)$, which can be approximated by the 
free propagator:

\begin{figure}[htb]
\makebox[0.5cm]{\phantom b}
\put(50,1){\epsfxsize=3truecm \epsfbox{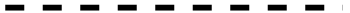}}
\put(150,1){$
\label{Gc0}
\equiv G_c^{(0)}(E)=\displaystyle{
{1 \over \displaystyle{{\bf p}^2/m-E}}} \,,$}
\end{figure}
\noindent 
as far as the computation of the
$\ln \nu_p$ ultraviolet divergences is concerned.  Moreover, each
$G_c^{(0)}$ produces a potential loop and one extra power of $m$ in the
numerator, which kills the powers of $1/m$ in the different
potentials. This allows the mixing of potentials with different powers
of $1/m$. In principle, this
would be a never ending story unless there is an small parameter
that tells us how far we have to go in the calculation in order to achieve
some given accuracy. The suppression factor is $\al$. One typical example is the diagram in Fig.~\ref{obs12}, 
which corresponds to  ($D^{(2)}_{d,s} \simeq \al(\nu)$ and $\al_{V_s} \simeq \al(\nu)$)
\be
G_c^{(0)}(E)
{\pi C_f D^{(2)}_{d,s} \over m^2}\delta^{(3)}({\bf r})
G_c^{(0)}(E)
C_f {\al_{V_s} \over r}
G_c^{(0)}(E)
{\pi C_f D^{(2)}_{d,s} \over m^2}\delta^{(3)}({\bf r})
G_c^{(0)}(E)
\,.
\ee
The relevant computation reads
\bea
&&
\langle{\bf r}=0|
G_c^{(0)}(E)
C_f {\al_{V_s} \over r}
G_c^{(0)}(E)
|{\bf r}=0\rangle
\\
\nn
&&
\qquad
\sim 
\int \frac{
{\rm d}^d p' }{ (2\pi)^d } \int \frac{ {\rm d}^d p }
{ (2\pi)^d } \frac{ m }{{\bf p}'^2 - mE } 
C_f
\frac{ 4\pi\alpha_{V_s} }{ {\bf q}^2 } \frac{ m }{{\bf p}^2-m E } 
\sim
- C_f\frac{m^2\alpha_{V_s}}{16\pi}  
\frac{ 1 }{\epsilon },
\eea
where ${\bf q}={\bf p}-{\bf p}'$. This divergence can be absorbed in the Wilson coefficient of the delta potential
$D_{d,s}^{(2)}$ as follows 
\be
\label{eqDdexample}
\delta D_{d,s}^{(2)} \sim 
\frac{1}{\epsilon}\alpha_{V_s}\left[D_{d,s}^{(2)}\right]^2+\cdots
\simeq 
\frac{1}{\epsilon}\al^3(\nu)+\cdots
\,.
\ee

It is particularly appealing how the EFT framework solves the problem of the ultraviolet divergences one finds in standard NR
quantum-mechanical perturbation theory calculations. When potential
divergences are found it can be more convenient to work in a momentum
representation (see for instance \cite{Czarnecki:1999mw}).  Nevertheless, it
is also possible to handle the ultraviolet divergences in position space
\cite{Yelkhovsky:2001tx}.  Either way, the computation should be performed
in the same scheme used to compute the potentials (see
Sec.~\ref{pNRmatchingI} for details).

\subsection{RG: Potential (and final) running}
\label{sec:RGpot}
Having set up the problem in a "quantum field theory" way with standard dimensional regularization, we are much closer to get 
a complete set of RG equations. We already have them for the soft and ultrasoft.
The final step is to obtain the RG equation for $\nu_p$.

As we have mentioned in sec \ref{QMPT}, integrals over ${\bf p}$ (or ${\bf x}$) appear when
solving the Schr\"odinger equation that dictates the dynamics of the
Heavy Quarkonium near threshold.  At low orders, these integrals are
finite, no dependence on $\nu_p$ occurs and 
one has $|{\bf p}| \sim 1/r \sim m\al$ and
${\bf p}^2/m \sim m\al^2$.  Therefore, one can lower $\nu_{us}$ down to
$\sim m\al^2$ reproducing the results obtained by \cite{Pineda:2001ra}.
Nevertheless,
at higher orders in NR quantum mechanics perturbation theory and/or if
some singular enough operators are introduced (as it is the case
of the heavy quarkonium production currents) the integrals over ${\bf
p}$ are divergent and singularities
proportional to $\ln\nu_p$ appear. These must be absorbed by the potentials
or by the Wilson coefficients of the currents. 
The log structure is dictated by the ultraviolet
behavior of ${\bf p}$ and $1/r$. This means that we can not replace
$1/r$ and $\nu_{us}$ by their physical expectation values but rather by
their cutoffs within the integral over ${\bf p}$, i.e. $\nu_p$. Therefore, besides the explicit dependence on $\nu_p$  of the
potential, which appears in $d$, the potential also implicitly depends on $\nu_p$
through the requirement $1/r \sim |{\bf p}| \ll \nu_p$, and also through
$\nu_{us}$, since $\nu_{us}$ has to fulfill ${\bf p}^2/m \ll \nu_{us}\ll |{\bf p}|$ 
in order to ensure that only soft degrees of freedom
have been integrated out for a given $|{\bf p}|$. This latter requirement holds 
if we fix the final point of the evolution of the ultrasoft RG equation to
be $\nu_{us}=\nu_p^2/m$. At this stage, a single cutoff,
$\nu_p$, exists and the correlation of cutoffs becomes manifest. 
The importance of the idea that the cutoffs of the NR
effective theory should be correlated was first realized in Ref. 
\cite{Luke:1999kz}. 
Finally, for the RG equation of $\nu_p$, the anomalous dimensions
of $\al^{(n,s)}_V(r;c(1/r),d(\nu_p,1/r);1/r,\nu_p^2/m)$ is at LO the same as
the one of $\al^{(n,s)}_V(\nu_p;c(\nu_p),d(\nu_p,\nu_p);\nu_p,\nu_p^2/m) \equiv \al^{(n,s)}_V(\nu_p)$.
At low orders, this discussion is equivalent to expanding $\ln r$ around $\ln \nu_p$ 
in the potential i.e. 
\be 
\al^{(n,s)}_V (r;c(1/r),d(\nu_p,1/r);1/r,\nu_p^2/m)
\simeq \al^{(n,s)}_V(\nu_p;c(\nu_p),d(\nu_p,\nu_p);\nu_p,\nu_p^2/m)
+\ln(\nu_pr)r{d \over d r} \al^{(n,s)}_V\bigg|_{1/r=\nu_p} + \cdots 
\,.
\ee 
The $\ln(\nu_pr)$ terms give subleading contributions to the
anomalous dimension when introduced in divergent integrals over ${\bf p}$. 
For most potentials, this expansion is quite trivial. In the case of $V_r$  and $V_{S^2}$ one 
has
\bea
D^{(2)}_{d/S^2} (q;c(q),d(\nu_p,q);q,\nu_p^2/m)
&\simeq&
D^{(2)}_{d/S^2} (\nu_p;c(\nu_p),d(\nu_p,\nu_p);\nu_p,\nu_p^2/m)
 \\
 \nn
 &&
 +(\ln q/\nu_p) 
q\frac{d}{dq}(
D^{(2)}_{d/S^2} (q;c(q),d(\nu_p,q);q,\nu_p^2/m))|_{q=\nu_p} +\cdots 
\eea
where
\bea
\nn
&&
q\frac{d}{dq}(
D^{(2)}_{d} (q;c(q),d(\nu_p,q);q,\nu_p^2/m))|_{q=\nu_p} 
=
-\frac{\al^2(\nu_p)}{\pi}\frac{16}{3}(\frac{C_A}{2}-C_f)
\left[
1+\ln\frac{\al(\nu_p)}{\al(\nu_p^2/m)}
\right]
\\
&&
\quad
+ \frac{\al^2(\nu_p)}{\pi}
\left[
-\frac{\beta_0}{2}
+\frac{2}{3}T_Fn_f(c_D(\nu_p)+c_1^{hl}(\nu_p))+(\beta_0-\frac{13}{3}C_A)c_F^2(\nu_p)
+(\frac{14}{3}C_f-\frac{2}{3}C_A)c_k^2
\right]
\,,
\eea
and
\bea
\label{VS2derivative}
&&
q\frac{d}{dq}( 
D^{(2)}_{S^2} (q;c(q),d(\nu_p,q);q,\nu_p^2/m))|_{q=\nu_p} 
=
\frac{1}{\pi}
\left(
-\frac{\beta_0}{2}+\frac{7}{4}C_A
\right)\al^2(\nu_p)c_F^2(\nu_p)
\,.
\eea
Actually, Eq. (\ref{VS2derivative}) is a necessary piece for the 
computation of the hyperfine splitting of the Heavy Quarkonium at NLL \cite{Kniehl:2003ap,Penin:2004xi}.

Note that in position space $V_{r/S^2,\MS}^{(2),RG}(r;\nu_p,\nu_{us})$ would read
\bea
\label{Vrexpanded}
V_{r/S^2,\MS}^{(2),RG}(r;\nu_p,\nu_{us})
&=&
\delta^3({\bf r})
\left(
\tilde V^{(2)}_{r/S^2,RG}(\nu_p;\nu_p,\nu_p,\nu_{us})
-(\ln \nu_p) 
q\frac{d}{dq}(\tilde V^{(2)}_{r/S^2,RG}(q;\nu_p,q,\nu_{us}) )|_{q=\nu_p} 
\right)
\nn
\\
&&
-\frac{1}{4\pi}
\left({\rm reg}\frac{1}{r^3}\right)q\frac{d}{dq}(
\tilde V^{(2)}_{r/S^2,RG}(q;\nu_p,q,\nu_{us})
)|_{q=\nu_p}+\cdots
\,,
\eea
and only the term proportional to ${\rm reg}\frac{1}{r^3}$ contributes to the $l\not=0$ states mass. Higher order terms in the Taylor 
expansion of the Fourier transform of $\ln q$ are subleading. 

$\al^{(n,s)}_V(\nu_p)$ appears through the divergences induced by the
iteration of the potentials in the way explained in Ref. \cite{Pineda:2001et}
and Sec.~\ref{QMPT}. In particular, the computation of the 
anomalous dimension can be organized within an expansion in $\al$ and using the free propagators $G_c^{(0)}$. 
Finally the running will go from $\nu_p \sim m$ down to $\nu_p \sim m\al$.
 A similar discussion applies to the running of the Wilson coefficients
of the currents (or, in other words, of the imaginary terms of the potential).
This completes the procedure to obtain the RG equations in pNRQCD.

The above discussion applied to the example discussed in Sec. \ref{QMPT} would give a correction to the running of the 
delta potential of the following class
\be
\label{eqDd}
\nu_p {d \over d\nu_p}D_{d}^{(2)}(\nu_p) \sim 
\alpha_{V_s}(\nu_p)[D_{d}^{(2)}(\nu_p)]^2+\cdots
\,.
\ee
Nevertheless, the complete running for the spin-independent delta potential is at present unknown. Even though the spin-dependent running is known \cite{Kniehl:2003ap,Penin:2004xi} it is quite lengthy to explain. Instead we will illustrate the procedure with the running of the electromagnetic current at NLL in Sec. \ref{secRGcurrent}.

\subsection{Renormalization (group) of the current }
\label{secRGcurrent}

The NR currents in pNRQCD have formally the same structure than in NRQCD (see Eqs. (\ref{vcurr}) and (\ref{gcurr}))
but changing the Wilson coefficient. They read (up to ${\cal O}(1/m)$ corrections that we do not discuss here):
\be
\bfm{j}=B_1 \psi^\dagger\bfsigma\chi(0)\bigg|_{\rm pNRQCD} \,,\qquad 
O_{\gamma\gamma}=B_0\psi^\dagger\chi(0)\bigg|_{\rm pNRQCD}
\,.
\ee
The NRQCD Wilson coefficients $b_s$ are functions of $\nu_p$ and $\nu_s$,
i.e. $b_s(\nu_p,\nu_s)$.  If we compare with the previous 
discussion of the potentials, the four-fermion 
Wilson coefficients $d$ play the role of $b_s$. In this case, unlike for the 
$d$'s, there is no running due to $\nu_s$ at the order of interest.
Integrating out the soft scale when matching local currents produces scaleless integrals, which are zero in dimensional 
regularization. This can be easily 
seen in the Coulomb gauge. 
This means that the Wilson coefficients are equal at the matching scale. 
Formally, 
\be
b_1\psi^\dagger\sigma^i\chi(0)\bigg|_{\rm NRQCD}=B_1\psi^\dagger\sigma^i\chi(0)\bigg|_{\rm pNRQCD}\,,
\ee
or, in other words, the matching condition reads 
$B_1(b_1(\nu_p,\nu_p),\nu_{us}=\nu_p)=b_1(\nu_p,\nu_p)$,  
and similarly for the pseudoscalar: $B_0(b_0(\nu_p,\nu_p),\nu_{us}=\nu_p)=b_{\rm 0, NR}(\nu_p,\nu_p)$. 
So, finally, the initial matching condition is nothing but  $b_s(\nu_p,\nu_p)\equiv b_s(\nu_p)$, which we already know. 

The running
of $\nu_{us}$ is also trivial as there is none at the order of interest 
(this has to do with the
fact that we are dealing with an electromagnetic annihilation process). Therefore, we
finally have $B_s(\nu_p) \equiv
B_s(b_s(\nu_p),\nu_p^2/m)=b_s(\nu_p)$.  We can see that we 
are in the analogous
situation to the running of $D_{d,s}^{(2)}(\nu_p)$ versus the running of
$d(\nu_p,\nu_p)$. 

\begin{figure}[h]
\vspace{-0.5cm}
\hspace{-0.1in}
\epsfxsize=2.8in
\centerline{\epsffile{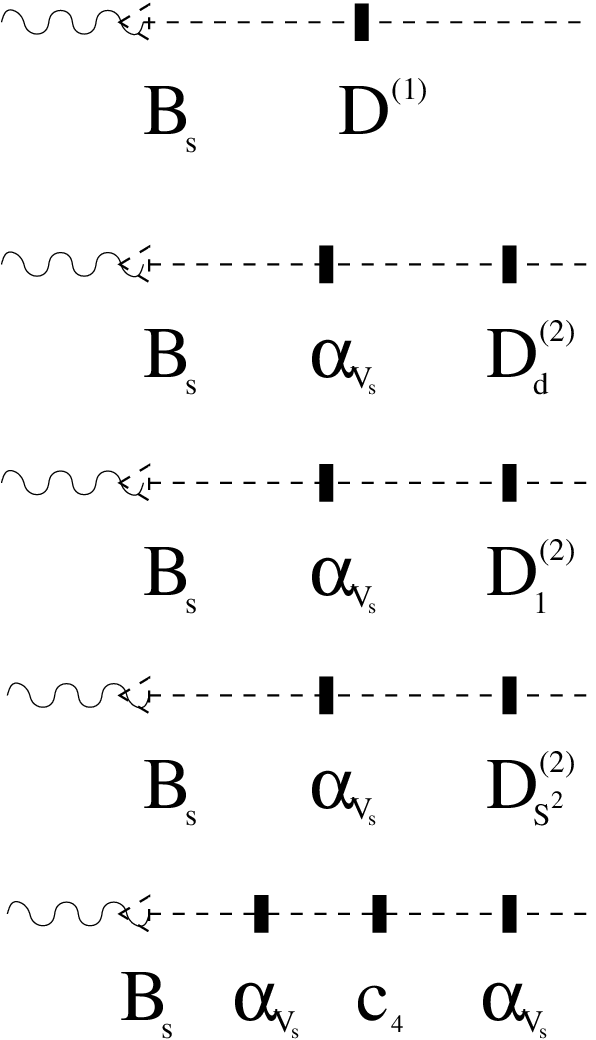}}
\caption {{\it Diagrams, up to permutations, that contribute to the
running of $B_s$ at NLL.}}
\label{Bs}
\end{figure}

The only thing left is the potential RG equation for $B_s(\nu_p)$, which is determined 
from the ultraviolet corrections to the current due to potential loops. 
The computation goes along the same lines than in the example of 
Fig. \ref{obs12}. The explicit diagrams to be computed for the RG running of
$B_s(\nu_p)$ are given in Fig. \ref{Bs} and the relevant information can be extracted from
Refs. \cite{Kniehl:1999ud,Kniehl:1999mx}. From this
figure, we can clearly illustrate the structure of the
computation. $O(1/m)$ corrections from $h_s$ only need one potential loop to kill
the $1/m$ coefficient. $O(1/m^2)$ corrections from $h_s$  need two
potential loops to kill the $1/m^2$ coefficient and so on. In the
situation with more than one potential loop, the additional potential
loops can be produced without additional $1/m$ factors coming from the
potential only if Coulomb potentials are introduced. This explains why the
$1/m$ potential needs zero Coulomb potential insertions, the $1/m^2$
potentials need one
Coulomb potential insertions and the $1/m^3$ term needs two Coulomb
potential insertions (for the running of $D_{d}^{(2)}$ and
$D_{S^2}^{(2)}$ we expect a similar structure). In principle, this
would be be a never ending story unless there is an small parameter
that tells us how far we have to go in the calculation in order to achieve
some given accuracy. This is indeed so. The $1/m$ potential is a NLL
effect and therefore higher powers in $D_{s}^{(1)}$ produce NNLL effects
or beyond. On the other hand, the introduction of Coulomb potentials
brings powers in $\al$, which suppresses the order of the calculation. In
our case, for a NLL calculation, the maximum power of the anomalous
dimension should be $\al^2$. This means that with zero $\al_{V_s}$
insertions ($O(1/m)$ potentials) the Wilson coefficient
$D_{s}^{(1)}$ has to be known with NLL accuracy, with one $\al_{V_s}$ insertion ($O(1/m^2)$ potentials) the
Wilson coefficients $D^{(2)}$ have to be
known with LL accuracy and with two $\al_{V_s}$ insertions ($O(1/m^3)$
potentials) the Wilson coefficients must have no running (this
explains why only $c_4$ is considered at this order). 

From the above discussion, the RG equation reads (in a different basis of potentials it was first obtained 
in Ref. \cite{Luke:1999kz})
\bea 
\label{cseq}
&&
\nu_p {d  \over d\nu_p} B_s(\nu_p) =B_s(\nu_p)
\left[
-{C_AC_f \over 2}D_s^{(1)}(\nu_p)
\right.
\\
\nn
&&
\qquad
\left.
-{C_f^2 \over 4}\al_{V_s}(\nu_p)
	\left\{\al_{V_s}(\nu_p)-{4\over 3}s(s+1)D_{S^2,s}^{(2)}(\nu_p)
		-D_{d,s}^{(2)}(\nu_p)+4D_{1,s}^{(2)}(\nu_p)
	\right\} 
\right]
\,,
\eea
where the RG-improved Wilson coefficients of the potentials 
can be read from previous sections or from Ref. \cite{Pineda:2000gza,Pineda:2001ra} at LL and
\cite{Brambilla:2006wp,Pineda:2011aw} at NNL (except for $D_d^{(2)}$) with the assignment $1/r \rightarrow \nu_p$ and 
$\nu_{us} \rightarrow \nu^2_p/m$. 

In an strict expansion in $\al$ the solution of Eq. (\ref{cseq}) at NLL reads
\bea 
B_s(\nu_p)&=&b_s(m)+ 
	A_1 {\al(m) \over w^{\beta_0}} \ln(w^{\beta_0})+ 
	A_2 \al(m) \bigg[z^{\beta_0}-1\bigg] + 
	A_3 \al(m) \bigg[z^{\beta_0-2 C_A}-1  \bigg]
\nn
\\
&&+
	A_4 \al(m) \bigg[z^{\beta_0-13C_A/6}-1 \bigg]+
	A_5 \al(m) \ln(z^{\beta_0})
\,,
\label{cssln}
\eea
where  $z=\left[{\al(\nu_p) \over 
\al(m)}\right]^{1 \over
\beta_0}$ and $w=\left[{\al(\nu_p^2/m) \over \al(\nu_p)}\right]^{1 \over
\beta_0}$. The coefficients $A_i$ in Eq.~(\ref{cssln}) read
\begin{eqnarray} 
\label{Acoeffs}
  A_1 &=& {8\pi C_f \over 3\beta_0^2 }\left(C_A^2 +2C_f^2+3 C_fC_A
  \right) \,,\nn\\ 
  A_2 &=& { \pi C_f [3 \beta_0(26C_A^2+19C_A C_f-32C_f^2)-
     C_A(208 C_A^2+651 C_A C_f+116 C_f^2)]\over 78\,\beta_0^2\, C_A } \,, \nn\\
  A_3 &=& -{\pi C_f^2 \Big[  \beta_0 (4s(s+1)-3)+ C_A (15-14s(s+1)) \Big]
     \over 6 (\beta_0-2 C_A)^2 }\,,\nn\\
  A_4 &=& {24 \pi C_f^2 (3\beta_0-11 C_A)(5 C_A+8 C_f) \over 13\, C_A
     (6\beta_0-13 C_A)^2}\,,\nn\\
  A_5 &=& {-\pi C_f^2 \over \beta_0^2\, (6\beta_0-13C_A) (\beta_0-2C_A)}
      \, \Bigg\{ C_A^2(-9C_A+100C_f)
	\nn\\ 
	&&+\beta_0 C_A(-74C_f+C_A(42-13s(s+1)))
	+6\beta_0^2(2C_f+C_A(-3+s(s+1)))\Bigg\}
     \,.
\end{eqnarray}
This result for $B_s (\nu_p )$ at NLL was first obtained using pNRQCD
in \cite{Pineda:2001et} and later reproduced using vNRQCD in Ref. \cite{Hoang:2002yy}. For the $B_1/B_0$ ratio 
the complete NNLL expression was computed in Ref. \cite{Penin:2004ay}. Some partial results at NNLL 
for $B_1$ and $B_0$ can be found in Ref. \cite{Pineda:2006ri}. Within vNRQCD some partial results at 
NNLL order can be found in Refs.  \cite{Hoang:2001mm,Hoang:2002yy,Hoang:2003ns}. At NNNLO the double logarithmic contributions were calculated in
\cite{Kniehl:1999mx} and the single logarithmic ones in
\cite{Kniehl:2002yv}. 

\section{Observables}
\label{Observables}

We are now in the position to compute observables. We only review 
observables (being the theoretically cleanest ones) that   
involve the calculation of the NR propagator of a quark-antiquark pair $G(E,{\bf r}, {\bf r}')$ only. 
Therefore, 
besides the heavy quarkonium spectrum (i.e. the poles of the Green function),
we consider inclusive (electromagnetic) decay widths, NR sum rules and
$t$-$\bar t$ production near threshold. For these only the normalization at
the origin is important, i.e. (the imaginary part of) $G(E,{\bf 0}, {\bf 0})$ has to be computed.

\subsection{Spectrum}
\label{sec:spec}
Given a bound state with with quantum number $n$, where $n$ generically denotes the quantum number of the bound state: 
$n \rightarrow$ ($n$ (principal quantum number), $l$ (orbital angular momentum), $s$ (total spin), $j$ (total angular momentum)), 
its  spectrum, $E_n$, and NR wave function, $\phi_n({\bf r})$,  are determined from the behavior of the NR Green function near the pole $E_n$
\be
G(E,{\bf r},{\bf r}') =
 \phi_n({\bf r})\phi^*_n({\bf r}')\frac{1}{E_n-E-i\eta}+{\cal O}((E_n-E)^0)
\ee
and can be recursively determined from the Coulomb (or other LO) solution: \\
\be
G(E,{\bf r},{\bf r}') 
 \simeq
 \frac{\phi_n({\bf r})\phi^{*}_n({\bf r}')}{E_n^C-E}-\frac{\phi_n({\bf r})\phi^{*}_n({\bf r}')\delta E_n}{(E_n^C-E)^2}+
 {\cal O}((E^C_n-E)^0,(E^C_n-E)^{-3})
\,,
\ee
where $E_n=E_n^C+\delta E_n$ and $E_n^C=-mC_f^2\al^2/(4n^2)$ is the solution of $h_s^C\phi_n^C=E_n^C\phi_n^C$. This formulation is general. Nevertheless, if we want to connect with the solution of a pure NR Schrödinger equation, we may consider the solution of the pure potential problem (with no ultrasoft effects):
\be
\label{EnpotMS}
h_s\phi^{pot}_{\MS,n}({\bf r})=\left(\frac{p^2}{2m_r}+V^{RG}_{s,\MS}\right)\phi^{pot}_{\MS,n}({\bf r})
=E_{\MS,n}^{pot}\phi^{pot}_{\MS,n}({\bf r})
\,,
\ee
where $V^{RG}_{s,\MS}$ has been defined in Eq. (\ref{VsMS})\footnote{Do not forget that Eq. (\ref{deltahm3}) also has to be added.}.
Then, 
\be
\label{Enjls}
E_{n}=\left(E^{pot}_{\MS,n}+\delta E^{us}_{\MS,n}\right)
\,,
\ee
where the ultrasoft effects are encoded in $\delta E_n^{us}$. Note that 
 at high enough orders $\delta E_n^{us}$ will include potential 
loops beside ultrasoft loops (it is important then to keep the potential in $D$ dimensions). We have also made explicit
 that, in general, $E_n^{pot}$, $\phi^{pot}_n$ will depend on the factorization scale and renormalization scheme the ultrasoft 
 (and potential) computation has been performed with.

The exact solution of Eq. (\ref{EnpotMS}) correctly produces all necessary soft and potential terms at N$^3$LL accuracy for $l\not=0$ and $s=0$ states (otherwise the precision is NNLL), as well as some subleading terms. Such exact solution would only be possible to obtain through numerical methods (which on the other hand could actually be more easy to implement in practice). If we want to restrict ourselves to an strict N$^3$LL computation (in particular if seeking for an explicit analytical result), Eq. (\ref{EnpotMS}) should be computed within quantum mechanics perturbation theory up to NNNLO for general quantum numbers. Up to NNLO such computation was performed in Ref. \cite{Pineda:1997hz}. The lengthy N$^3$LO computation is 
known for $l=0$ and $n=1$ \cite{Penin:2002zv} but missing for general quantum numbers.

The energy shift due to the ultrasoft correction can be 
written in the following compact form
\bea
\label{deltaEusMSnl}
&&
\delta E^{us}_{\MS,n}
=
C_f V_A^2\langle n |
{\bf r}(h_o-E_{n})^3
\Biggr[
-\frac{\al }{9 \pi } \left(6 \ln \left(\frac{h_o-E_{n}}{\nu_{us} }\right)+6\ln 2-5\right)
\\
&& 
\nn  
+\frac{\al^2 }{108 \pi ^2} 
\left(18 \beta_0 \ln ^2\left(\frac{h_o-E_{n}}{\nu_{us}
   }\right)-6 \left(N_c \left(13+4 \pi ^2\right)-2 \beta_0 (-5+3\ln 2)\right) \ln \left(\frac{h_o-E_{n}}{\nu_{us}
   }\right)
   \right.
   \\
   &&
   \nn
   \left.
+2C_A \left(84-39 \ln 2+4 \pi ^2 (2-3\ln 2)-72 \zeta (3)\right)+\beta_0 \left(67+3 \pi ^2-60 \ln 2+18\ln^2 2\right)\right)
   \Biggr]{\bf r}
   | n \rangle 
\,,
\eea 
where the states $|n \rangle $ and the energies $E_{n}$ used above are the solution of the Schrödinger potential 
including the 1-loop static potential (i.e. with NLO accuracy): 
\be
\left[
\frac{{\bf p}^2}{2m_r}-\frac{C_f\,\al(\nu)}{r}\,
\bigg\{1+\frac{\al(\nu)}{4\pi} a_1(\nu;r)\bigg\}
\right]| n,l \rangle=E_{n,l}| n,l \rangle
\,,
\ee
and $h_o$ is approximated to its NLO expression:
\be
h_o=\frac{{\bf p}^2}{2m_r}+\frac{1}{2N_c}\frac{\,\al(\nu)}{r}\,
\bigg\{1+\frac{\al(\nu)}{4\pi} a_1(\nu;r)\bigg\}
\,.
\ee
Eq. (\ref{deltaEusMSnl}) includes 
the complete LO ${\cal O}(m\al^5)$ and NLO  ${\cal O}(m\al^6)$ ultrasoft effects, as well as subleading effects. The LO expression would be 
enough for the N$^3$LL precision and reads (for $l=0$)
\begin{eqnarray}
\delta E^{us}_{\MS,n}&=&-{2\al^3\over 3\pi}E^C_n\left\{
\left[{1\over 4}C_A^3+{2\over n}C_A^2C_f+
\left({6\over n}-{1\over n^2}\right)C_AC_f^2+
{4\over n}C_f^3\right]
\left(\ln{\nu_{us}\over E^C_1}+{5\over6}-\ln{2}\right)+C_f^3L^E_n\right\}\,,
\nn
\\
\eea
where $L_n^E$ stands for the non-abelian Bethe logarithms. A semianalytic expression exists for $l=0$ states \cite{Kniehl:1999ud}
but is missing for general quantum numbers. Some expressions can be found in the Appendix. 

In a strict fixed order computation one should expand the wave functions and $h_o-E_{n}$ to the appropriated order in Eq. (\ref{deltaEusMSnl}), but  in some situations it could be more convenient to handle this expression numerically.
 
Eq. (\ref{deltaEusMSnl}) is only valid when $m\al^2 \gg \lQ$, In this situation the first nonperturbative effects can be written in terms of local condensates and 
were first computed by Voloshin and Leutwyler \cite{Leutwyler:1980tn,Voloshin:1979uv}. The subleading corrections can be found in Ref. \cite{Pineda:1996uk}.

\subsection{Coupling to hard photons. Vacuum polarization in the NR limit}

The vacuum polarization due to the heavy quark-antiquark pair reads
\be
(q_\mu q_\nu-g_{\mu\nu}q^2)\Pi(q^2)=i\int d^4x e^{iq\cdot x}\langle 
0|T\{j_\mu(x)j_\nu(0)\}|0\rangle
\,,
\ee
where $q=(\sqrt{s},0)$ and  $j_\mu(x)\equiv \bar Q \gamma_\mu Q(x)$.  The NR limit of its imaginary term, or more specifically:
\begin{equation}
R(E)\equiv 
\frac{\sigma(e^+e^-\to Q\bar Q) }{\sigma(e^+e^-\to \mu^+\mu^-)} = \frac{4\pi}{s}e_Q^2\,  {\rm Im} \left(-i \int d^4 x\,  e^{i q\cdot x} 
\langle 0 | T\{j^\mu(x)\, j_\mu(0)\} | 0\rangle \right)
\,,
\label{Rcorr}
\end{equation}
where $E=\sqrt{s}-2m$ and $e_Q$ is the electric charge of the heavy quark ($e_b=-1/3$, $e_c=2/3$, $e_t=2/3$),  
 is the crucial ingredient to describe several physical processes, like NR Sum rules ($b$-$\bar b$, $c$-$\bar c$),
$t$-$\bar t$ production near threshold, and inclusive electromagnetic decay widths. These processes are all mediated through the coupling of Heavy Quarkonium to hard photons (those with energy of the order of the heavy quark mass). 
The NR expression for $R(E)$ can be related with the (spin one) NR Green function defined in Eq. (\ref{Gsfull}) through the equality
\begin{equation}
\label{RE}
R(E)=\frac{24\pi e_Q^2N_c}{s}
\left(B_1^2-B_1d_1\frac{E}{3m}\right)
{\rm Im}\,G_{s=1}(E,{\bf 0},{\bf 0})
\,,
\end{equation}
which is valid with NNLL accuracy. Hence, the full QCD
calculation can be split into calculating the Wilson coefficients of the current, $B_1$ and $d_1$, and the NR Green function. $B_1$ has been discussed in Sec. \ref{secRGcurrent} and is known at NLL. $d_1$ has been computed at LL in Refs. \cite{Hoang:2001mm,Pineda:2006ri}. 
The spectral decomposition of the NR 
Green function would be the standard one of NR quantum mechanics\footnote{We have also included 
a decay width $\Gamma_t$, relevant for $t$-$\bar t$ production near threshold.}:
$$
G(E,{\bf 0},{\bf 0})=\sum_{m=0}^{\infty}
{|\phi_{m}({\bf 0})|^2 \over E_{m}-E+i\eta-i\Gamma_t} 
+
{1 \over \pi} \int_0^{\infty}
dE'{|\phi_{E'}({\bf 0})|^2 \over E_{E'}-E+i\eta-i\Gamma_t}
$$
where the wave function and energies are the same as those appearing in Sec. \ref{sec:spec}. 
On the other hand, the 
LO, Coulomb, NR Green function reads
\begin{eqnarray}
G^{(D)}_c(E,{\bf 0},{\bf 0}) &=&\frac{m_r}{2\pi}
\bigg[A_{\MS}^{(D)}(\epsilon;\nu)+B_{V_s^{C}}^{\MS}(E;\nu)+{\cal O}(\epsilon)\bigg]\,,
\label{eq:GDCoulomb}
\end{eqnarray}
where
\begin{equation}
A^{(D)}_{\MS}(\epsilon;\nu)
=-\frac{g^2\, C_f\, m_r}{8\pi\epsilon}
\left(\frac{\nu^2 e^{\gamma_E}}{4\pi}\right)^{2\epsilon}
\, ,
\label{eq:AD}
\end{equation}
\begin{equation}
\label{eq:BMSC}
B^{\MS}_{V_s^C} = 2m_r\,C_f\,\al
\left(
-\frac{1}{2\lambda}
- \frac{1}{2} \ln\left(\frac{-8 m_r E}{\nu^2}\right)
       + \frac{1}{2} - \gamma_E - \psi(1-\lambda)
	   \right)
\,,
\end{equation}
and $\lambda\equiv C_f\, \al/( \sqrt{-2E/m_r})$, though in this section we are only interested in the imaginary part of 
$G(E,{\bf 0},{\bf 0})$. For a decomposition of the Coulomb Green function in terms of partial waves for a general 
${\bf x}$, and ${\bf x}'$ see the Appendix. For a discussion on the three-point Coulomb Green function see Ref. \cite{Kiselev:1999sc}.

Beyond LO, there is and has been an ongoing effort in obtaining the NR Green function, Im$G(E,{\bf 0},{\bf 0})$,  
with higher degree of accuracy (either at finite order or with RG improvement). 
Actually the first computation of this object comes back to \cite{Fadin:1988fn} at LO and 
\cite{Strassler:1990nw} at NLO (though both relied on potential models). 
At  NNLO (${\cal O}(\al^ 2)$ corrections), 
it reduces to a purely quantum-mechanical calculation along the lines
of Sec.~\ref{QMPT} (ultrasoft gluons do not play any role).  This calculation has been carried out by
several groups and
the final outcome is summarized in \cite{Hoang:2000yr}. At  NNNLO precision there are some partial results
 \cite{Kniehl:1999mx,Kniehl:2002yv,Hoang:2003ns,Penin:2005eu,Beneke:2005hg,Beneke:2007gj,Beneke:2007pj,Smirnov:2008pn}. The NLL and NLO expression for the NR Green function are equal, as at this level the resummation of logarithms only 
affect the current Wilson coefficient $B_s$, the NLL expression of which have been shown in 
 Sec. \ref{secRGcurrent} \cite{Pineda:2001et}. The NNLL expression can also easily deduced (see Ref. \cite{Pineda:2006ri}), as it only requires the introduction of the potential Wilson coefficients. Therefore, the missing term for the complete 
 NNLL expression of $R(E)$ is the NNLL running of $B_1$.
 
\medskip

\noindent
{\bf $t$-$\bar t$ production near threshold}.\\
The $t$-$\bar t$ pair will be dominantly produced via $e^+ e^- \rightarrow
\gamma^\ast \, ,\, Z^\ast \rightarrow t\bar t$ with the centre of mass energy
$\sqrt{q^2}=\sqrt{s}\sim 2 m$. 
In order to simplify the discussion we ignore the
$Z$ exchange in what follows. The cross section can then be written as
\begin{equation}
\sigma^\gamma (s) = \frac{4 \pi \alpha^2_{\rm EM}}{3\, s}\, R(E)
\,,
\label{sigma}
\end{equation}
where $\alpha_{\rm
EM}=\frac{e^2}{4\pi}$ is the electromagnetic coupling and should be computed at the hard scale in this observable. Therefore, $\sigma^\gamma (s) $ can be directly read from the results discussed
above. Nowadays the precision is NNLO/NLL, with partial NNLL/NNNLO results, as far as QCD effects is concerned. A phenomenological analysis is given in Sec. \ref{phen:hardphotons}.

\medskip

\noindent
{\bf NR Sum rules}.\\
Using causality and the optical 
theorem one obtains
\be
\label{moments}
M_n\equiv {12\pi^2e_Q^2 \over n!}\left({d \over dq^2}\right)^n\Pi(q^2)|_{q^2=0} =
\int_{\sqrt{s_{min}}}^\infty {ds \over s^{n+1}}R(E) \,.
\ee 
For large values of $n$, new scales appear in the problem, besides $m$ and $\lQ$,
like $m/\sqrt{n}$, $m/{n}$ and so on. As usual in this review we take $n$ such that those scales are much larger 
than $\lQ$, and neglect any nonperturbative effect. On the other hand, for $n$ large
enough, one will have $\sqrt{n}\al \sim 1$ and a complete resummation of
these terms should be achieved\footnote{In practical applications the boundary for 
applying both conditions is usually taken around $n \sim 10$ for the bottom case.}. The quantity $\sqrt{n}\al$ appears in the
computation through the ratio of two different scales: $(m\al)/(m/\sqrt{n})$.
Hence, we  see the following analogy with the NR situation: $1/\sqrt{n}$ plays
the same role as $v$, the velocity of the heavy quark, and by taking
$\sqrt{n}\al \sim 1$  we are considering the NR limit.

The theoretical expressions for the moments $M_n^{th}$ can be computed order 
by order in the NR expansion in $1/\sqrt{n}$ and $\al$, which at each order
resums all the terms proportional to $\al \sqrt{n}$ to any power. Nowadays
they are known in the on-shell scheme at NNLO, which
includes all corrections up to order $1/n$, $\al/\sqrt{n}$ and $\al^2$
\cite{Kuhn:1998uy,Penin:1998zh,Hoang:1998uv,Melnikov:1998ug,Beneke:1999fe}, and NNL.
The NNLL expression is also known except for the NNLL running of $B_1$. 
With this accuracy, the dispersion integration for the moments
$M_n$ takes the form
\bea
\label{Pnexpression1}
M_n &=& 48 \pi e_Q^2 N_c 
\int_{-\infty}^\infty \frac{dE}{(E+2m)^{2n+3}}
\left(B_1^2 - B_1 d_1 \frac{E}{3m}\right) 
{\rm Im}\, G_{s=1}(E,{\bf 0},{\bf 0}) 
\\
\nn
& \simeq &
\frac{3\pi N_ce_Q^2}{2^{2n-1}\,m^{2n+2}}\,\int\limits_{E_1}^\infty 
\frac{d E}{m} \,\exp\bigg (\,
-\frac{E}{m}\,n
\,\bigg )\,\bigg(\,
1 - \frac{E}{2\,m} + \frac{E^2}{4\,m^2}\,n
\,\bigg)
\left(
B_1^2 - B_1 d_1 \frac{E}{3m}
\right){\rm Im}\, G_{s=1}(E,{\bf 0},{\bf 0}) 
\,,
\eea
where $E_1$ is the binding
energy of the lowest lying resonance. The exponential form of the LO
NR contribution to the energy integration has to be
chosen because $E$ scales like $v^2\sim 1/n$.
For some analytic expressions of the moments we refer to \cite{Hoang:1998uv}.

\medskip

\noindent
{\bf Inclusive electromagnetic decay widths}.\\
The spin-one decay is known with NNLL accuracy (except for $B_1$):
\begin{eqnarray}
\label{vector}
\Gamma(V(nS) \rightarrow e^+e^-) = 
16\pi \, 
{C_A \over 3} \left[ { \alpha_{\rm EM}\, e_Q \over M_{V(nS)}} \right]^2
\left|\phi^{(s=1)}_n({\bf 0}) \right|^2
\left\{ B_1-d_1{M_{V(nS)}-2m \over 6m} \right\}^2
\,.
\end{eqnarray}

The corrections to the wave function at the origin are obtained by
taking the residue of the Green function at the position of the poles
\begin{equation}
\left|\phi^{(s=1)}_n({\bf 0}) \right|^2
=
\left|\phi^{C}_n({\bf 0})\right|^2
\left(1+\delta \phi^{(s=1)}_n\right)
=
{\
\vbox{\hbox{\rlap{Res}\lower9pt\vbox
{\hbox{$\scriptscriptstyle{E=E_n}$}}}}\
} \! 
 G_{s=1}(E,{\bf 0},{\bf 0})
\,,
\end{equation}
where the LO wave function is given by
\begin{equation}
\left|\phi^{C}_n({\bf 0})\right|^2=
{1 \over \pi} \left({mC_f\al \over 2n }\right)^3.
\end{equation}
The corrections to $\delta \phi_n^{(s=1)}$ 
produced by $\delta h_s$ have already been calculated with NNLO 
accuracy \cite{Melnikov:1998ug,Penin} in the direct matching scheme. One can 
also obtain them in the dimensional regularized $\MS$ scheme with 
NNLL accuracy by incorporating the RG improved Wilson coefficients. One obtains 
then the following correction to the wave function \cite{Pineda:2006ri}\footnote{We thank Yuichiro Kiyo for pointing out that $3 \rightarrow 5$ in the $c_4$ term.}
\beqa
&&
\delta \phi_n^{(s=1)}
= \frac{\al^2C_A^3}{2\beta_0}
\log{\left[\al(\nu_p)\over \al(\nu_p/(2m))\right]}
\\
\nn
&&
+ \, {\al \over \pi}
\left(\frac{3\, a_1}{4} +
{\beta_0\over 2}
\left(3L[n]
    + S[1,n] + 2 n S[2,n]-1-\frac{n \pi^2}{3}\right) \right)
\\
\nn
&&
+ \, C_fC_A D_s^{(1)}\,
\left(L[n]
- S[1,n] + {2\over n}+{5 \over 4} \right)
\\
\nn
&&
+ \, 2 C_f^2 \al  D_{1,s}^{(2)}
\left(L[n] - S[1,n] 
- \frac{5}{8n^2} + \frac{2}{n} + \frac{3}{2}  \right)
\\
\nn
&&
- \, {C_f^2 \al  D_{S^2,s}^{(2)} \over 3}
\left(L[n] - S[1,n]
+{2\over n} + {11 \over 12} \right)
\\
\nn
&&
- \, {3C_f^2 \al  D_{d,s}^{(2)} \over 2}
\left(L[n] - S[1,n]
+{2\over n}+{1 \over 2} \right)
\\
\nn
&&
- \, {C_f^2 \over 4}D_{2,s}^{(2)}\al
\\
\nn
&&
+ \, c_4{C_f^2  \al^2 \over 2}
\left(L[n] - S[1,n]
- \frac{5}{4n^2} + \frac{2}{n} + \frac{3}{2} \right)
\\
\nn
&&
+\frac{\al^2}{(4\pi)^2}
\bigg(
3a_1^2 + 3a_2 - 14a_1\beta_0 + 4\beta_0^2 - 2\beta_1 + \beta_0^2\pi^2 
- \frac{8a_1\beta_0n\pi^2}{3} + \frac{4\beta_0^2n\pi^2}{3} 
- \frac{2\beta_1n\pi^2}{3} 
+ \frac{\beta_0^2n^2\pi^4}{9} \\
\nn
&&
+ 24a_1\beta_0
  L[n] 
- 28\beta_0^2L[n] 
+ 6\beta_1L[n] 
- \frac{16\beta_0^2n\pi^2}{3}L[n] 
+ 24\beta_0^2L[n]^2 
\\
\nn
&&
+ 8a_1\beta_0S[1, n] - 20\beta_0^2S[1, n] + 2\beta_1S[1, n] 
- \frac{12\beta_0^2S[1, n]}{n} - \frac{8\beta_0^2n\pi^2S[1, n]}{3} 
\\
\nn
&&
+ 16\beta_0^2L[n]S[1, n] 
+ 8\beta_0^2S[1, n]^2 
+ 8\beta_0^2S[2, n] + 16a_1\beta_0nS[2, n] 
\\
\nn
&&
- 8\beta_0^2 n S[2, n] 
+ 4\beta_1 n S[2, n] - \frac{4\beta_0^2n^2\pi^2S[2, n]}{3} 
+ 32\beta_0^2nL[n]S[2, n] 
\\
\nn
&&
+ 
   16\beta_0^2nS[1, n]S[2, n] + 4\beta_0^2n^2S[2, n]^2 + 28\beta_0^2nS[3, n] 
   - 20\beta_0^2n^2S[4, n] 
   \\
\nn
&&
- 24\beta_0^2nS_2[2, 1, n] + 16\beta_0^2n^2S_2[3, 1, n] + 
   20\beta_0^2n\, \zeta(3)\bigg)
\,,
\end{eqnarray}
where 
\begin{equation}
L[n]= \log{\left[\frac{\mu_s n}{mC_f\al}\right]}
\,,
\quad
S[a,n]=\sum_{k=1}^n \frac{1}{k^a}
\,,
\quad
S_2[a,b,n]=\sum_{k=1}^n \frac{1}{k^a}S[b,k]
\,.\end{equation}

From these expressions one can obtain the wave-function correction 
for the spin zero case
\begin{equation}
\delta \phi_n^{(s=0)}=\delta \phi_n^{(s=1)}+\delta \phi_n^{\Delta s}
\,,
\end{equation}
where
\begin{equation}
\delta \phi_n^{\Delta s}
=
-\frac{2}{3}C_f^2  D_{S^2,s}^{(2)} \al
\left(-2 L[n]+2 S[1,n]
-\frac{4}{n} -\frac{7}{3}\right)
\end{equation}
by using the results from Ref.~\cite{Penin:2004ay}. Therefore, the 
decay of the pseudoscalar Heavy Quarkonium to two photons reads
($d_0=d_1$)
\begin{eqnarray}
\label{pseudo}
\Gamma(P(nS) \rightarrow \gamma\gamma) = 
16 \pi\,  C_A \left[ { \alpha_{\rm EM}\, e_Q^2 \over M_{P(nS)}} \right]^2
\left|\phi^{(s=0)}_n({\bf 0}) \right|^2
\left\{
B_0-d_0{M_{P(nS)}-2m \over 6m}
\right\}^2
\,.
\end{eqnarray}

\medskip

\noindent
{\bf Ratios}\\
For the previous observables the complete NNLL expression is at present unknown. This is not 
so for the Heavy Quarkonium production and annihilation spin ratio, which we define as
\begin{equation}
{\cal R}_Q=
{\sigma(e^+e^-  \rightarrow {\cal Q}(n^3S_1) )\over
\sigma(\gamma\gamma \rightarrow {\cal Q}(n^1S_0))}=
{\Gamma({\cal Q}(n^3S_1)\to e^+e^-)\over
\Gamma({\cal Q}(n^1S_0)\to  \gamma\gamma)}\,,
\end{equation}
and its effective theory expression reads
\begin{equation}
{\cal R}_Q={1\over 3e_Q^2}\frac{B_1^2}{B_0^2}
{|\phi_n^{(s=1)}({\bf 0})|^2\over|\phi_n^{(s=0)}({\bf 0})|^2}+{\cal O}(\al v^2)\,.
\label{Rdef}
\end{equation}
This object is known with NNLL accuracy \cite{Penin:2004ay}.

\section{Summary for the practitioner}
\begin{figure}[htb]
\vspace{-0.1in}
\hspace{-0.3in}
\epsfxsize=4.1in
\centerline{\epsffile{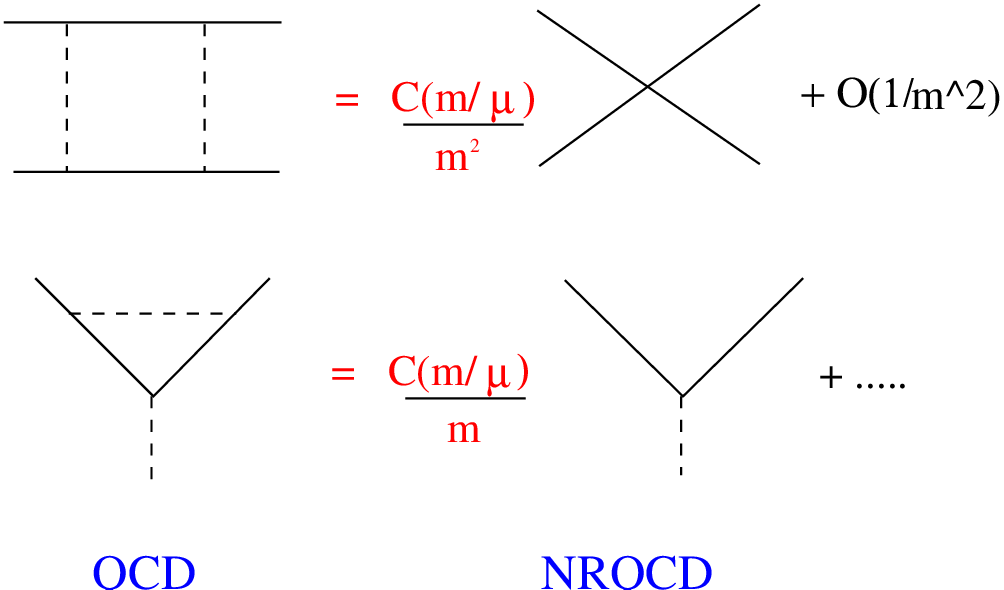}}
\caption{\it Examples of matching between QCD and NRQCD.}
\label{fig:NRQCDmatching}
\end{figure}

In this section, we summarize the four main techniques needed in order 
to efficiently perform high-precision perturbative computations 
in weakly-coupled NR bound state systems:  
\begin{enumerate}
\item
Matching QCD to NRQCD: Relativistic Feynman diagrams
\item
Matching NRQCD to pNRQCD (getting the potential): 
NR (HQET-like) Feynman diagrams
\item
Observable: Quantum mechanics perturbation theory 
\item
Observable: Ultrasoft loops 
\end{enumerate}
The first two points explain the techniques needed to obtain pNRQCD 
from QCD, whereas the last two explain the kind of computations faced in 
the EFT when computing observables. All the computations can be performed in dimensional regularization 
and only one scale appears in each type of integral, which becomes homogeneous. 
This is a very strong simplification 
of the problem. In practice this is implemented in the following way:

{\bf Point 1)}.  One analytically expands over the three-momentum and residual energy
in the integrand before the integration is made in both the full and
the effective theory \cite{Manohar:1997qy,Pineda:1998kj}. 
\bea
\nn
 {\bf QCD} \qquad
\int d^4q f(q,m,|{\bf p}|,E)&=& \int d^4q f(q,m,0,0)+
{\cal O}\left({E \over m},
{|{\bf p}|\over m} \right)
\sim C(\frac{\nu_{\rm NR}}{m})({\rm tree\ level})|_{NRQCD}
\\
{\bf NRQCD} \qquad
 \int d^4q f(q,|{\bf p}|,E)&=& \int d^4q f(q,0,0)= 0\,!!
\eea
Therefore, the computation of loops in the effective theory just gives {\it zero} and 
{\it
one matches loops in QCD with only one scale (the mass) to tree level diagrams in NRQCD}, which we 
schematically draw in Fig. \ref{fig:NRQCDmatching}.

\begin{figure}[htb]
\makebox[0.0cm]{\phantom b}
\put(100,153){${\bf p}$}\put(120,162){$>$}
\put(170,153){${\bf p}'$}\put(190,162){$>$}
\put(155,130){${\bf k}={\bf p}-{\bf p}'$}
\put(415,130){$=\tilde V^C$}
\put(415,50){$=\delta \tilde h$}
\put(100,100){\epsfxsize=10truecm \epsfbox{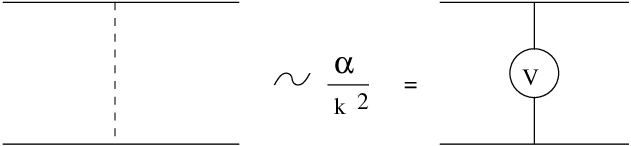}}
\put(100,20){\epsfxsize=10truecm \epsfbox{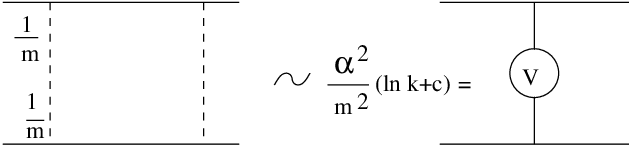}}
\put(145,0){NRQCD}
\put(320,0){pNRQCD}
\vspace{1mm}
\caption{\it Examples of matching between NRQCD and pNRQCD.}
\label{fig:matpNR}
\end{figure}
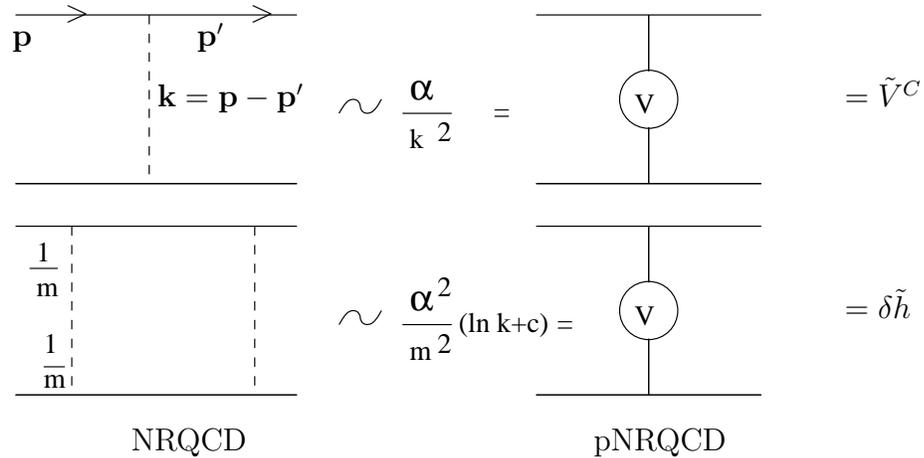

{\bf Point 2)} works analogously \cite{Pineda:1998kn}. One expands in the
scales that are left in the effective theory. We integrate out the
scale {\bf k} (transfer momentum between the quark and antiquark) or 
its Fourier transform variable ${\bf r}$. 
Again loops in the EFT are zero and only tree-level 
diagrams have to be computed in the EFT: 
\bea
{\bf NRQCD}  \qquad \int d^4q f(q,k,|{\bf p}|,E)&=& 
\int d^4q f(q,k,0,0)+{\cal O}\left({E \over k},
{|{\bf p}|\over k} \right) \sim \delta \tilde h ({\rm potential})
\\
{\bf pNRQCD}
\qquad
\int d^4q f(q,|{\bf p}|,E)&=& \int d^4q f(q,0,0)= 0\,!!
\eea
We illustrate the matching in Fig. \ref{fig:matpNR}.  Formally the one-loop diagram 
is equal to the QCD diagram shown before. The difference is that it  
has to be computed with HQET quark propagators ($1/(q^0+i\eta)$) and 
that the vertices are also different.

Once we have obtained the potentials we have all the ingredients of the pNRQCD Lagrangian. In order 
to write it in a more compact form, with gauge invariance and the multipole expansion explicit, 
is convenient to project to the quark-antiquark sector and to express the quark-antiquark state  
in terms of a single bilinear field, which, by means of field redefinitions, is decomposed in $S$ and $O$, two 
fields that transform as a singlet and octet under ultrasoft gauge transformations. Finally, 
\be
{\cal L}={\rm Tr}\left[S\left(i\partial^0-h_s^{C}-\delta h_s\right)S+O\left(iD^0-h_o\right)O+V_AS{\bf r}\cdot {\bf E}O+\cdots\right]
\ee
where $h_s^{C} = {{\bf p}^2 \over m}+V_s^{C}(r)$ and $\delta h_s$ schematically represents the corrections to the 
potential.

{\bf Observables}. Once the pNRQCD Lagrangian has been obtained one can 
compute observables. A key quantity in this respect is 
the Green function. In order to go beyond the LO 
description of the bound state one has to compute 
corrections to the Green Function 
($H_I \sim {\bf x}\cdot {\bf E}$ schematically represents 
the interaction with ultrasoft gluons of the singlet and octet field):
$$
G(E) \sim {1 \over h^{C}_s+\delta h_s-H_I-E}=G_c+\delta G
\qquad
G_c(E)={1 \over \displaystyle{h_s^C-E}}
\ .
$$
These corrections can be organized as an expansion
in $1/m$, $\al$ and the multipole
  expansion. Two type of integrals appear then, which correspond to points 3) and 4) above.
  
{\bf Point 3)}. For example, if we were interested in computing the spectrum at $O(m\al^6)$ 
(for QED see \cite{Czarnecki:1999mw}), one
should consider the iteration of subleading potentials ($\delta h_s$) in the
propagator:

\bigskip

\begin{figure}[h]
\makebox[0.0cm]{\phantom b}
\put(10,1){$\displaystyle{\delta G^{pot.}=}$}
\put(60,1){\epsfxsize=3truecm \epsfbox{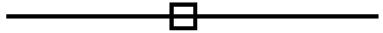}}
\put(180,1){\epsfxsize=3truecm \epsfbox{pot_insert.eps}}
\put(220,1){\epsfxsize=3truecm \epsfbox{pot_insert.eps}}
\put(93,12){$\delta h_s$}
\put(213,12){$\delta h_s$}
\put(253,12){$\delta h_s$}
\put(320,1){$+ \cdots$}
\put(160,1){$+$}
\put(30,-24){$\displaystyle{\sim {1 \over h_s^{C}-E}\delta h_s {1 \over h_s^{C}-E}
+
{1 \over h_s^{C}-E}\delta h_s {1 \over h_s^{C}-E}\delta h_s {1 \over h_s^{C}-E}
+\cdots}$.}
\end{figure}

At some point, these corrections produce divergences. For example, a correction of the 
type: $\delta(r)G_c(C_f\al/r)G_c\delta(r)$, would produce the following divergence
\bea
&&
\langle{\bf r}=0|
{1 \over \displaystyle{E-{\bf p}^2/m}}
C_f {\al \over r}
{1 \over \displaystyle{E-{\bf p}^2/m}}
|{\bf r}=0\rangle
\\
\nn
&&
\qquad
\sim 
\int \frac{
{\rm d}^d p' }{ (2\pi)^d } \int \frac{ {\rm d}^d p }
{ (2\pi)^d } \frac{ m }{{\bf p}'^2 - mE } 
C_f
\frac{ 4\pi\al }{ ({\bf p-p'})^2 } \frac{ m }{{\bf p}^2-m E } 
\sim
- C_f\frac{m^2\al}{16\pi}  
\left(\frac{ 1 }{\epsilon }+2\ln(\frac{mE}{\nu_p})+\cdots\right).
\eea
Nevertheless, the existence of divergences in the effective theory is not a problem, since 
they get absorbed in the potentials ($\delta h_s$). 

\noindent
{\bf Point 4)}. The same happens 
with ultrasoft gluons, \cite{Pineda:1997ie,Kniehl:1999ud,Brambilla:1999xj}:
\begin{figure}[htb]
\makebox[0.0cm]{\phantom b}
\put(0,1){$\delta G^{\rm us}=$}
\put(50,-2){\epsfxsize=5truecm \epsfbox{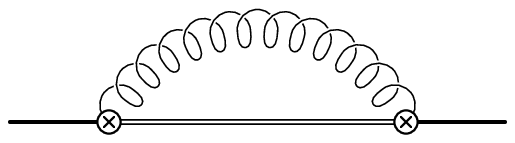}}
\put(82,-3){$\underbrace{\hbox{~~~~~~~~~~~~~~~~~~~~~}}$}
\put(93,-19){$1/(E-h_o)$}
\put(200,1){$\displaystyle{\sim  G_c(E)\int { d^{d}{\bf k} 
\over (2\pi)^{d}}
{\bf r}\,{k \over k+h_o-E}
{\bf r}\,G_c(E)}$}
\end{figure}
\be
\sim G_c(E)\,
{\bf r}\, (h_o-E)^3
\left\{
{1 \over \epsilon}+\gamma+\ln{(h_o-E)^2 \over \nu_{us}^2}+C
\right\}\,
{\bf r}\, G_c(E)
\,,
\ee
which also produces divergences that get absorbed in $h_s$. Overall, we get a 
consistent EFT.

\medskip

By obtaining the poles of the Green function one obtains the spectroscopy of the bound 
state. From the normalization of the Green function one can obtain inclusive electromagnetic 
decays, NR sum rules, and, in general, describe heavy quarkonium production 
near threshold. All these observables can be obtained from the vacuum polarization
$$
(q_\mu q_\nu-g_{\mu\nu})\Pi(q^2)=i\int d^4x e^{iqx}\langle {\rm vac}|
T\{j_{\mu}(x)j_{\nu}(0)\}|{\rm vac} \rangle
\,,
$$
which in the NR limit schematically reads
$$
j^{i}=\bar Q \gamma^{i} Q=B_1\psi^\dagger\bfsigma^i \chi+\cdots
\,,
\qquad
B_1=1+a_1\al+a_2\al^2+\cdots
\,,
$$
$$
\Pi(q^2)\sim B_1^2\langle {\bf r}={\bf 0}|G(E)| {\bf r}={\bf 0}\rangle
=B_1^2G(E,{\bf 0},{\bf 0})
$$
$$
G(E,{\bf 0},{\bf 0})=\sum_{m=0}^{\infty}
{|\phi_{m}({\bf 0})|^2 \over E_{m}-E+i\eta-i\Gamma_t} 
+
{1 \over \pi} \int_0^{\infty}
dE'{|\phi_{E'}({\bf 0})|^2 \over E_{E'}-E+i\eta-i\Gamma_t}
\,.
$$
For instance, for inclusive electromagnetic decays we would have
\be
\Gamma(V \rightarrow e^+e^-) \sim {1 \over m^2}B_1^2|\phi_n({\bf 0})|^2
\ee
\begin{equation}
\label{decay}
\left|\phi_n({\bf 0}) \right|^2
=
\left|\phi^{C}_n({\bf 0})\right|^2
\left(1+\delta \phi_n\right)
=
{\
\vbox{\hbox{\rlap{Res}\lower9pt\vbox
{\hbox{$\scriptscriptstyle{E=E_n}$}}}}\
} \! 
 G(E, {\bf  0},{\bf  0})
\,.
\end{equation}
where {\bf $
\displaystyle{\left|\phi_n({\bf 0}) \right|^2}
$}
is scheme and scale dependent.

For heavy quarkonium production we would have 
\be
\sigma_{t-\bar t}(s) \sim B_1^2{\rm Im}G_{s=1}(\sqrt{s}-2m,{\bf 0},{\bf 0})+\cdots
\ee
and for NR sum rules 
\be
\label{Mn}
M_n \equiv \frac{12\pi^2 e_Q^2}{n!} 
\left(\frac{d}{dq^2}\right)^n \Pi(q^2)\big|_{q^2=0} 
\simeq 
 48 \pi e_Q^2 N_c 
\int_{-\infty}^\infty \frac{dE}{(E+2m)^{2n+3}}
\left(B_1^2 - B_1 d_1 \frac{E}{3m}\right) 
{\rm Im}\, G_{s=1}(E,{\bf 0},{\bf 0})
\,.
\ee

\medskip

\noindent
{\bf Threshold expansion}:\\
It is quite instructive to look for the connection between pNRQCD and the threshold
expansion \cite{Beneke:1997zp}. The latter study the behavior of QCD diagrams in perturbation 
theory near the energy threshold region. The procedure consists in taking one specific diagram
and splitting it in the different existing regions of momenta. The main outcome of this study was that, 
near threshold, there are four momentum 
modes in a given diagram depending on the energy and momentum of quarks and gluons:

\medskip

\noindent
(i) {\bf hard modes}. Quarks and gluons with energy and three-momenta of
${\cal O}(m)$.\\
(ii) {\bf soft modes}. Quarks and gluons with energy and three-momenta of
${\cal O}(mv)$ (the quarks are off-shell in this situation).\\
(iii) {\bf potential modes}. Quarks and gluons with energy of ${\cal O}(mv^2)$
and three-momenta of ${\cal O}(mv)$ (the gluons are off-shell in this 
situation).\\
(iv) {\bf ultrasoft modes}. Quarks and gluons with energy 
and three-momenta of ${\cal O}(mv^2)$ (in practice, it does not seem there are 
quarks in this situation).

\medskip

There is a nice correspondence (as it could not be otherwise) between the construction of pNRQCD 
and the separation of modes shown above:

\begin{enumerate}
\item
Matching QCD to NRQCD: Integrating out hard modes.
\item
Matching NRQCD to pNRQCD: Integrating out soft modes and potential gluons.
\item
Dynamical degrees of freedom of pNRQCD: Potential quarks and ultrasoft gluons\footnote{Actually the degrees of 
freedom of pNRQCD are not only those, but any with energy smaller than the ultrasoft scale. In principle such degrees of freedom may show up at higher orders in perturbation theory in some kinematical situations.}.
\end{enumerate}

Note that the threshold expansion shows the existence of the ultrasoft and potential modes within perturbation theory. 
Nevertheless, it is within the effective theory, which gives power counting rules, where one realizes that those modes can not 
be treated within perturbation theory and a resummation of diagrams is needed. Finally, 
pinch singularities also appear in computations using the threshold expansion.
We have seen here that understanding the pinch singularities within the EFT 
framework provides a consistent prescription to eliminate them in each case.

\section{Phenomenological Analysis}

Very detailed reviews on the phenomenology of Heavy Quarkonium can be found in Refs. 
\cite{Brambilla:2004wf,Asner:2008nq,Brambilla:2010cs}. In Ref. \cite{Brambilla:2004jw} some applications of pNRQCD both at weak and 
strong coupling were considered. Here we focus on the weak coupling limit and update over this reference. 
Most formulas can be found in Sec. \ref{Observables}. Here we will only perform the phenomenological analysis 
and discuss the comparison with experiment (when possible).

The phenomenology on Heavy Quarkonium is immense. The restriction to the weak coupling limit narrows the applicability of the 
theory considerable, yet there is still plenty of room for applications. Even among these we have restricted to 
those which are the cleanest from the theoretical point of view: 
First, to the computation of the spectrum at weak coupling, second to the interaction of Heavy Quarkonium with 
ultrasoft photons, which can be measured in radiative transitions. Finally we have considered the interaction 
with hard photons, which can be measured in $t$-$\bar t$ production near threshold, heavy quarkonium NR 
sum rules, and inclusive electromagnetic decays (and in particular their ratios). 

\subsection{Spectroscopy}
\label{spectroscopy}
\begin{figure}[h]
\hspace{-0.1in}
\epsfxsize=3.85in
\centerline{
\epsffile{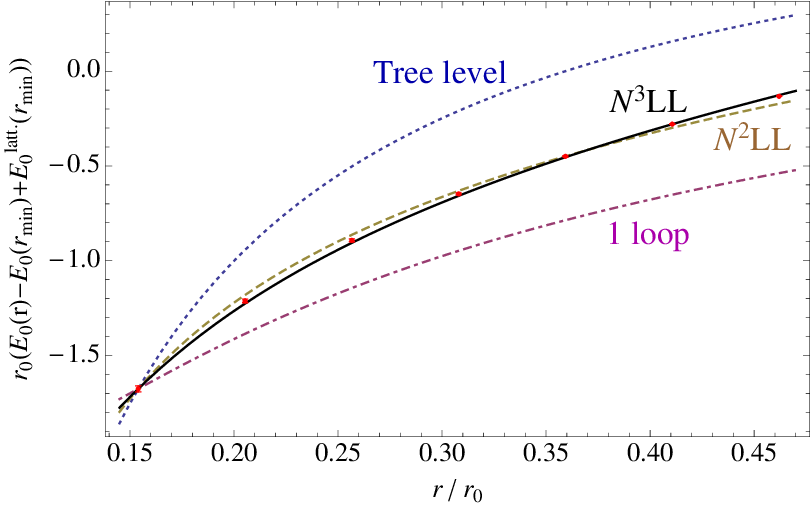}
}
\caption{\it Static potential in the RS scheme \cite{Pineda:2001zq} at different orders in perturbation theory plus its 
comparison with lattice simulations \cite{Necco:2001xg}
in the quenched approximation. From X. Garcia-Tormo, based on Refs. \cite{Brambilla:2009bi,Brambilla:2010pp}. }
\label{staticpot}
\end{figure}

We first would like to discern which states (and 
to which extent) can be described with this formalism. 
The cleanest place to address this question is the static potential, by checking 
up to which scale it can be described by a convergent perturbative series expansion. 
The outcome is that, once the renormalon cancellation is implemented, 
its convergence greatly improves and, in the cases 
when the comparison is possible, it agrees with lattice simulations (at least up to around 1 GeV) 
\cite{Recksiegel:2001xq,Pineda:2002se,Lee:2002sn,Brambilla:2009bi,Brambilla:2010pp}. 
We show a recent analysis in Fig. \ref{staticpot} for illustration. Note that, even though perturbation theory is convergent, in order to accurately describe lattice data, one has to go high orders. We believe that this effect is specially important for the wave function at the origin. We further 
discuss this issue in Secs. \ref{phen:hardphotons} and \ref{phen:decayratio}.

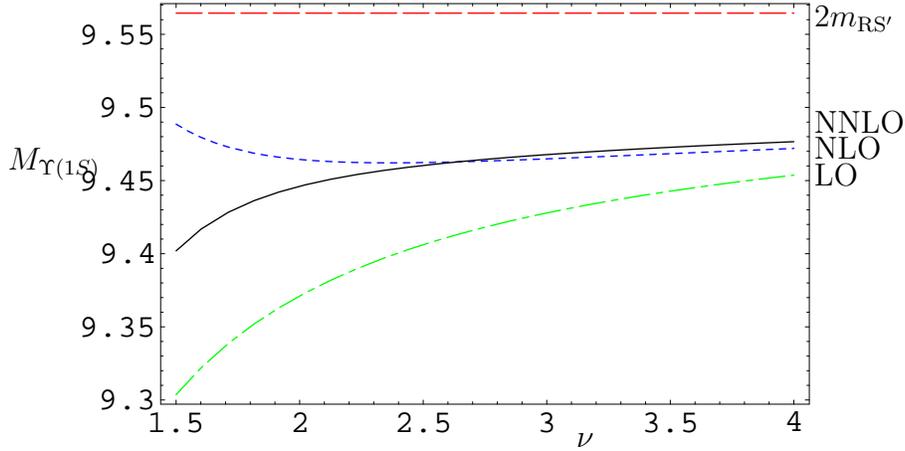
\begin{figure}[h]
\begin{center}
\hspace{-0.1in}
\epsfxsize=3.85in
\put(-25,106){$M_{\Upsilon(1S)}$}
\put(190,1){$\nu$}
\put(280,160){$2m_{\RS'}$}
\put(280,110){NLO}
\put(280,120){NNLO}
\put(280,100){LO}
\epsffile{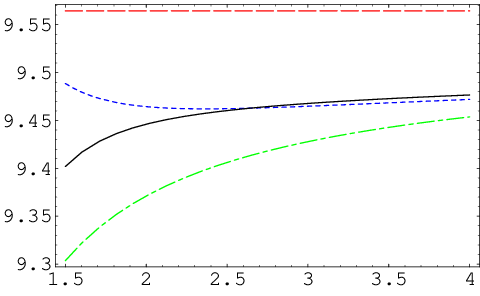}
\end{center}
\caption{\it $M_{\Upsilon(1S)}$ at different orders in perturbation theory in the RS' scheme using 
$m_b(m_b)=4.214$ MeV. To be compared with the experimental number:  $M_{\Upsilon(1S)}=9460$ MeV.
From Ref. \cite{Pineda:2001zq}. }
\label{figmass} 
\end{figure}

These results encourage the use of the weak coupling version of pNRQCD for 
spectroscopy. Its use for the $M_{\Upsilon(1S)}$ has lead to competitive 
determinations of the bottom mass $m_b(m_b) \sim 4.2$ with relative good convergence 
\cite{Beneke:1999fe,Brambilla:2001fw,Pineda:2001zq,Brambilla:2001qk,Lee:2003hh}. See Fig. \ref{figmass} for illustration. The situation has not significantly improved since the early 
years of the new millennium. At present the accuracy appears to be limited by non-perturbative effects.  

If the bottomonium ground state can be described with the weak coupling version of pNRQCD, 
it should also be possible to describe its pseudoscalar partner, the $\eta_b$. Nevertheless 
the predicted value for the hyperfine splitting 
$\sim 40$ MeV \cite{Recksiegel:2003fm,Kniehl:2003ap} does not agree very 
well with the recent experimental determination $\sim 70$ MeV \cite{:2008vj,Bonvicini:2009hs}, with a two sigma deviation.
Nevertheless, the experimental situation is still not settled, see \cite{QWG11} where a preliminary value $\sim 60$ MeV was 
quoted.

With respect to other quarkonium states, the $B_c(1^1S_0)$ system has been studied in Refs. 
\cite{Brambilla:2000db,Brambilla:2001fw,Brambilla:2001qk} obtaining reasonable 
results: $M_{B_c}(1S)=6307\pm 17$ MeV. Actually, this figure was a prediction of the theory prior that  
the experimental number was obtained: $6287\pm 4.8 \pm 1.1$ MeV \cite{Aaltonen:2007gv,Abazov:2008kv}.

For higher excitations of bottomonium, and charmonium, the situation is not conclusive. 
There are different claims. Whereas in Refs. \cite{Beneke:2005hg,GarciaiTormo:2005bs,DomenechGarret:2008vk}
it is claimed that it is not possible to describe bottomonium higher excitations 
in perturbation theory, an opposite stand is taken in Refs. 
\cite{Brambilla:2001fw,Brambilla:2001qk,Recksiegel:2002za,Recksiegel:2003fm}. 
At this respect we can not avoid mention that Ref. \cite{Recksiegel:2003fm} produced a number 
for the $\eta_c(2S)$ mass before, and consistent with, the last experimental figures by 
Babar \cite{Aubert:2003pt} and Cleo III \cite{Asner:2003wv} (before there were two 
excluding experimental numbers by Bell \cite{Choi:2002na}
and Crystal Ball \cite{Edwards:1981mq}). From the Charmonium ground state there is a 
determination of $m_c$ \cite{Brambilla:2001fw} using the 
Upsilon scheme \cite{Hoang:1998ng} with a relatively convergent series.
 
A more detailed discussion on the heavy quarkonium spectroscopy and in particular for the states mentioned here can be found in Refs.  \cite{Brambilla:2004jw,Brambilla:2004wf,Asner:2008nq,Brambilla:2010cs}.

\subsection{Ultrasoft photons. Radiative transitions}
\label{sec:phen}

In Ref. \cite{Brambilla:2005zw}
 the magnetic dipole transitions between two weakly-coupled heavy quarkonia states were computed and the results applied to the transitions between spin-one and spin-zero states for the ground state 
 of bottomonium and charmonium. For the first excitation ($n=2$) of Heavy Quarkonium only bottomonium was considered, computing 
 their transitions between $n=2$ and $n=1$ states, as well as the different possible transitions between $n=2$ states.  Here 
 we only review results concerned with the ground state of bottomonium and charmonium to which is more likely that a weak coupling approach 
 could be applied.
 
The LO operator contributing to the M1 transitions reads (at LO $V^{\frac{\sigma\cdot B}{m}}_S = 1$)
 \bea
\delta {\cal L}_{\gamma\, \rm pNRQCD} = \int d^3 r \;  {\rm Tr} \, \Bigg\{ 
\cdots
+ \frac{
1}{2 m}
\; V^{\frac{\sigma\cdot B}{m}}_S 
\; \left\{{\rm S}^\dagger , \bfsigma \cdot e e_Q {\bf B}^{\em}\right\} {\rm S} 
\cdots 
\Bigg\}\,.
\label{gammapNRQCD:Lag}
\eea
This term produces the following M1 transition width between two $S$-wave states
\be
\Gamma_{n^3S_1\to n^\prime{^1S_0}\,\gamma} =  
\frac{4}{3} \, \alpha_{\rm EM}\, e_Q^2
\, \frac{k^3_\gamma}{m^2}\, \left|\int_0^\infty dr \, r^2 \, R_{n^\prime 0}(r) \, R_{n0}(r) \;
j_0\left( \frac{k_\gamma r}{2}\right)
\right|^2 \,,
\label{M1:NRlimit}
\ee
where $R_{nl}(r)$ is the radial Schr\"odinger wave function.
The photon energy $k_\gamma$ is approximately the difference between the masses of the two
quarkonia, therefore, it is of order $mv^2$ or smaller. Since $r\sim 1/(mv)$, we may expand the spherical Bessel function 
$j_0(k_\gamma r/2) = 1 - (k_\gamma r)^2/24 + \cdots$.
At LO in the multipole expansion, for $n=n'$, 
the overlap integral is 1. Such transitions are usually referred to 
as  {\it allowed}. At LO, for $n\neq n'$, the overlap integral is 0. 
These transitions  are usually referred to as {\it hindered}. 
The widths of hindered transitions are entirely given by higher-order and
relativistic corrections. 

Equation (\ref{M1:NRlimit}) is not sufficient to explain the observed 
transition widths. In the case of allowed ones, for instance, it overpredicts the observed
$J/\psi\to \eta_c \,\gamma$ transition rate by a factor 2 to 3.
A large anomalous magnetic moment or large relativistic corrections 
have been advocated as a solution to this problem.
Hence, it is crucial to supplement Eq.~(\ref{M1:NRlimit})
with higher-order corrections, which were computed in Ref. \cite{Brambilla:2005zw}.

\medskip 

\noindent
$J/\psi \to \eta_c \,\gamma$.\\
The transition $J/\psi \to \eta_c \,\gamma$ 
has been problematic to accommodate in potential models because 
its LO width is about $2.83$ keV (for $m_c = M_{J/\psi}/2 = 1548$
MeV), relatively far away from the experimental 
value of  $(1.7 \pm 0.4)$ keV \cite{PDG2010} (thought the experimental situation is not completely stable. The 2010 PDG value has an S factor 1.6; 
In 2004 the value was $1.18 \pm 0.36$ keV 
\cite{PDG2004} and recently CLEO has reported the value 1.84 from the branching fraction $1.98\pm0.09\pm0.30\%$).
In Ref. \cite{Brambilla:2005zw} the transition width was computed in the weak coupling limit up to order $k_\gamma^3\,v^2_c/m^2$: 
\bea
\Gamma_{J/\psi \to \eta_c \,\gamma} \!\! 
&=& \!\! 
\frac{16}{3} \alpha_{\rm EM}e_c^2 \frac{k_\gamma^3}{M_{J/\psi}^2}
\left[ 1 + C_f\frac{\al(M_{J/\psi}/2)}{\pi}
+\frac{2}{3}\frac{\braQM{1S} 
3 V_s^{C} - r V_s^{C\,\prime} 
\ketQM{1S}}{M_{J/\psi}}
\right]
\nn\\
\!\! 
&=& \!\! 
\frac{16}{3} \alpha_{\rm EM} e_c^2 \frac{k_\gamma^3}{M_{J/\psi}^2}
\left[ 1 + C_f\frac{\al(M_{J/\psi}/2)}{\pi} 
- \frac{2}{3} (C_f\al(p_{J/\psi}))^2
\right]
\,,
\label{Gamma1Sc}
\eea
where in the first line the charm mass has been reexpressed in terms of the $J/\psi$ mass,
$$
M_{J/\psi} = 2m_c + \braQM{1S} 
\left(
\frac{{\bf p}^2}{m_c} +  V_s^{C}(r)  
\right)
\ketQM{1S},
$$
and has been made use of the virial theorem to get rid of the kinetic energy.
In Eq.~(\ref{Gamma1Sc}) it has made explicit that the normalization scale for the $\al$ 
inherited from $c_F^\em$ (the Wilson coefficient analogous to $c_F$ but changing the chromomagnetic by the electromagnetic field) is the charm mass ($\al(M_{J/\psi}/2) \approx 0.35$), 
and for the $\al$ coming from the Coulomb potential the typical momentum 
transfer is $p_{J/\psi} \approx m C_f \al(p_{J/\psi})/2  \approx 0.8$ GeV.
Numerically one obtains:
\be
\Gamma_{J/\psi \to \eta_c \,\gamma} =  (1.5 \pm 1.0)\;\hbox{keV}.
\label{Gamma1Scnum}
\ee
The uncertainty has been estimated by assuming the next corrections to be suppressed 
by a factor $\al^3(p_{J/\psi})$ with respect to the transition width in the NR limit.

Some comments are in order. First, the uncertainty in Eq. (\ref{Gamma1Scnum}) is 
large. According to Ref. \cite{Brambilla:2005zw}, it fully accounts for the large uncertainty coming from
higher-order relativistic corrections, which may be large if one considers that 
those of order $k_\gamma^3\,v^2_c/m^2$ have reduced the LO result by about 50\%, and 
for the uncertainties in the normalization scales of the strong-coupling constant.
Both uncertainties may only be reduced by higher-order calculations. 

Despite the uncertainties, the value given in Eq.~(\ref{Gamma1Scnum}) is perfectly 
consistent with the experimental one. This means that assuming the ground-state charmonium 
to be a weakly-coupled system leads to relativistic corrections to the 
transition width of the right sign and size. This is not trivial.
If we look at the expression after the first equality in Eq.~(\ref{Gamma1Sc}), 
we may notice that $3 V_s^{C} - rV_s^{C\,\prime}$ is negative in the 
case of a Coulomb potential (i.e. it lowers the transition width), 
but positive in the case of a confining linear potential (i.e. it increases
the transition width). This may explain some of the difficulties met by potential 
models in reproducing $\Gamma_{J/\psi \to \eta_c \,\gamma}$. In any rate, it should be remembered that 
Eq.~(\ref{Gamma1Sc}) is not the correct expression to be used in the
strong-coupling regime. 

\medskip

\noindent
$\Upsilon(1S) \to \eta_b \,\gamma$.\\
For the allowed M1 transitions in the ground state bottomonium system one has $\Upsilon(1S) \to \eta_b \,\gamma$:
\bea
\Gamma_{\Upsilon(1S) \to \eta_b \,\gamma} &=& 
\frac{16}{3} \alpha_{\rm EM} e_b^2 \frac{k_\gamma^3}{M_{\Upsilon(1S)}^2}
\left[ 1 + C_f\frac{\al(M_{\Upsilon(1S)}/2)}{\pi} 
- \frac{2}{3} (C_f\al(p_{\Upsilon(1S)}))^2
\right]
\,,
\label{Gamma1Sb}
\eea
where the $b$ mass has been expressed in terms of the $\Upsilon(1S)$ mass.
We have made explicit that the renormalization scale for the $\al$ 
inherited from $c_F^{\em}$ is the bottom mass ($\al(M_{\Upsilon(1S)}/2) \approx 0.22$), 
while for the $\al$ coming from the Coulomb potential in the $\Upsilon(1S)$
system the typical momentum transfer is $p_{\Upsilon(1S)} \approx m C_f \al(p_{\Upsilon(1S)})/2  \approx 1.2$ GeV. 

The most recent determination using this expression can be found in Ref. \cite{Brambilla:2010cs}, where a branching fraction of $(2.85 \pm 0.30) \times 10^{-4}$ was quoted. This value translates to the following decay width 
\be 
\Gamma_{\Upsilon(1S) \to \eta_b \,\gamma} =  (15.1 \pm 1.6)\;\hbox{eV}.
\ee 
There is not experimental data for bottomonium to compare with. Therefore this number is a prediction. 

\medskip

\begin{figure}[ht]
 \centering
  \includegraphics[scale=0.4, angle=-90]{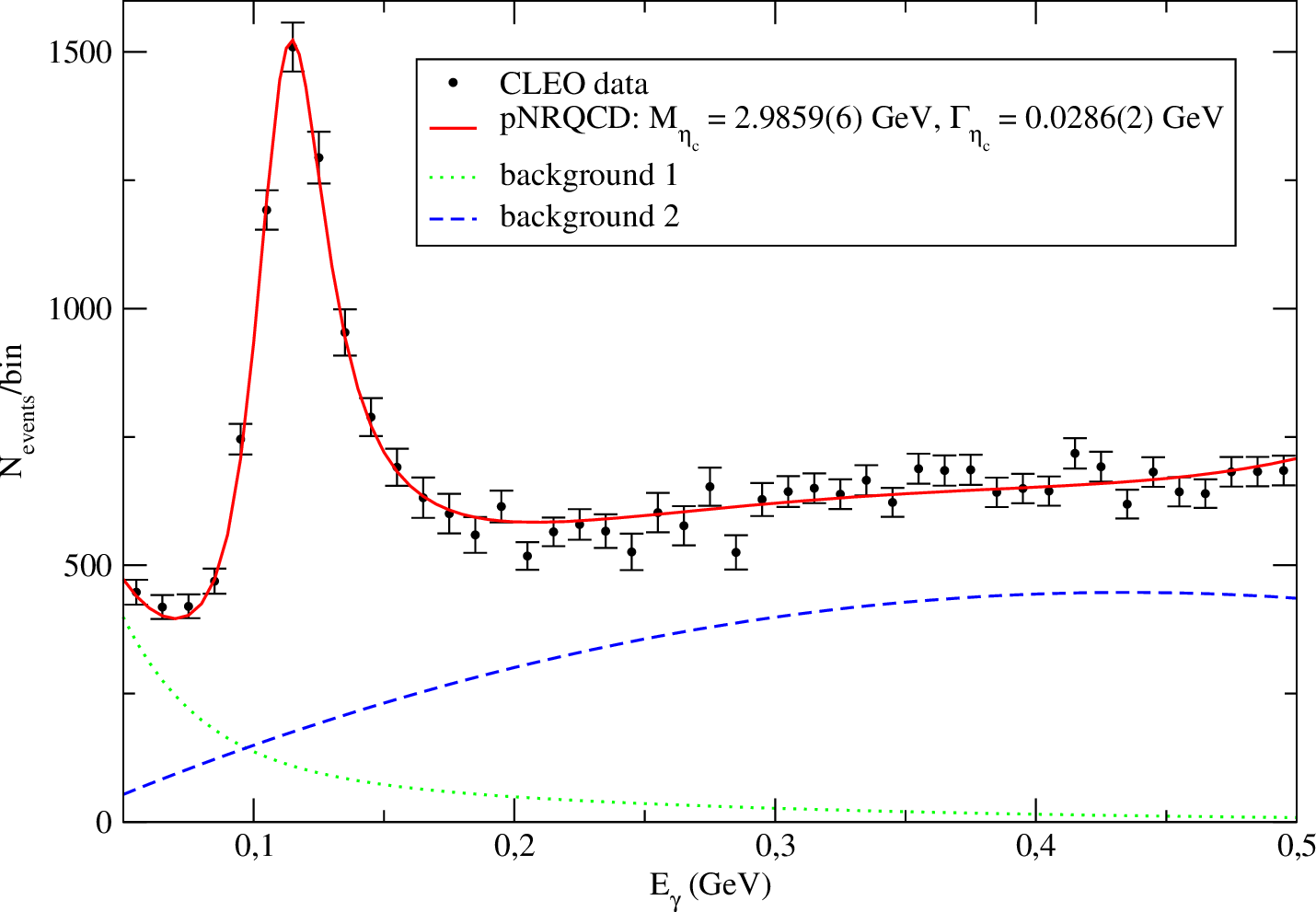}
\caption{\it Best fit to CLEO's data for the photon spectrum in $J/\psi\to\eta_c \gamma$ 
using Eqs. (\ref{M1}) and (\ref{E1}) for the theoretical signal together with the two background 
sources used in CLEO. From Ref. \cite{Brambilla:2010ey}.}
 \label{bestfit}
\end{figure}

\noindent
{\bf Photon-line shape}.\\
Finally we mention another interesting study done in Ref. \cite{Brambilla:2010ey} concerning the determination of the 
$\eta_c$ mass and its total decay width.
In that work, the photon line shape was considered in the NR limit using the 
sum of the magnetic (it has been pointed out in Ref.~\cite{:2008fb} that the dependence on $E_\gamma^3$ 
is responsible for the asymmetric shape of the photon spectrum)
\begin{equation}
\label{M1}
\frac{\mathrm{d\Gamma^{M1}_{J/\psi\to\eta_c\gamma}}}{\mathrm{d}E_\gamma}
=\frac{64}{27}\frac{\alpha_{\rm EM}}{\pi}\frac{E_\gamma^3}{M_{J/\psi}^2}
\frac{{\Gamma_{\eta_c}}/{2}}{\left(
M_{J/\psi}-M_{\eta_c}-E_\gamma\right)^2+\frac{\Gamma_{\eta_c}^2}{4}}.
\end{equation} 
and electric
\begin{eqnarray}
\label{E1}
\frac{\mathrm{d\Gamma^{E1}_{J/\psi\to\eta_c\gamma}}}{\mathrm{d}E_\gamma}=
\frac{448}{243} \alpha_{\rm EM}\frac{E_\gamma}{m_c}
\alpha^2\frac{\bigg|\phi_{J/\psi}({\bf 0})\bigg|^2}{m_c^3}
\bigg|a_e(E_\gamma)\bigg|^2,
\end{eqnarray} 
dipole contribution.
The function $a_e(E_\gamma)$ has been discussed in
Ref.~\cite{Manohar:2003xv}, and 
a closed analytical form derived in \cite{Voloshin:2003hh}.
The best fit was in good agreement with CLEO's
experimental determination \cite{:2008fb}, see Fig. \ref{bestfit}.

\subsection{Hard photons. Production and decays}
\label{phen:hardphotons}

We now consider observables proportional to the NR heavy quark vacuum polarization: inclusive decay widths, 
NR sum rules and heavy quark-antiquark production near threshold. They are induced by the 
coupling of the heavy quarks with hard photons. 

\medskip

\begin{figure}[h!!]
   \epsfxsize=12cm
   \centerline{\epsffile{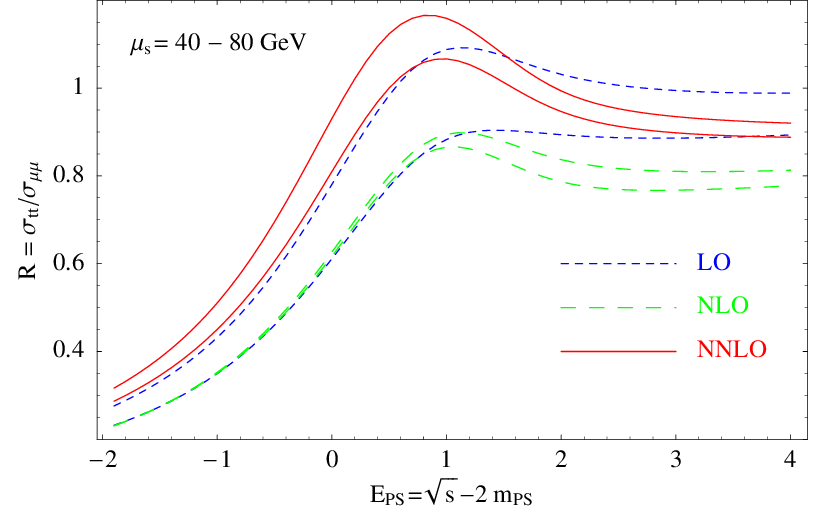} }
   \centerline{\epsffile{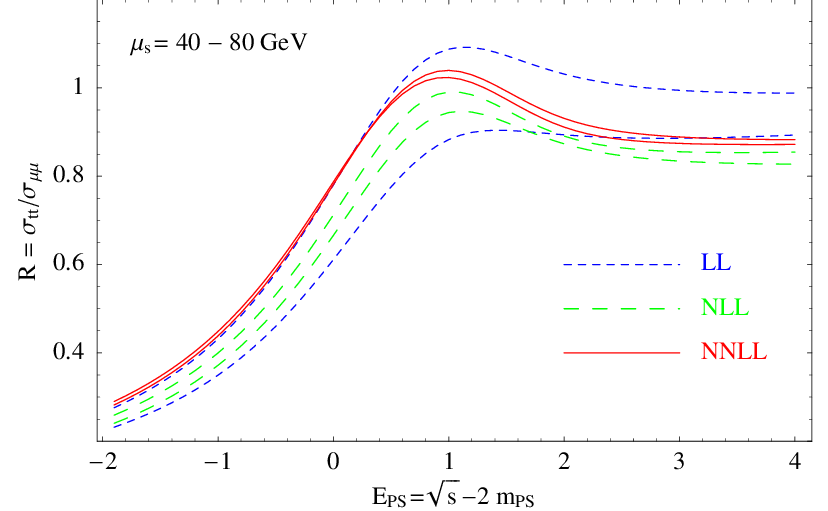} }
   \caption{\it Threshold scan for $t\bar{t}$. The upper figure shows the LO, NLO and NNLO fixed order results, 
whereas in the figure below the LL, NLL and NNLL RG improved results are displayed. The factorization scale is varied from
   $\nu\equiv\mu_s$=40~GeV to $\mu_s$=80~GeV. From Ref. \cite{Pineda:2006ri}. }
   \label{fig:scan1}
\end{figure}

\noindent
{\bf $t$-$\bar t$ production near threshold} (one of the main physical cases for the construction of future electron-positron linear 
colliders \cite{Djouadi:2007ik}) is the ideal place to test pNRQCD in the weak coupling limit. 
The large energy transfer between top and antitop, produced by their large masses ($m_t\sim$ 175 GeV), 
makes this system the place on 
which the weak coupling expansion should work better. At present, as we have already mentioned, the theoretical expression is completely known with NNLO/NLL accuracy, with many 
partial results at the NNNLO/NNLL level.
 Since the decay width $\Gamma_t\sim 1.5$ GeV $\sim m_t\al^2$, which is the ultrasoft scale, a
remnant of the would-be toponium $1S$ state is expected to show up as a bump
in the total cross section. From the threshold scan around this bump  it is possible to obtain the top quark mass with
a high accuracy working with schemes where the renormalon cancellation is incorporated. 
Yet, as we can see from Fig. \ref{fig:scan1}, 
finite order computations still suffer from large corrections. The resummation of logarithms (first advocated in Ref. \cite{Hoang:2000ib}) produces a huge impact in the convergence of the normalization. In the plot, factorization scales below 40 GeV are not considered, as they produce large variations of the theoretical prediction. It has been argued \cite{Beneke:2005hg}
that this scale dependence is due to several insertions of the static potential, and could be solved by treating the static 
potential exactly (see the discussion in Sec. \ref{phen:decayratio}) or restricting to large factorization scales.
Even after that, the resulting series \cite{Hoang:2003ns,Pineda:2006ri} had larger uncertainties than expected 
(even if the absolute value of the corrections is small).
This, however, may be due to the scheme dependence of the result.
Therefore, it is premature to
draw any definite conclusion about the convergence of the series before getting
the complete NNLL evaluation, which, even if difficult, is
within reach.  This is of utmost importance for future determinations of
the top mass and the Higgs-top coupling at a future Linear Collider
\cite{Martinez:2002st}.

\medskip

\begin{figure}
   \epsfxsize=12cm
   \centerline{\epsffile{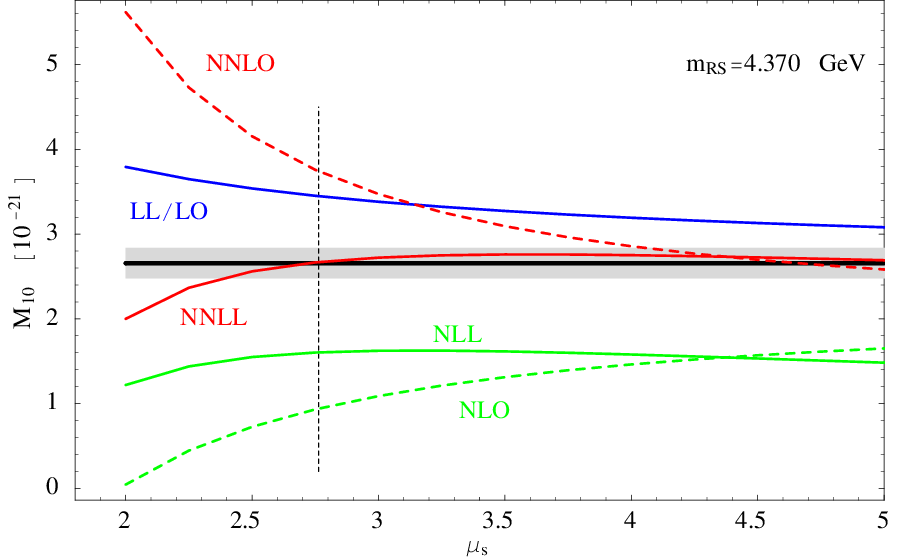} }
   \caption{\it The moment $M_{10}$ as a function of
 $\nu \equiv \mu_s$ at LO/LL, NLO, NLL, NNLO and NNLL for $m_{b {\rm
 RS}}$(2~GeV) = 4.370~GeV in the RS scheme. The
 experimental moment with its error is also shown (grey band). From Ref. 
 \cite{Pineda:2006gx}. }
 \label{fig:Mom10}
\end{figure}

\noindent
{\bf NR Sum rules}.\\
The problem of convergence of finite order computations is even more acute in $b$ physics. Even, 
if we consider one of the optimal observables, like NR sum rules, the convergence is poor and 
the factorization scale large (see Fig. \ref{fig:Mom10}). 
This was one of the main problems for accurate determinations of the bottom mass from NR sum rules. 
The implementation of the RG greatly diminishes the factorization scale dependence and somewhat 
also improves the convergence (see Fig. \ref{fig:Mom10}). Using the PS \cite{Beneke:1998rk} 
and RS renormalon subtraction schemes, 
this has led to one of the most accurate determination of the bottom mass \cite{Pineda:2006gx}: 
\begin{eqnarray*}
\vspace{-0.4in}
\,\left.
\begin{array}{ll}
&
\displaystyle{m_{b,{\rm PS}}(2 {\rm GeV}) = 4.52 \pm 0.06 \;{\rm GeV}}
\\
&
\displaystyle{m_{b,{\rm RS}}(2 {\rm GeV}) = 4.37 \pm 0.07 \;{\rm GeV}}
\end{array} \right\} \rightarrow
{ \overline{m}_b(\overline{m}_b)=4.19\pm 0.06 \,{\rm GeV}}
,
\end{eqnarray*}
and also a competitive determination of the charm mass \cite{Signer:2008da}: $\overline{m}_c(\overline{m}_c)=1.25\pm0.04$ GeV.
Note that the perturbative series is sign-alternating, the opposite than for  
electromagnetic decays (that we will consider next). The convergence of the perturbative series in 
sum rules is also better than in electromagnetic decays. 

On the experimental side NR sum rules are ideal. 
By taking $n$ large on the right-hand side of Eq.~(\ref{moments})  the contribution from 
high momenta (the continuum region) is suppressed. Actually, this is the region which is
less well known on the experimental side. Therefore, the experimental errors are 
significantly reduced using NR sum rules. In practice, the following parameterization is used
\be
M_n^{ex}=\sum_{k=1}^6{9\pi \over 
\alpha_{\rm EM}^2}{\Gamma_{\Upsilon(k)} \over 
M_{\Upsilon(k)}^{(2n+1)}}+\int_{\sqrt{s_{B\bar B}}}{ds \over s^{n+1}}r_{cont}(s)
\,,
\ee
where $s_{B\bar B}$ is the $B$-$\bar B$ threshold, and $\alpha_{\rm EM}$ should be computed at the hard scale.

\medskip

\begin{figure}[h!]
   \begin{center}
    \includegraphics[width=.73\textwidth]{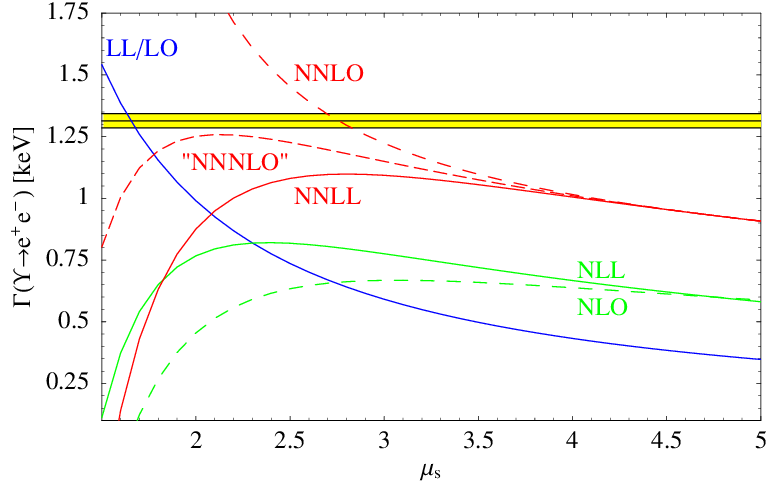}
   \caption{\it Prediction for the $\Upsilon(1S)$ decay rate to
   $e^+e^-$ in the RS' scheme. The "NNNLO" result is obtained by re-expanding the NNLL result and
keeping only terms that are NNNLO. From Ref. \cite{Pineda:2006ri}.}
   \label{fig:decayee}
   \end{center}
\end{figure}

\noindent
{\bf Decay Widths}.\\
The electromagnetic inclusive decay widths are also known with NNLO/NLL precision, with several partial results at 
NNNLO/NNLL order (see Sec.~\ref{Observables}). 
Nevertheless, they suffer from relatively large scale uncertainties, even after the resummation of logarithms. The corrections are huge, producing a bad convergence,
which have so far prevented their use for phenomenological analysis. 
We show the plot of the $\Upsilon(1S)$ decay rate to
   $e^+e^-$ in Fig. \ref{fig:decayee}. 

\medskip

\begin{figure}[th]
\makebox[1cm]{\phantom b}
\put(0,0){\epsfxsize=9truecm \epsfbox{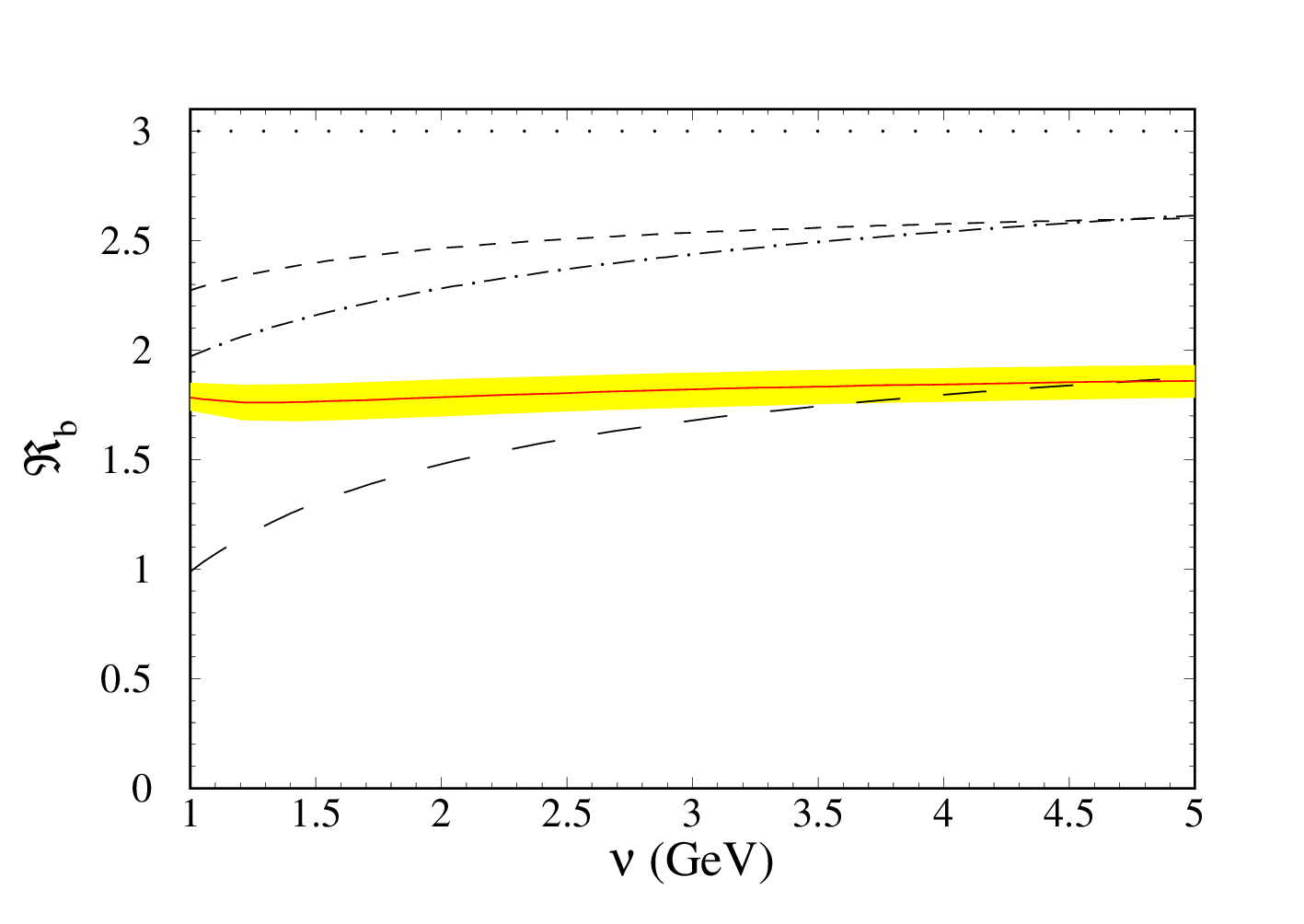}}
\put(240,0){\epsfxsize=9truecm \epsfbox{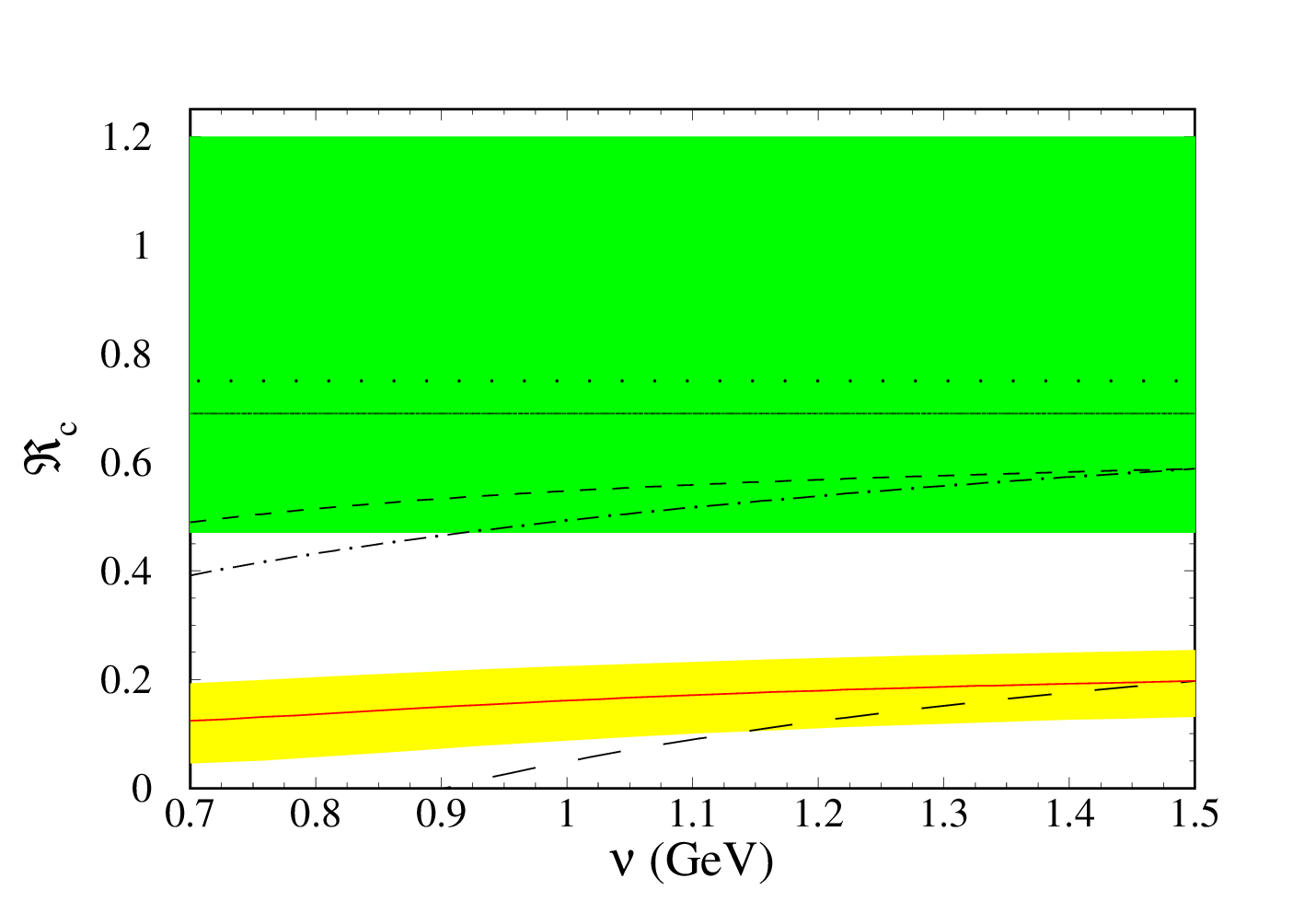}}
\put(-5,130){$(a)$}
\put(235,130){$(b)$}
\caption{\label{figbc} {\it The spin  ratio  as a function of the
renormalization scale $\nu$ in LO$\equiv$LL (dotted line), NLO (short-dashed
line), NNLO (long-dashed line), NLL (dot-dashed
line), and NNLL (solid line) approximation.
For the NNLL result the band reflects the errors 
due to $\al(M_Z)=0.118\pm 0.003$. 
Panel $(a)$ shows the bottomonium ground state case.
Panel $(b)$ shows the charmonium ground state case.
In the charmonium case, the upper  band
represents the experimental error of the ratio \cite{PDG2004} where the central 
value is given by the horizontal solid line. From Ref.\cite{Penin:2004ay}.}}
\end{figure}

\noindent
{\bf Decay Width Ratios}.\\
For all the previous observables the NNLL plots were partial, not complete. Therefore, one may wonder 
about the error associated to those analysis. At present, the only place where the complete NNLL expression is known 
is the decay ratio \cite{Penin:2004ay}. The outcome was an almost complete factorization scale independence of the result, 
as we illustrate in Fig. \ref{figbc} for the charm and bottom case. On the other hand the convergence was quite bad in 
the charm case. The series for the bottom case looked convergent though with large corrections. For the top case, which 
we do not show here, the convergence was much better. 

\subsection{Improved perturbation theory}
\label{phen:decayratio}

This section goes slightly away from the main trend of the review. 
We have so far worked in the strict weak-coupling regime, where one approximates the static potential by the
Coulomb potential, $V^C_s = -C_f\, \al(\nu)/r$, and include
higher-order terms in $\al(\nu)$ perturbatively. As we have already mentioned, 
this has worked fine for the mass of the 
$\Upsilon(1S)$, but other properties of
the bottomonium ground state like the hyperfine splitting or
electromagnetic decay widths have shown either problems of convergence or poor agreement 
with experiment. Even the theoretical expressions for the $t$-$\bar
t$ production near threshold and bottomonium sum rules suffer from a large factorization scale dependence in fixed 
order computations. In principle, the novel feature of these observables
compared to the heavy quarkonium ground state mass is a
bigger sensitivity to the shape of the wave function and to its behavior at the origin (the hyperfine splitting is 
also quite sensitive to the wave function at the origin, $|\phi_n({\bf 0})|^2$). It may well be that the present precision of finite order calculations is not 
enough to properly reproduce the shape of the wave function (in the same way that one has to go to high orders in 
perturbation theory in order to properly reproduce the shape of the static potential, see Fig. \ref{staticpot}).  This problem could be solved by 
performing even higher order computations or by a reorganization of the perturbative expansion.

\begin{figure}[htb]
\hspace{-1.2in}
   \epsfxsize=15.2cm
   \centerline{\epsffile{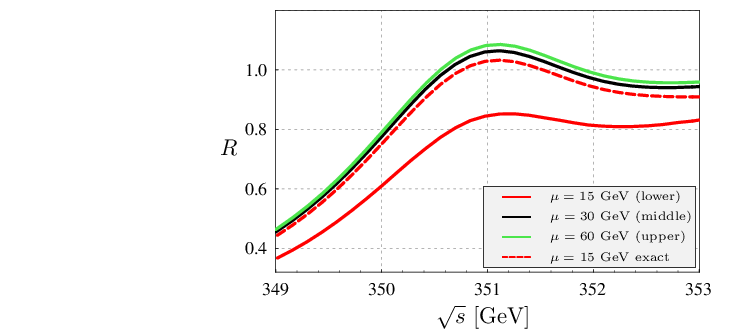}}
\caption{\it Top quark pair production cross section with the static potential only. Scale dependence 
of the third-order approximation. From Ref. \cite{Beneke:2005hg}.}
\label{BenekeKiyo}
\end{figure}

A first modification of perturbation theory comes from including the resummation of logarithms, 
which we have already discussed in the previous sections. 
This resummation significantly diminishes the factorization scale dependence in  $t$-$\bar
t$ production near threshold and bottomonium sum rules, making the result more stable. For the inclusive decay widths a 
significant scale dependence remains and not a good convergence is found. In all three cases there were a remaining large 
factorization scale dependence at low scales; In the bottomonium case for scales below around 2 GeV and in the 
$t$-$\bar t$ case  
below 20 GeV.  If one forgets about relativistic corrections, it is possible to exactly (albeit numerically) solve the Schr\"odinger 
equation with the exact static potential: 
\begin{eqnarray}
\label{H0}
h_s^{(0)}\equiv -\frac{{\bf \nabla}^2}{2m_r}+V^{(0)}_s(r).
\end{eqnarray}
The result is finite and no need for handling divergences. This program has been undertaken in  
Ref. \cite{Beneke:2005hg}. In this reference it has been show that using this exact solution most of the scale dependence at low 
scales vanishes for $t$-$\bar t$ (for $b$-$\bar b$ not such analysis exists). We show their analysis in Fig. \ref{BenekeKiyo}.

Leaving aside this factorization scale dependence at low scales, the convergence is relatively good 
for $t$-$\bar t$ production near threshold, yet the uncertainty was somewhat larger than expected. For NR bottomonium sum rules, the series is also convergent but the size of the corrections is much larger. In all these observables the complete NNLL is not yet known. Therefore, one could cast some doubts on these results, but not 
for the decay ratios, which have been computed with NNLL accuracy in Ref. \cite{Penin:2004ay}. 
The scale
dependence greatly improved over fixed-order computations and the
result was much more stable.  The convergence could be classified as
good for the top case, reasonable for the bottom, and not good for the
charm, although in all three cases the scale dependence of the
theoretical result was quite small, as can be seen in Fig. \ref{figbc}.  
For the charm case there is
experimental data available, and the agreement with experiment
deteriorates when higher order corrections are introduced.  On the
other hand, in
Ref. \cite{Czarnecki:2001zc},  a potential model was considered (a
Cornell-like one, yet compatible with perturbation theory at short
distances) for the bound-state dynamics, but a tree-level perturbative potential for the
spin-dependence. The matching in the
ultraviolet with QCD was performed along the lines of what would-be pNRQCD in the
strong coupling regime. Their net result was that they were able to obtain
consistency with experiment albeit with large errors. Unfortunately,
this result suffers from model dependence. In particular, since a
perturbative potential has been used for the spin-dependent potential,
it would have been more consistent to treat the static potential also
in a perturbative approach. In this respect, we have already discussed in Sec. \ref{spectroscopy}
and illustrated in Fig. \ref{staticpot}
that  the
inclusion of perturbative corrections to the static potential leads to
a convergent series, which gets closer to the lattice
values up to scales of around 1 GeV. It
is then natural to ask whether the inclusion of these effects may lead
to a better agreement in the case of charmonium and for sizable
corrections in the case of bottomonium and $t$-$\bar t$ production
near threshold.  Note that in this
comparison between lattice and perturbation theory one has to go to
high orders to get good agreement. Therefore, a computation of the
relativistic correction based on the LO expression for the
static potential, i.e.  the Coulomb potential, as the one used in an
strict NNLL computation of the decays, $t$-$\bar t$, ..., may lead to large corrections, since these
corrections, as well as the wave function at the origin, could be
particularly sensitive to the shape of the potential.

The wave function at the origin is divergent once relativistic corrections are included.
These divergences have to
be absorbed by the Wilson coefficients of the effective theory:
potentials and current Wilson coefficients. If one considers the
decay ratio, the dependence on the wave function associated to $V_s^{(0)}$ drops out and only the relativistic correction
survives  (therefore the analysis of Ref. \cite{Beneke:2005hg} does not apply here). 
This makes the decay ratio the cleanest possible place on
which to quantify the importance of the relativistic corrections to
the wave function. Therefore, in Ref. \cite{Kiyo:2010jm}
the perturbative expansion was reorganized considering 
the static potential exactly, whereas the relativistic terms were treated as corrections. 
By doing so it is expected to have an
effect similar to the one observed in Ref.~\cite{Czarnecki:2001zc}.
Including also the RG improved expressions, it is
expected to obtain results with only a modest scale dependence.  The
explicit computation confirmed to a large extent these
expectations.   Note that this computation is
completely based on a weak-coupling analysis derived from QCD 
and no non-perturbative input is introduced.

One then reorganizes the perturbative expansion. The LO Hamiltonian is now Eq. (\ref{H0}). 
On the other hand, the spin-dependent
potential 
\begin{equation}
\label{DeltaH}
\delta h_{S^2}\equiv
- \frac{4\pi C_f D^{(2)}_{S^2}}{d\, m_1m_2}\,
   [{\bf S}_1^i,{\bf S}_1^j][{\bf S}_2^i,{\bf S}_2^j]
   \delta^{(d)}({\bf r})
\end{equation}
is considered to be a perturbation to the result obtained with
$h_s^{(0)}$.  Therefore, we distinguish between an expansion in $v$ and
$\al$. $v$ has an expansion in $\al$ itself but this expansion does
not converge quickly for these relativistic corrections. 
 This is the reason we choose to take the static potential
exactly. 
  We now turn to the computation of
\begin{eqnarray}
\label{rho_n}
  \frac{|\phi_n^{(s=1)}({\bf 0})|^2}{|\phi_n^{(s=0)}({\bf 0})|^2} &\equiv&
  \rho_n(\nu)\,\,\equiv\,\,1+\delta \rho_n(\nu)
  \,.
\end{eqnarray}
Applying Rayleigh-Schr\"odinger perturbation theory to the problem we
obtain
\begin{eqnarray}
\phi^{(s)}_n({\bf 0})
&=&
\phi_{n}^{(0)}({\bf 0})
-
\widehat{G}(E^{(0)}_n)\delta \tilde h_{S^2}
\, 
\phi_n^{(0)}({\bf 0})
+ {\cal O}\left(\tilde h_{S^2}\right)^2, \,
\label{eq:psi_at_0}
\end{eqnarray}
where $\phi_n^{(0)}({\bf 0})$ is the wave function for the LO Hamiltonian
$h_s^{(0)}$ and $\widehat{G}(E^{(0)}_n)$ is the reduced Green function
at $E=E^{(0)}_n$, which is defined by
\begin{equation}
\widehat{G}(E^{(0)}_n)
\equiv
\sum_m{}^{\prime}
\frac{|\phi^{(0)}_m({\bf 0})|^2}{E^{(0)}_m-E^{(0)}_n}
=
\lim_{E\rightarrow E^{(0)}_n}
\bigg(
G(E,{\bf 0},{\bf 0})-\frac{|\phi^{(0)}_n({\bf 0})|^2}{E^{(0)}_n-E}
\bigg)\, .
\label{eq:Gred}
\end{equation}
The prime indicates that the sum does not include the state $n$ and (ultrasoft effects can 
be neglected with this precision)
\begin{equation}
G(E,{\bf 0},{\bf 0})
\equiv
\lim_{r \rightarrow 0}G(E,r,r)
\simeq
\lim_{r \rightarrow 0}\langle {\bf r}| \frac{1}{h_s^{(0)}-E-i 0}|{\bf r} \rangle
\,.
\end{equation}
The short distance behavior of the static potential $V_s^{(0)}(r)\sim
1/r$ makes $G(E,{\bf 0},{\bf 0})$ and, therefore, $\delta\rho_n$ divergent. Thus we
 need to regularize the Green function. We do it in two
different ways: dimensional regularization, and finite-$r$
regularization. 

The divergences in $\delta \rho_n$ are cancelled by divergences in the
Wilson coefficient $B_1/B_0$. Since the latter has been computed
in dimensional regularization, we will need $G(E,{\bf 0},{\bf 0})$ in dimensional
regularization as well. We denote the corresponding bare and reduced
Green functions by $G^{(D)}(E) =G^{(D)}(E,{\bf 0},{\bf 0})$ and
$\widehat{G}^{(D)}(E^{(0)}_n)$ respectively:
\begin{eqnarray}
G^{(D)}(E)&=&\frac{m_r}{2\pi}
\bigg[A_{\MS}^{(D)}(\epsilon;\nu)+B_{V_s^{(0)}}^{\MS}(E;\nu)\bigg]\,,
\label{eq:GD}
\\
\widehat{G}^{(D)}(E^{(0)}_n)&=&\frac{m_r}{2\pi}
\bigg[A_{\MS}^{(D)}(\epsilon;\nu)+\widehat{B}_{V_s^{(0)}}^{\MS}(E^{(0)}_n;\nu)\bigg]\,,
\label{eq:GDhat}
\end{eqnarray}
where $B_{V^{(0)}_s}^{\MS}(E;\nu)$ and
$\widehat{B}_{V^{(0)}_s}^{\MS}(E^{(0)}_n;\nu)$ are finite in 4 dimensions.
$A_{\MS}^{(D)}$ will be removed by renormalization (note that, unlike the Coulomb case, $A_{\MS}^{(D)}$ now has ${\cal O}(\al^2)$ corrections).  The divergences
are then absorbed in $B_1/B_0$ and we can write
\begin{eqnarray}
&&
\delta \rho_n^{\MS}(\nu)
=
-\frac{8 m_r C_f}{3 m_1m_2}
D_{S^2,s}^{(2)}(\nu)
\left(\widehat{B}_{V^{(0)}_s}^{\MS}(E^{(0)}_n;\nu) 
+
\frac{4}{3}m_rC_f\al+{\cal O}(\al^2)\right).
\label{eq:rho_formulaMS}
\end{eqnarray}
This will have to be combined with the $\MS$ subtracted matching
coefficient $B_1/B_0$ in \Eqn{Rdef} to obtain the decay ratio.

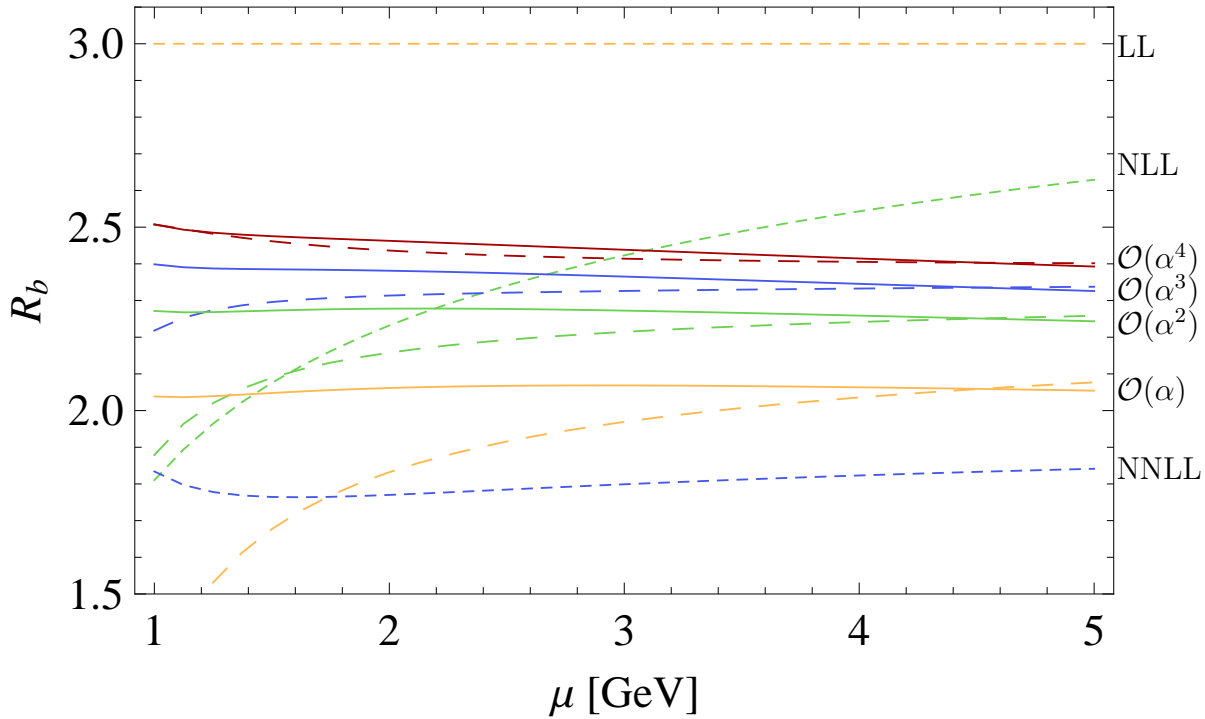
\begin{figure}[t]
\epsfxsize=0.8\textwidth
\epsffile{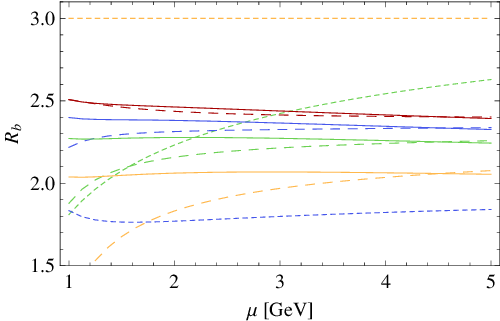}
\put(1,90){NNLL}
\put(1,120){${\cal O}(\al)$}
\put(1,145){${\cal O}(\al^2)$}
\put(1,158){${\cal O}(\al^3)$}
\put(1,170){${\cal O}(\al^4)$}
\put(1,205){NLL}
\put(1,250){LL}
\caption{\label{fig:ratioRSp2_b} \it Decay ratio in the PS scheme at NNLOI
  (dashed) and NNLLI (solid) at different orders in $\al$
  in the static potential (${\cal O}(\al)$: yellow; ${\cal
    O}(\al^2)$: green; ${\cal O}(\al^3)$: blue; ${\cal O}(\al^4)$:
  red). For reference we also include the LL, NLL, and NNLL results
  (short-dashed).  From Ref. \cite{Kiyo:2010jm}. }
\end{figure}

We are then faced with the computation of $\widehat{G}^{(D)}(E^{(0)}_n)$ or, equivalently,
$\widehat{B}_{V^{(0)}_s}^{\MS}(E^{(0)}_n;\nu)$, with the effect of the
static potential included exactly. This calls for a numerical
evaluation of the Green function rather than pursuing an analytic
approach.  Numerical calculations are most conveniently performed in
coordinate space. It is here where finite-$r$ regularization comes
into play obtaining
\begin{eqnarray}
 \lim_{r_0\rightarrow 0}G(r_0, r_0; E)
&=&
\frac{m_r}{2\pi}
\bigg[ \frac{1}{r_0} -2m_rC_f\al \ln\left(\nu\, e^{\gamma_E}r_0\right)
 +{\cal O}(\al^2)+B_{V_s^{(0)}}^{(r)}(E;\nu)\bigg]\, ,
\label{eq:Gr}
\end{eqnarray}
where $B_{V_s^{(0)}}^{(r)}(E;\nu)$ is finite and amenable to a numerical analysis (for the details see
Refs. \cite{Strassler:1990nw,CzaMel2,Kiyo:2010jm}). We can also obtain the change of scheme from $r$ to 
$\MS$ regularization by computing the same object at finite orders in $\al$.
We finally obtain
\bea
\label{deltaMSKiyo}
\delta \rho_n^{\MS}(\nu) &=&
-\frac{8 m_r C_f}{3 m_1m_2}
D_{S^2,s}^{(2)}(\nu)
\left(\widehat{B}_{V^{(0)}_s}^{(r)}(E^{(0)}_n;\nu) 
+
\frac{1}{3} m_rC_f\,\al+{\cal O}(\al^2)\right).
\eea
which we can use in the decay ratios. We are now ready for the
numerical evaluation of the decay ratios using different approximations for 
$V^{(0)}_s$. 
Using the numerical results obtained for ${\hat
  B}^{(r)}_{V_s^{(0)}}(E^{(0)}_n;\nu)$  in Ref. \cite{Kiyo:2010jm} one can get improved
determinations of the decay ratio.   The main source of uncertainties in the evaluation of
${\hat B}^{(r)}_{V_s^{(0)}}(E^{(0)}_n;\nu)$ is reflected by the
computations at different orders in $\al$ in the static potential
and, to a lesser extent, by the dependence on $\nu$ (see Fig. \ref{fig:ratioRSp2_b} with $\nu=\mu$). 
In comparison, the dependence on other parameters is
small (see Ref. \cite{Kiyo:2010jm} for details).  The scheme dependence for renormalon subtraction is also small
compared with the uncertainty due to the computation at different
orders.  

In order to explore different power counting expansions different approximations are considered. The
results obtained within a strict perturbative expansion 
are labelled as LO, NLO and NNLO respectively and, after log resummation, as
 LL, NLL and NNLL (see \cite{Penin:2004ay}). 
Taking into account the static potential exactly, we obtain improved predictions for the relativistic corrections 
that we label by including "I" to the previous labelling: NLLI
(including $B_1/B_0$ with NLL precision and the improved relativistic
correction $\delta \rho_n$) and NNLLI ($B_1/B_0$ with NNLL precision and
the improved relativistic correction $\delta \rho_n$). For comparison we will
also consider the result without resummation of the logarithms in the
Wilson coefficient, NNLOI ($B_1/B_0$ with NNLO precision and the improved 
relativistic correction $\delta \rho_n$). For both, NNLLI and NNLOI we
will consider the results taking the RG improved static potential at LO, NLO, NNLO
and NNNLO.

From the point of view of a double counting in $\al$ and $v$, the NLL
result (with NLL precision for $B_1/B_0$) can be considered as ${\cal
  O}(\al, v^0)$, whereas NLLI is ${\cal O}(\al, v^2)$, and NNLLI is
${\cal O}(\al^2, v^2)$.  As a general trend, moving from NLL to NLLI
improves the scale dependence.  This is due to the fact that, by using
the RG improved expressions, NNLO ${\cal O}(\al^2)$ logarithms count as NLL and can be
matched with a part of the scale dependence of the relativistic ${\cal
  O}(v^2)$ correction.  Note as well that the inclusion of $B_1/B_0$ with
NNLL precision accounts for ${\cal O}(\al^3)$ leading logarithms and
beyond. Those should be cancelled by the inclusion of the subleading
scale dependence of the relativistic correction. Most of it is
actually built in by the numerical evaluation of the relativistic
correction with $h_s^{(0)}$, Eq. (\ref{H0}). In principle, this should be
reflected in an improvement in the scale dependence in going from NLLI
to NNLLI.  On the other hand, this double counting in $\al$ and $v$ scheme 
produces an unmatched scheme dependence, which can only be eliminated by working at the same
order in $\al$ and $v$.

We now focus on the bottom case. In Fig.~\ref{fig:ratioRSp2_b} we show the
decay ratio in the PS scheme at NNLOI and NNLLI at different orders in
$\al$ in the static potential. For reference we also include the LL,
NLL, and NNLL results.  We can see that the inclusion of the RG
Wilson coefficients has a significant impact in reducing the scale
dependence. There is a sizable gap when moving
from NNLL to NNLLI. The bulk of it is already obtained by taken the 
NLO(LO) RG improved static potential in the PS(RS') scheme. 
The inclusion of subleading corrections to the potential
produces a smaller, yet sizable, effect.  

This analysis was used to obtain an updated prediction for
$\Gamma(\eta_b(1S) \rightarrow \gamma\gamma)$.  
The NNLLI result with $\nu=2$~GeV was used for the central value.
The theoretical error was estimated considering the difference between the NLLI 
and NNLLI result for $\nu=2$ GeV. Other experimental and theoretical uncertainties were much smaller. 
After rounding they obtained 
\begin{equation}
\label{GammaEta}
\Gamma(\eta_b(1S) \rightarrow \gamma\gamma)=0.54 \pm 0.15\, {\rm keV}.
\end{equation} 

For the top and charm case a similar pattern was found (see \cite{Kiyo:2010jm}
for details), what it changes is the magnitude of the corrections: larger for charm and smaller for top. 
Whereas in the case of top the new scheme improves over an already quite convergent 
series, in the case of charm the improvement in the agreement with experiment 
is dramatic. For charmonium this scheme brings consistency between the weak coupling computation
and the experimental value of the decay ratio, but the theoretical
error is large. 

These results call for a reanalysis in this new scheme of previous studies. In particular, 
it is an open question its impact in the case of the hyperfine splitting. 

\section{Conclusions}

The EFT named pNRQCD aims at describing the Heavy Quarkonium. It smoothly connects potential models and relativistic quantum field theories. The problem is formulated in a 
NR quantum mechanical fashion in terms of Schr\"odinger equations.
  In this review we have focused on pNRQCD in the weak coupling limit. In this limit the construction of the effective Lagrangian (the determination of its Wilson coefficients) can be done within perturbation theory, order by order in $\al$ and $1/m$. The construction gets greatly simplified by following a step-by-step procedure, dealing with one scale at each step. The final effective theory resembles very much a Schrödinger equation, yet ultrasoft gluons are incorporated in a second-quantized, systematic and gauge-invariant way. Even though the computation of physical observables requires the summation of an infinite number
of diagrams, all the steps of the computation can be performed in dimensional 
regularization. One can then go through the renormalization in the NR bound states problem using the very same 
techniques that one uses in standard renormalization of relativistic quantum field theories. Then, one can naturally obtain the 
RG equations of the Wilson coefficients of the effective theory, which can also be computed in perturbation theory, 
order-by-order in $\al$. 
By solving them the resummation of the large logarithms arising due to the various scales in the problem 
is readily obtained.  This problem is
non-trivial because all scales (hard, soft and ultrasoft) play a role. We have provided some 
explicit examples on the construction of the theory (matching) and also on the renormalization, and in some 
cases provided the most up-to-date expressions available in the literature. 

On the experimental side one major issue is to clarify which bound states (i.e. range of energies) belong to the
weak coupling regime, and to which extent they can be described by this theory. There are places where we 
expect the theory to work better. These are the golden modes of the theory, and should be studied in detail to answer these questions. 
Among those the ideal place is clearly the production of $t$-$\bar t$ near threshold, due to the large mass of the top. 
Yet, its study has not been free of problems. The proper treatment of renormalon effects is compulsory for accurate determinations of the 
top mass from the scan of the $t$-$\bar t$ production near threshold in a future Next Linear Collider. The normalization remained a challenge, as the size of the corrections and the factorization scale dependence were large. The resummation of logarithms has improved the situation considerable, yet the remaining uncertainties were somewhat larger than
expected. At this respect there has been recent developments considering a modification of the LO Hamiltonian (therefore a rearrangement of the perturbative series) that may lead to a more convergent series. First results along this line are encouraging.

The main drawback of $t$-$\bar t$ production near threshold is that there is not experimental data to confront theory. 
This is clearly not so for the next natural place on which to apply the formalism, $b$-$\bar b$ systems. Among all possible 
observables, one should look upon those where the energy transfer between the $b$ quarks is the largest. 
Those would be optimal for applicability of the weak coupling version of pNRQCD. This means $b$-$\bar b$ 
NR sum rules and some observables related with the 
bottomonium ground state. In fact two of the most competitive determinations of the $b$ quark mass comes from 
NR sum rules and the 
$\Upsilon(1S)$ mass (again the correct treatment of the renormalon is crucial). We have seen that the $\Upsilon(1S)$ 
mass perturbative series 
shows a convergent pattern and the precision is set by non-perturbative effects. NR sum rules show 
somewhat a similar behavior than 
$t$-$\bar t$ production near threshold, but more extreme, since $\al$ is bigger in this case. The resummation of logarithms significantly diminish the scale dependence and considerable helps to give an stable result. The size of the corrections are large though. It remains to be seen whether the rearrangement of the perturbative expansion mentioned in the previous paragraph has a significant effect as well here, in the hyperfine splitting, or in inclusive decay widths (which even show a worse behavior). For the decay ratio such analysis exists and there is a huge improvement, which is also seen (even more dramatically) 
for Charmonium. Yet, the applicability of the weak coupling version of pNRQCD to Charmonium is quite inconclusive. In many instances, LO results give reasonable numbers (albeit with huge uncertainties).  Beyond LO, 
a determination of $m_c$ has been obtained using the $J/\Psi$ mass in the Upsilon scheme with a relatively convergent series. A determination of $m_c$ in the PS and RS scheme using NR sum rules also exists.
 
Overall, the study of Heavy Quarkonium at weak coupling has already provided (or could provide) with 
good determinations of some of the parameters of the standard model like the heavy quark masses: $m_t$, $m_b$ and $m_c$. 
It has also improved our understanding of the dynamics of those states through the study of transitions, decays, production, NR 
sum rules, etc... At this respect a better understanding of the convergence pattern of the wave function at the origin would certainly help, which could be achieved from higher order computations, completing the NNLL and NNNLO expressions, and/or 
using improved rearrangement of the perturbative series. 

Even though we did not have space to discuss them, semiinclusive radiative decays of Heavy Quarkonium (see Ref. \cite{GarciaiTormo:2005ch}) have been used to obtain determinations of $\alpha_s(M_z)$  \cite{Brambilla:2007cz} using  $\Upsilon(1S)$ data. We also had no space for the incorporation of finite temperature effects, which have received quite some attention recently, or the generalization of the effective theory if the heavy particles are not stable. 

\bigskip

{\bf Acknowledgments}. 
This work was partially supported by the spanish 
grants FPA2010-16963 and FPA2011-25948, and by the catalan grant SGR2009-00894. We thank C. Peset for pointing out some of the misprints corrected in this version. 

\vfill
\eject

\appendix

\section{NRQCD Feynman rules}
The propagator of the $\psi$ and $\chi_c$ field are equal (as both represent particles): $\displaystyle{\frac{i}{q^0-\frac{{\bf q}^2}{2m}+i\eta}}$. Note though that in the computation of the potential the static version should be used: $\displaystyle{\frac{i}{q^0+i\eta}}$. Moreover, identity matrices in spin and colour are implicit.

The gluon propagator depends on the gauge. Usually one uses either the Feynman: $\displaystyle{-\frac{i\delta_{ab}}{q^2+i\eta}g^{\mu\nu}}$
or the Coulomb gauge:
\begin{figure}[htb]
\makebox[0.0cm]{\phantom b}
\put(0,1){Longitudinal gluon ($\langle A^a_0A^b_0\rangle$)=}
\put(180,1){\epsfxsize=4.5truecm \epsfbox{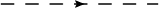}}
\put(330,1){$
\displaystyle{=\frac{i}{{\bf q}^2}\delta^{ab}}$,}
\label{figA0propagator}
\end{figure}
\begin{figure}[htb]
\makebox[0.0cm]{\phantom b}
\put(0,1){Tranverse gluon ($\langle A^a_iA^b_j\rangle$)=}
\put(180,1){\epsfxsize=4.5truecm \epsfbox{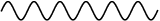}}
\put(330,1){$
\displaystyle{=\frac{i}{q^2+i\eta}\left(\delta_{ij}-\frac{q_iq_j}{{\bf q}^2}\right)\delta^{ab}}$,}
\label{figAijpropagator}
\end{figure}

The Feynman rules for the NRQED vertices can be found in Refs \cite{Kinoshita:1995mt,Labelle:1996en} and for the 
NRQCD vertices in Ref. \cite{Bodwin:1998mn}. Here we display them in a notation consistent with this review 
(for the $1/m^2$ operators the set of Feynman rules is not complete, we have displayed only those that appear more often). We take $k=p-p'$. 
For the interaction of gluons with particles we have
\begin{align}
\lefteqn{\raisebox{-7 ex}{\includegraphics[width=0.25 \textwidth]{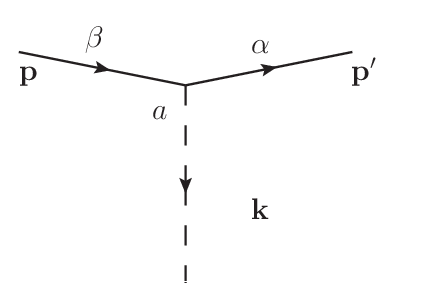} } =   \qquad \begin{array}{l}
{\rm Coulomb\;vertex}
\\[3.7 ex]
-igT^a_{\alpha\beta}
\\[2 ex]
\\[2 ex]
 \end{array}
}&&   
\\[4 ex]
\lefteqn{\raisebox{-7 ex}{\includegraphics[width=0.25 \textwidth]{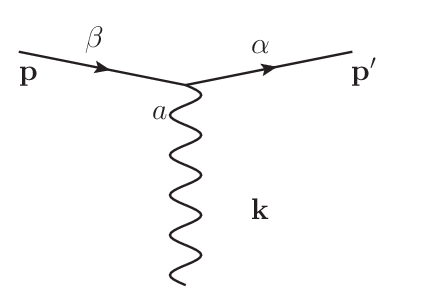} } =   \qquad \begin{array}{l}
{\rm Dipole\;vertex}
\\[3.7 ex]
\frac{ig}{2m}({\bf p}+{\bf p}')T^a_{\alpha\beta}
\\[2 ex]
\\[2 ex]
 \end{array}
}&&   
\\[4 ex]
\lefteqn{\raisebox{-7 ex}{\includegraphics[width=0.25 \textwidth]{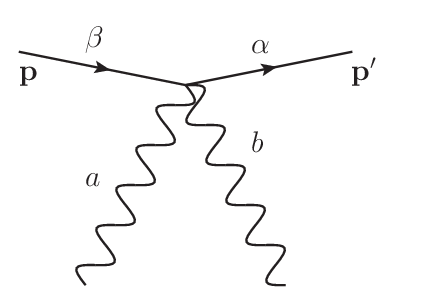} } =   \qquad \begin{array}{l}
{\rm Seagull\;vertex}
\\[3.7 ex]
-\frac{ig^2}{2m}(T^aT^b+T^bT^a)_{\alpha\beta}\delta^{ij}
\\[2 ex]
\\[2 ex]
 \end{array}
}&&   
\end{align}
\begin{align}
\lefteqn{\raisebox{-7 ex}{\includegraphics[width=0.25 \textwidth]{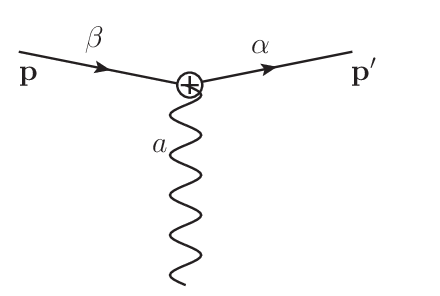} } =   \qquad \begin{array}{l}
{\rm Fermi\;vertex}
\\[3.7 ex]
\frac{c_Fg}{2m}(\bfsigma \times {\bf k}) T^a_{\alpha\beta}
\\[2 ex]
\\[2 ex]
 \end{array}
}&&   
\\[4 ex]
\lefteqn{\raisebox{-7 ex}{\includegraphics[width=0.25 \textwidth]{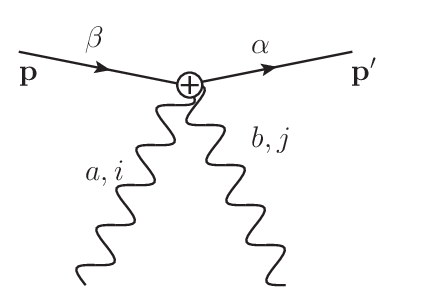} } =   \qquad \begin{array}{l}
{\rm Fermi\;vertex\; (non-abelian)}
\\[3.7 ex]
\frac{c_Fg^2}{2m}\bfsigma^k\epsilon^{kij}[T^a,T^b]_{\alpha\beta}
\\[2 ex]
\\[2 ex]
 \end{array}
}&&   
\\[4 ex]
\lefteqn{\raisebox{-7 ex}{\includegraphics[width=0.25 \textwidth]{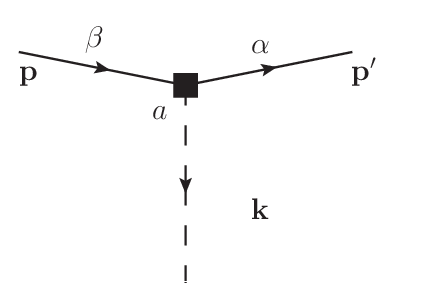} } =   \qquad \begin{array}{l}
{\rm Darwin\;vertex\; (A_0)}
\\[3.7 ex]
i\frac{c_Dg}{8m^2}{\bf k}^2T^a_{\alpha\beta}
\\[2 ex]
\\[2 ex]
 \end{array}
}&&   
\\[4 ex]
\lefteqn{\raisebox{-7 ex}{\includegraphics[width=0.25 \textwidth]{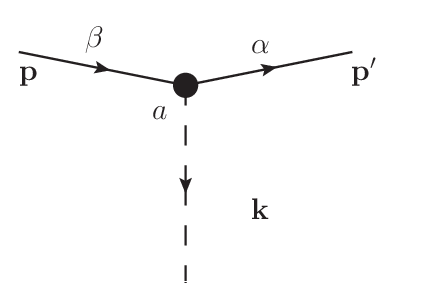} } =   \qquad \begin{array}{l}
{\rm c_S\;vertex\; (A_0)}
\\[3.7 ex]
i\frac{c_Sg}{4m^2}\bfsigma \cdot ({\bf p}'\times{\bf p})T^a_{\alpha\beta}
\\[2 ex]
\\[2 ex]
 \end{array}
}&&   
\end{align}
{\bf Antiparticles}.
The Feynman rules for the antiparticle can be easily deduced from the ones of the particle by the exchange $g \rightarrow -g$ and 
$T^a \rightarrow (T^a)^{T}$.

For the four-fermion operators we have
\begin{align}
\lefteqn{\raisebox{-7 ex}{\includegraphics[width=0.25 \textwidth]{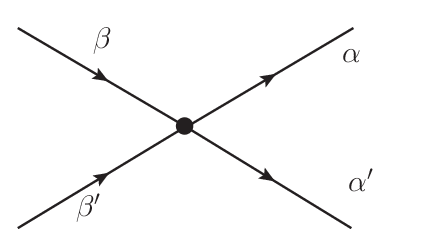} } =   \qquad \begin{array}{l}
d_{ss} {\rm \;vertex}
\\[3.7 ex]
-i\frac{d_{ss}}{m_1m_2}\delta_{\alpha\beta}\delta_{\alpha'\beta'}
\\[2 ex]
\\[2 ex]
 \end{array}
}&&   
\\[4 ex]
\lefteqn{\raisebox{-7 ex}{\includegraphics[width=0.25 \textwidth]{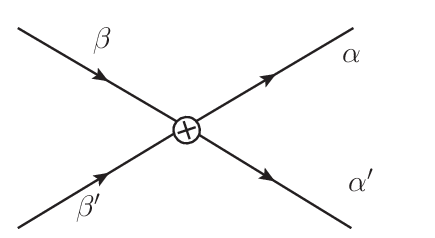} } =   \qquad \begin{array}{l}
d_{sv} {\rm \;vertex}
\\[3.7 ex]
i\frac{d_{sv}}{m_1m_2}\bfsigma_1\cdot\bfsigma_2\delta_{\alpha\beta}\delta_{\alpha'\beta'}
\\[2 ex]
\\[2 ex]
 \end{array}
}&&   
\end{align}

\begin{align}
\lefteqn{\raisebox{-7 ex}{\includegraphics[width=0.25 \textwidth]{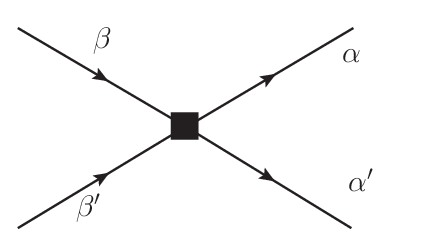} } =   \qquad \begin{array}{l}
d_{vs} {\rm \;vertex}
\\[3.7 ex]
-i\frac{d_{vs}}{m_1m_2}T^a_{\alpha\beta}T^a_{\beta'\alpha'}
\\[2 ex]
\\[2 ex]
 \end{array}
}&&   
\\[4 ex]
\lefteqn{\raisebox{-7 ex}{\includegraphics[width=0.25 \textwidth]{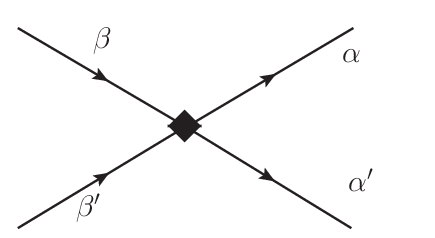} } =   \qquad \begin{array}{l}
d_{vv} {\rm \;vertex}
\\[3.7 ex]
i\frac{d_{vv}}{m_1m_2}\bfsigma_1\cdot\bfsigma_2T^a_{\alpha\beta}T^a_{\beta'\alpha'}
\\[2 ex]
\\[2 ex]
 \end{array}
}&&   
\end{align}

The QED Feynman diagrams can be obtained by eliminating the non-abelian diagrams and replacing $T^a \rightarrow 1$.

\section{pNRQCD Feynman rules}
\label{sec:pNRFR}

\begin{figure}[htb]
\makebox[6.5cm]{\phantom b}
\put(-150,1){\epsfxsize=13truecm \epsfbox{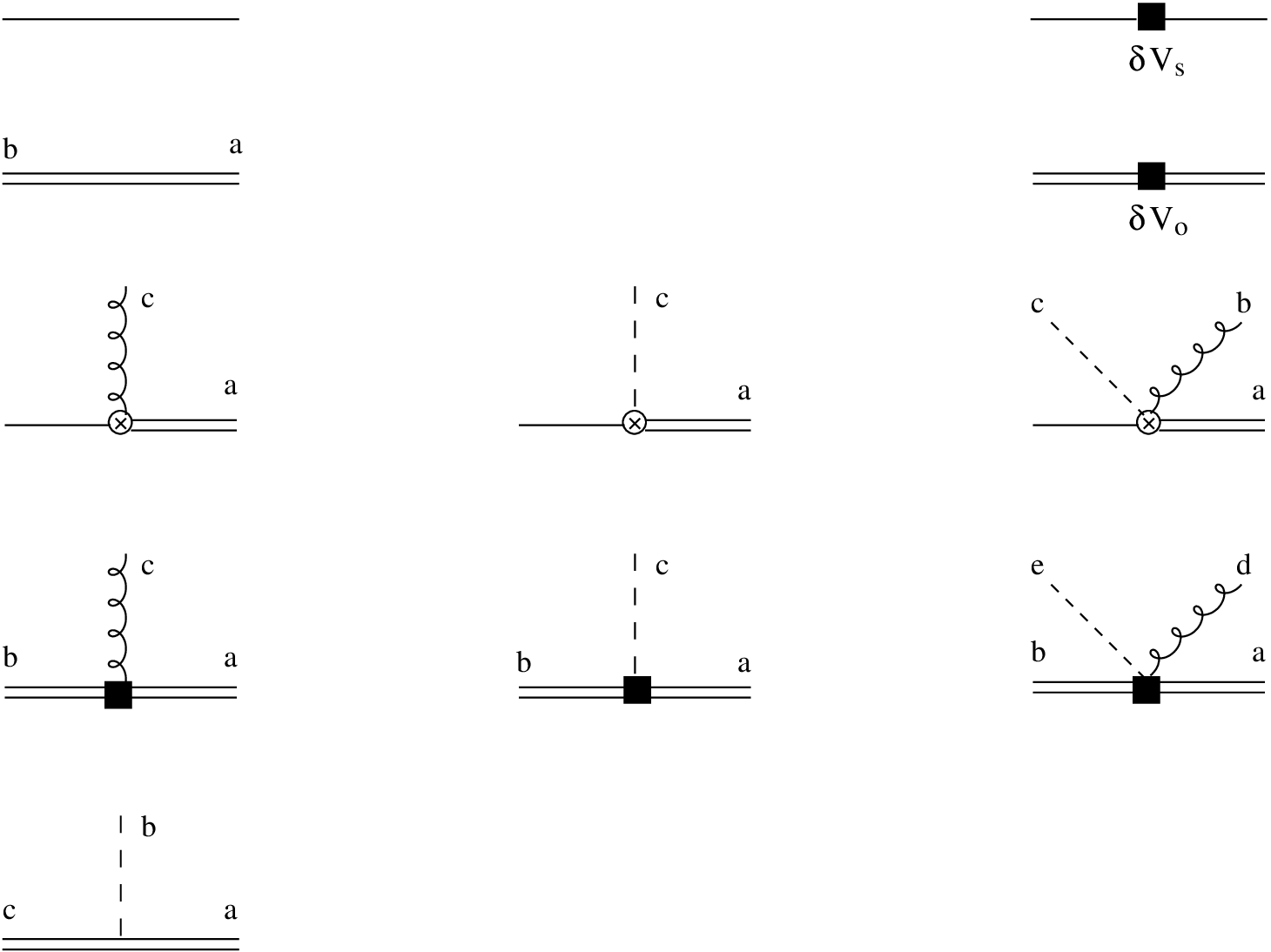}}
\put(-75,270){\tiny $=-iG_c(E)=\displaystyle{{i \over 
\displaystyle{E-h_s^C}}
=
{i \over \displaystyle{E-{\bf p}^2/m-C_f \al/ r}}}$}
\put(-75,225){\tiny $=-iG_c^{o}(E)\delta_{ab}=\displaystyle{{i 
\delta_{ab}\over 
\displaystyle{E-h_o^C}}
={i \delta_{ab} \over \displaystyle{E-{\bf p}^2/m-(1/(2N_c)) \al/ r}}}
$}
\put(225,270){\tiny $= - i \delta V_s $}
\put(225,225){\tiny $= - i \delta V_o $}
\put(75,153){\tiny $= g V_A \displaystyle\sqrt{T_F\over N_c} 
\delta_{ca}{\bf r} \cdot{\bf P}$}
\put(-75,153){\tiny $= -g V_A \displaystyle\sqrt{T_F\over N_c} 
\delta_{ca}{\bf r} P^0$}
\put(225,153){\tiny $=-ig^2V_A \displaystyle\sqrt{T_F\over N_c} {\bf 
r}f_{abc}$}
\put(75,75){\tiny $= g \displaystyle{V_B\over 2} d^{abc} {\bf r} \cdot 
{\bf P}$}
\put(-75,75){\tiny $=-g \displaystyle{V_B\over 2} d^{abc} {\bf r} P^0$}
\put(225,75){\tiny $=-ig^2\displaystyle{V_B\over 2} d^{abc} f_{cde}{\bf 
r}$}
\put(-75,0){\tiny $=  g f^{abc}$}
\caption{ \it Propagators and vertices of the pNRQCD Lagrangian 
(\ref{pnrqcd0}). Dashed lines represent longitudinal gluons and 
curly lines transverse gluons. $P^\mu$ represents the gluon 
incoming momentum.}
\label{pnrqcdfig}
\end{figure}

The propagator of the singlet reads 
\be
{i \over \displaystyle{E-h_s}} \,.
\ee
This expression contains subleading terms in the velocity expansion. 
In order to have homogeneous power counting, it is 
convenient to expand it about the Coulomb Green function, $G_c$, 
defined in Fig.~\ref{pnrqcdfig}, which scales as $1/(mv^2)$, 
and similarly for the octet. The complete set of Feynman rules 
at the order displayed in (\ref{pnrqcd0}) is shown in Fig.~\ref{pnrqcdfig}.

\section{Constants and useful Formulae}
\be
T_F= {1 \over 2}; \quad C_A=N_c; \quad C_f = {N_c^2 - 1 \over 2\,N_c}\,.
\ee
\be
\beta_0=11\frac{C_A}{3}-{4 \over 3}T_Fn_f;
\qquad
\beta_1=34\frac{C_A^2}{3}-{20 \over 3}C_AT_Fn_f-4C_fT_Fn_f
;\ee
\be
\beta_2= {2857 \over 54}C_A^3-{1415 \over 27}C_A^2T_Fn_f+\frac{158}{27}C_AT_F^2n_f^2
-\frac{205}{9}C_AC_fT_Fn_f+\frac{44}{9}C_fT_F^2n_f^2+2C_f^2T_Fn_f
.\ee

\medskip

\noindent
{\bf NRQCD Lagrangian Wilson coefficients}: \\
$\al(m)$ has 
$n_f$ active light flavours and we define $z=\left[{\al(\nu) \over \al(m)}\right]^{1 \over
\beta_0}\simeq 1 -1/(2\pi)\al(\nu)\ln ({\nu \over m})$:
\bea
c_F(\nu)&=&c_F(m)-1+z^{-C_A}
\,,
\nn\\
c_D(\nu)&=&c_D(m)-1+
{9C_A \over 9C_A+8T_Fn_f}
\left\{
-\frac{5 C_A + 4 T_F n_f}{4 C_A + 4
T_F n_f} z^{-2 C_A} +
\frac{C_A +16 C_f - 8 T_F n_f}{2(C_A-2T_F n_f)}
\right.
\nn
\\
&&\qquad
+ \frac{-7 C_A^2 + 32 C_A C_f - 4 C_A T_F n_f +32 C_f T_F
n_f}{4(C_A + T_F n_f)(2 T_F n_f-C_A)} z^{4 T_F n_f/3 - 2C_A/3}
\nn
\\
&&
\qquad
\left.
+{8T_Fn_f \over 9C_A}
\left[
z^{-2C_A}+\left({20 \over 13}+{32 \over 13}{C_f \over
C_A}\right)\left[1-z^{-13C_A \over 6}\right]
\right]
\right\}
\,,
\nn\\
d_{ss}(\nu)&=&d_{ss}(m)+4C_f\left(C_f-{C_A \over 2}\right){\pi\over\beta_0}\al(m)\left[z^{\beta_0}-1\right]
\,,
\nn\\
d_{sv}(\nu)&=&d_{sv}(m)
\,,
\nn\\
d_{vs}(\nu)&=&d_{vs}(m)-\left(C_f-C_A\right){8\pi\over\beta_0}\al(m)
\left[z^{\beta_0}-1 \right]
\nn
\\
&&
-{27C_A^2 \over 9C_A+8T_Fn_f}{\pi \over \beta_0}\al(m)
\left\{
-\frac{5 C_A + 4 T_F n_f}{4 C_A + 4
T_F n_f}{\beta_0 \over \beta_0-2C_A}\left(z^{\beta_0-2 C_A}-1\right) 
\right.
\nn
\\
&&
\qquad
+
\frac{C_A +16 C_f - 8 T_F n_f}{2(C_A-2T_F n_f)}\left(z^{\beta_0}-1\right)
\nn
\\
&&
\qquad
+ \frac{-7 C_A^2 + 32 C_A C_f - 4 C_A T_F n_f +32 C_f T_F
n_f}{4(C_A + T_F n_f)(2 T_F n_f-C_A)}
\nn
\\
&&
\qquad
\qquad
\times
{3\beta_0 \over 3\beta_0+4T_Fn_f-2C_A}\left( z^{\beta_0+4 T_F n_f/3 - 2C_A/3}-1\right)
\nn
\\
&&
\qquad
+{8T_Fn_f \over 9C_A}
\left[{\beta_0 \over \beta_0-2C_A}\left(z^{\beta_0-2C_A}-1\right)
+\left({20 \over 13}+{32 \over 13}{C_f \over C_A}\right)
\right.
\nn
\\
&&
\qquad\qquad
\left.
\left.
\times
\left(
\left[z^{\beta_0}-1\right]-{6\beta_0 \over 6\beta_0-13C_A}
\left[z^{\beta_0-{13C_A \over 6}}-1\right]
\right)
\right]
\right\}
\,,
\nn\\
d_{vv}(\nu)&=&d_{vv}(m)+{C_A \over
\beta_0-2C_A}\pi\al(m)\left\{z^{\beta_0-2C_A}-1\right\}
\label{RGeqhs}
\,,
\eea
where
\be
d_2(m)=\frac{\al(m)}{60\pi}T_F
,
\ee
\be
c_F(m)=1+\frac{\al(m)}{2\pi}(C_f+C_A)
,
\ee
\be
c_D(m)=1+\frac{\al(m)}{2\pi}C_A-16d_2(m)
.
\ee
\begin{eqnarray}
c_1^{g} &=& { \al(m)\over 360 \pi} T_F  \,,
\end{eqnarray}

\bea
d_{ss}^{a}(m)&=& \al^2(m) C_{f}\left({C_{A}\over 2}-C_{f}\right)
               \left(2-2\ln2 + i \pi\right) \,,\\
d_{sv}^{a}(m)&=& 0 \,,\\
d_{vs}^{a}(m)&=& {\al^2(m) \over 2}
               \left(-{3 \over 2}C_A+4C_f \right)
                \left(2-2\ln2 + i \pi\right) \,,\\
d_{vv}^{a}(m)&=& -\pi\al(m)\Biggl[1+{\al(m)\over\pi}\Biggl(
T_{R} \left[{1\over 3}n_{f}\left( 2\ln2-{5\over 3}-i\pi\right)
-{8\over
9}\right] \nonumber\\
&  &+C_{A}{109\over 36} - 4C_{f}\Biggr)\Biggr]
\,.
\eea

\bea
\nonumber
d_{ss}(m) &=& -{d_{ss}^a(m) \over 2N_c}- {3 d_{sv}^a(m)  \over 2N_c}
      -{N^2_c-1 \over 4N_c^2}d_{vs}^a(m)  - 3{N^2_c-1\over 4N_c^2} d_{vv}^a(m)
  + {2 \over 3}
C_{f}\left( {C_{A}\over 2}-C_{f}\right)
 \al^2(m)
\,,\\
\nonumber
d_{sv}(m)  &=& -{d_{ss}^a(m)  \over 2N_c}+ { d_{sv}^a(m)  \over 2N_c}
      -{N^2_c-1 \over 4N_c^2}d_{vs}^a(m)  + {N^2_c-1\over 4N_c^2} d_{vv}^a(m) +
C_{f}\left( {C_{A}\over 2}-C_{f}\right)
 \al^2(m)
\,,\\
\nonumber
d_{vs}(m)  &=& -d_{ss}^a(m)  - 3 d_{sv}^a(m) 
      +{d_{vs}^a(m)  \over 2N_c}+  {3 d_{vv}^a(m)  \over 2N_c}
+ \left({4 \over 3} C_f 
  + {11 \over 12} C_A \right)\al^2(m)
\,,\\
d_{vv}(m)  &=&  -d_{ss}^a(m)  +  d_{sv}^a(m) 
      +{d_{vs}^a(m)  \over 2N_c}- { d_{vv}^a(m)  \over 2N_c}
+\al^2(m)(2C_f-\frac{C_A}{2})
\,.\eea

The results displayed above for the NRQCD Wilson coefficients, $c$'s and $d$'s,
are correct with LL and NLO accuracy, but not beyond. At this order $d(\nu_p,\nu_s)\simeq d(\nu_s)$. 
We will also need
\bea
c_1^{hl}(\nu)&=&
{9C_A \over 9C_A+8T_Fn_f}
\left\{
\frac{5 C_A + 4 T_F n_f}{4 C_A + 4
T_F n_f} z^{-2 C_A} -
\frac{C_A +16 C_f - 8 T_F n_f}{2(C_A-2T_F n_f)}
\right.
\nn
\\
&&\qquad
- \frac{-7 C_A^2 + 32 C_A C_f - 4 C_A T_F n_f +32 C_f T_F
n_f}{4(C_A + T_F n_f)(2 T_F n_f-C_A)} z^{4 T_F n_f/3 - 2C_A/3}
\nn
\\
&&
\qquad
\left.
+
z^{-2C_A}+\left({20 \over 13}+{32 \over 13}{C_f \over
C_A}\right)\left[1-z^{-13C_A \over 6}\right]
\right\}
\,.
\eea
This is correct with LL accuracy. The complete ${\cal O}(\al^2)$ correction to $c_F$ is also known \cite{Amoros:1997rx}. Moreover $c_k=c_4=1$ and $c_S=2c_F-1$ due to reparameterization invariance \cite{Manohar:1997qy}.

Finally, we would like to remark that, since the basis of operators is not minimal, there are some ambiguities in the values of some Wilson coefficients, in 
particular the expressions of $d_{vs}$ and $c_D$ can depend on the gauge, yet the combination $\al c_D+d_{vs}/\pi$ 
is free of ambiguities. 

\medskip
{\bf NRQCD current Wilson coefficients}.\\
\begin{eqnarray}
  b_1^{(1)}=-2C_f\,,\qquad
  b_0^{(1)}=-\left(\frac{5}{2}-\frac{\pi^2}{8}\right)C_f\,,
\end{eqnarray}
\bea
b_1^{(2)}(\nu)&=&\left(-{151\over72}
+{89\pi^2\over144}
-{5\pi^2\over6}\ln2-{13\over4}\zeta(3)\right)C_AC_f
+\left({23\over8}-{79\pi^2\over36}\right.
\nn\\
&&
+\left.\pi^2\ln2-{1\over2}\zeta(3)\right)C_f^2
+\left({22\over9}-{2\pi^2\over9}\right)C_fT_F
\nn\\
&&
+{11\over18}C_fT_Fn_f+\left[\beta_0+\pi^2\left({C_A\over 2}
+{C_f \over 3}\right)\right]C_f\ln\left({m\over \nu}\right)\,,
\label{cv2}
\end{eqnarray}
\begin{eqnarray}
b_0^{(2)}(\nu)&=&
-4.79(5)C_AC_f-21.02(10)C_f^2+0.224(1)C_fT_F
+\left(\frac{41}{36}-\frac{13\pi^2}{144}-\frac{2}{3}\ln2 \right.
\nn\\
&&
\left.-\frac{7}{24}\zeta(3)\right)C_fT_Fn_f+\left[\left(\frac{5}{4}
-\frac{\pi^2}{16}\right)\beta_0+\pi^2\left({C_A
\over 2}+{C_f}\right)\right]C_f\ln\left({m\over \nu}\right)\,.
\label{cg2}
\end{eqnarray}

\medskip

\noindent
{\bf Static potential related constants}.
\be
a_1={31C_A-20T_Fn_f \over 9};
\ee
\bea
\nonumber
&a_2 =& 
{{400\,{{{\it n_f}}^2}\,{{{\it T_F}}^2}}\over {81}} -
     {\it C_f}\,{\it n_f}\,{\it T_F}\,
      \left( {{55}\over 3} - 16\,\zeta(3) \right) 
\\
&&
\nn
 +
     {{{\it C_A}}^2}\,\left( {{4343}\over {162}} +
        {{16\,{{\pi }^2} - {{\pi }^4}}\over 4} + {{22\,\zeta(3)}\over 3}
        \right)
 - {\it C_A}\,{\it n_f}\,{\it T_F}\,
      \left( {{1798}\over {81}} + {{56\,\zeta(3)}\over 3} \right)
	  ;
\eea
\begin{eqnarray}
  a_3 = a_3^{(3)} n_f^3 + a_3^{(2)} n_f^2 + a_3^{(1)} n_f + a_3^{(0)}
  \,,
\end{eqnarray}
where
\begin{eqnarray}
  a_3^{(3)} &=& - \left(\frac{20}{9}\right)^3 T_F^3
  \,,\nonumber\\
  a_3^{(2)} &=&
  \left(\frac{12541}{243}
    + \frac{368\zeta(3)}{3}
    + \frac{64\pi^4}{135}
  \right) C_A T_F^2
  +
  \left(\frac{14002}{81}
    - \frac{416\zeta(3)}{3}
  \right) C_f T_F^2
  \,,\nonumber\\
  a_3^{(1)} &=&
  \left(-709.717
  \right) C_A^2 T_F
  +
  \left(-\frac{71281}{162}
    + 264 \zeta(3)
    + 80 \zeta(5)
  \right) C_AC_f T_F
  \nonumber\\&&\mbox{}
  +
  \left(\frac{286}{9}
    + \frac{296\zeta(3)}{3}
    - 160\zeta(5)
  \right) C_f^2 T_F
 +
  \left(-56.83(1)
  \right) \frac{d_F^{abcd}d_F^{abcd}}{N_A} \,,
  \nonumber\\
  a_3^{(0)} &=&
  502.24(1) \,\, C_A^3
  -136.39(12)\,\, \frac{d_F^{abcd}d_A^{abcd}}{N_A}
  \,,
  \label{eq::a3}
\end{eqnarray}
and
\be
\frac{d^{abcd}_F d^{abcd}_A}{N_A}=\frac{N_c(N_c^2+6)}{48}
\,.
\ee

\medskip

\noindent
{\bf Fourier Transforms}.\\
It is convenient to have expressions for the potentials in $d$-dimensions both in position and momentum space. 
Therefore, the $d$-dimensional Fourier transform is needed. It can be found in several places, in particular in Ref. \cite{Pascual:1984zb}
\begin{equation}
\int\frac{d^dk}{(2\pi)^d}\frac{e^{-i\bk\cdot\br}}{\mbk^n}=
\frac{2^{-n}\pi^{-d/2}}{r^{d-n}}\frac{\Gamma\left(d/2-n/2\right)}{\Gamma\left(n/2\right)}.
\label{ftregdim}
\end{equation}
We also show other useful Fourier transform equation in three dimensions, quoted from Ref. \cite{Titard:1993nn}:
\begin{eqnarray*}
&\displaystyle{
 \Vt({\bf k})
\phantom{aaaaaaaaaaaaaaaa}} &
V({\bf r}) = \frac{1}{(2\pi)^{3}} \,\int\! d^3\! k \,e^{-i {\bf k} \cdot {\bf r}}\,\Vt({\bf k})
\\
&&
\\
&\displaystyle{
 \log k
\phantom{aaaaaaaaaaaaaaaa}} &
 - \frac{1}{4\pi} {\rm reg}\,\frac{1}{r^3}
\\
&\displaystyle{
 \frac{1}{k}
\phantom{aaaaaaaaaaaaaaaa}} &
\frac{1}{2\pi^2}\,\frac{1}{r^2}
\\
&\displaystyle{
\frac{\log k}{k}
\phantom{aaaaaaaaaaaaaaaa}} &
- \frac{1}{2\pi^2}\,\frac{\log r + \eul}{r^2}
\\
&\displaystyle{
\frac{1}{k^2}
\phantom{aaaaaaaaaaaaaaaa}} &
\frac{1}{4 \pi}\,\frac{1}{r}
\\
&\displaystyle{
\frac{\log k}{k^2}
\phantom{aaaaaaaaaaaaaaaa}} &
- \frac{1}{4\pi}\,\frac{\log r + \eul}{r}
\\
&\displaystyle{
\frac{\log^2 k}{k^2}
\phantom{aaaaaaaaaaaaaaaa}} &
\frac{1}{4\pi}\,\frac{(\log r + \eul)^2
+ \pi^2/12}{r}
\\
&\displaystyle{
\Lambda({\bf k})
\phantom{aaaaaaaaaaaaaaaa}} &
-\frac{1}{4\pi r^3} {\bf S} \cdot {\bf L}
\\
&\displaystyle{
\Lambda({\bf k})\log k
\phantom{aaaaaaaaaaaaaaaa}} &
-\frac{1- \log r}{4\pi r^3} {\bf S} \cdot {\bf L}
\\
&\displaystyle{
 \frac{1}{3} T(\bf k)
\phantom{aaaaaaaaaaaaaaaa}} &
\frac{1}{4\pi r^3}\,S_{12}(\bf r)
\\
&\displaystyle{
 \frac{\log k}{3} \,T(\bf k)
\phantom{aaaaaaaaaaaaaaaa}} &
\frac{4/3 - \log r}{4\pi r^3}
\,S_{12}(\bf r)
\end{eqnarray*}
where,
\begin{eqnarray*}
\Lambda
 &\equiv&
- i {\bf S} \cdot \frac{{\bf k} \times {\bf p}}{{\bf k}^2} \; , \;
\; T  \equiv
\frac{{\bf k}^2 \bfsigma_1 \cdot \bfsigma_2 -
3({\bf k} \cdot \bfsigma_1) ({\bf k} \cdot \bfsigma_2)}{{\bf k}^2}  \;.
\end{eqnarray*}
and
\begin{eqnarray*}
\int d^n r \phi({\bf r}) \,{\rm reg}\,\frac{1}{r^n}
 &\equiv&  \lim_{\eps \to 0}
\left\{
 \int\!d^n r \,\phi({\bf r})\,\frac{r^{\eps}}{r^n} - A(n,\eps)\phi(0)
\right\}
\;,
\\
A(n,\eps)
 &\equiv&
 \frac{2\pi^{n/2}}{\Gamma(n/2)}
\left\{
 \frac{1}{\eps} + \log 2 + \frac{\psi(n/2) - \eul}{2}
\right\}
\;.
\end{eqnarray*}

\medskip

\noindent
{\bf Coulomb Green function}.\\
One defines the partial-wave Green function in the standard way:
\be
\langle {\bf x} | {1 \over h_s^C-E} | {\bf y} \rangle =
\sum_{l=0}^{\infty} (2\,l+1) G_l^{(\al)} (x,y; E) P_l({ {\bf x}
\cdot {\bf y} \over x y})
\,.
\ee
For negative energy $E=-k^2/(2m_r)$ one can get
\bea
\label{gl}
&&G_l^{(\al)} (x,y; -k^2/(2m_r))
\\
&&
\nn
= {m_r\,k \over  \pi} (2\,k\,x)^l (2\,k\,y)^l
e^{-k(x+y)} \sum_{s=0}^{\infty}
{L_s^{2\,l+1}(2\,k\,x) L_s^{2\,l+1}(2\,k\,y)\,s! \over
(s+l+1-(m_r  C_f\al)/k)\,(s+2\,l+1)!}
\,.
\eea
These formulas are quoted from Ref. \cite{Voloshin:1979uv} (note that there is a
misprint in formula (15) there, and $(s+l+1)!$ must be changed to $(s +
2\,l+1)!$).

\medskip

\noindent
{\bf  Expectation Values}.\\
We also present here a few expectation values in (unperturbed)
Coulombic wave functions, which we take from Ref. \cite{Titard:1993nn}. We define $a$ as
the Bohr--like radius; i.e.,  $a= 1/(m_rC_f \alpha)$. We have:
\begin{eqnarray*}
 \langle {\rm reg}\,\frac{1}{r^3} \rangle_{n0}
 & = &
 \frac{4}{n^3 a^3}\left\{ \log \frac{n a}{2} -
 \sum_{k=1}^{n} \frac{1}{k} -\frac{n-1}{2 n} \right\} \;,
\\
 \langle {\rm reg}\,\frac{1}{r^3}  \rangle_{nl}
 & = &
 \frac{2}{n^3 a^3} \,\frac{1}{l(l+1)(2 l+1)} \;, \; l \neq 0 \;
\\
\langle \frac{\log r}{r} \rangle_{nl}
 & = &
\frac{\log n a/2 + \psi(n+l+1)}{n^2 a}
\\
\langle \frac{\log r}{r^2} \rangle_{nl}
 & = &
\frac{2}{n^3 (2 l+1) a^2}
 \left\{ \log \frac{n a}{2} - \psi(n+l+1)
+ \psi(2 l+2) + \psi(2 l+1) \right\}
\\
 \langle \frac{\log r}{r^3} \rangle_{nl}
 & = &
\frac{2}{n^3 l(l+1)(2 l +1) a^3}
\\
 && \ \ \ \ \ \times
 \left\{ \log \frac{n a}{2} - \psi(n+l+1)
+ \psi(2 l+3) + \psi(2 l)-\frac{n-l-1/2}{n} \right\} \; , \;
 l \neq 0
\\
\langle \frac{\log^2 r}{r} \rangle_{10}
 & = &
 \frac{1}{a} \left\{
 \log^2 \frac{a}{2} + 2(1 - \eul) \log \frac{a}{2} + (1-\eul)^2 +
 \frac{\pi^2}{6} - 1 \right\}
\\
 \langle \frac{1}{r}\Delta \rangle_{nl}
 & = &
 \frac{2 l +1 - 4 n}{n^4(2 l +1)a^3}
\\
 \langle \frac{\log r}{r}\Delta \rangle_{nl}
 & = &
 \frac{1}{n^4(2 l +1)a^3}
 \left\{ (2 l +1 - 4 n)\log\frac{n a}{2}
 \right.
\\
&& \ \ \ \ \ \ \
 \left.
+ (2 l +1 + 4 n)\psi(n+l+1)
 - 4 n \left[ \psi(2l+2) +\psi(2l+1)\right]
 \phantom{\frac{1}{2}} \!\!\!\!\! \right\} \;,
\end{eqnarray*}
\vskip0.6cm
\noindent
where
\bdm
 \psi(x) = \frac{{\rm dlog}\,\Gamma(x)}{{\rm d}x}
\,.
\edm
The formulas involving logarithms may be obtained
by differentiating the following expressions:
\begin{eqnarray*}
 && \langle r^p \rangle \,=\, a^p \,\frac{n^{p-1}}{2^{1+p}}\,
\frac{(n-l-1)!}{(n+l)!}
\\
&& \ \ \ \ \ \ \times \,\left\{
\begin{array}{ll}
 \displaystyle{
 \sum_{r=0}^{n-l-1} \, \frac{\Gamma(2l+3+p+r)\Gamma^2(2+p)}
{\Gamma(r+1)\Gamma^2(n-l-r)\Gamma^2(3+l+p-n+r) }}
 &\ \ , \, p > -2\;,
\\
 \displaystyle{
 \sum_{r=0}^{n-l-1} \, \frac{\Gamma(2l+3+p+r)\Gamma^2(n-l-2-p-r)}
{\Gamma(r+1)\Gamma^2(n-l-r)\Gamma(-1-p) }}
 &\ \ , \, p < -1\;,
\end{array}
\right.
\end{eqnarray*}
and can be deduced from Refs. \cite{bb:beth,bb:gali}.

\medskip

\noindent
{\bf Bethe logarithms}.\\
Finally, we take some expressions for the non-abelian Bethe logarithms (for $l=0$) that appear in weakly coupled heavy quarkonium from 
\cite{Kniehl:2002br}, see also Ref. \cite{Kniehl:1999ud}:
\begin{equation}
L^E_n={1\over C_f^2\al^2E^C_n}\int{{\rm d}^3{\bfm k}\over(2\pi)^3}
|\langle{\bfm r}\rangle_{{\bff k}n}|^2
\left(E^C_n-{{\bfm k}^2\over m}\right)^3
\ln{E^C_1\over E^C_n-{\bfm k}^2/m},
\end{equation}
where ${\bfm k}$ labels an eigenstate of $h_o^C$. $L^E_n$ can be reduced to one-parameter integrals of elementary functions.
For the reader's convenience, we list the relevant formulae here.
They read
\begin{equation}
L^E_n=\int_0^\infty{\rm d}\nu\,Y_n^E(\nu)X^2_n(\nu),
\end{equation}
where
\begin{eqnarray}
Y^E_n(\nu)&=&{2^6\rho_n^5\nu(\nu^2+1)\exp[4\nu\arccot(\nu/\rho_n)]\over
n^2(\nu^2+\rho_n^2)^3[\exp(2\pi\nu)-1]}\ln{n^2\nu^2\over\nu^2+\rho_n^2},
\nonumber\\
X_1(\nu)&=&\rho_1+2,
\nonumber\\
X_2(\nu)&=&{\nu^2(2\rho_2^2+9\rho_2+8)-\rho_2^2(\rho_2+4)\over
(\nu^2+\rho_2^2)},
\nonumber\\
X_3(\nu)&=&{\nu^4(8\rho_3^3+60\rho_3^2+123\rho_3+66)
-2\nu^2\rho_3^2(6\rho_3^2+41\rho_3+54)+3\rho_3^4(\rho_3+6)\over
3(\nu^2+\rho_3^2)^2},
\end{eqnarray}
with
\begin{equation}
\rho_n=n\left({C_A\over2C_f}-1\right)={n\over8}.
\end{equation}
For $n=1,2,3$, the following numerical values are obtained in Ref. \cite{Kniehl:2002br}:
\begin{equation}
L^E_1=-81.5379,\qquad L^E_2=-37.6710,\qquad L^E_3=-22.4818.
\end{equation}



\begin{thebibliography}{99}
\itemsep -2pt 

\bibitem{Yndurain} F.J. Yndur\'ain, ``The Theory of Quarks and Gluon
  Interactions'', 3rd edition, Springer, Heidelberg, 1999.
  
\bibitem{Caswell:1985ui}
  W.~E.~Caswell and G.~P.~Lepage,
  Phys.\ Lett.\  B {\bf 167}, 437 (1986).
  
\bibitem{Pineda:1997bj}
  A.~Pineda and J.~Soto,
  Nucl.\ Phys.\ Proc.\ Suppl.\  {\bf 64}, 428 (1998).

\bibitem{Brambilla:2004jw}
  N.~Brambilla, A.~Pineda, J.~Soto and A.~Vairo,
  Rev.\ Mod.\ Phys.\  {\bf 77}, 1423 (2005).
 
\bibitem{NRQCD1} B.A.~Thacker and G.P.~Lepage, Phys. Rev. {\bf D43}, 196
  (1991). 

\bibitem{Brambilla:1999xf}
  N.~Brambilla, A.~Pineda, J.~Soto and A.~Vairo,
  Nucl.\ Phys.\  B {\bf 566}, 275 (2000)
  [arXiv:hep-ph/9907240].

\bibitem{Brambilla:2000gk}
  N.~Brambilla, A.~Pineda, J.~Soto, A.~Vairo,
  Phys.\ Rev.\  {\bf D63}, 014023 (2001).
  [hep-ph/0002250].

\bibitem{HQET}
B. Grinstein, in {\it Proceedings of the Workshop on High
Energy Phenomenology}, Mexico City Mexico, Jul. 1-12, 1991, 161-216; 
T. Mannel, {\it Chinese Journal of Physics} {\bf 31} (1993);
M. Neubert, {\it Phys. Rep.} {\bf245} (1994) 259.
 
\bibitem{Pineda:1998kn}
  A.~Pineda and J.~Soto,
  Phys.\ Rev.\  D {\bf 59}, 016005 (1999).

\bibitem{Czarnecki:1998zv}
  A.~Czarnecki, K.~Melnikov, A.~Yelkhovsky,
  Phys.\ Rev.\ Lett.\  {\bf 82}, 311-314 (1999).
  [hep-ph/9809341].

\bibitem{Melnikov:1999uf}
  K.~Melnikov and A.~Yelkhovsky,
  Phys.\ Lett.\  B {\bf 458}, 143 (1999)
  [arXiv:hep-ph/9902276].
  
\bibitem{Melnikov:2000zz}
  K.~Melnikov and A.~Yelkhovsky,
  Phys.\ Rev.\ Lett.\  {\bf 86}, 1498 (2001)
  [arXiv:hep-ph/0010131].
    
\bibitem{Kniehl:2000cx}
  B.~A.~Kniehl and A.~A.~Penin,
  Phys.\ Rev.\ Lett.\  {\bf 85}, 5094 (2000)
  [arXiv:hep-ph/0010159].
 
\bibitem{Bodwin:1994jh}
  G.~T.~Bodwin, E.~Braaten and G.~P.~Lepage,
  Phys.\ Rev.\  D {\bf 51}, 1125 (1995)
  [Erratum-ibid.\  D {\bf 55}, 5853 (1997)]
  [arXiv:hep-ph/9407339].
 
\bibitem{Mannel:1994xh}
  T.~Mannel and G.~A.~Schuler,
  Z.\ Phys.\  C {\bf 67}, 159 (1995)
  [arXiv:hep-ph/9410333].

\bibitem{Luke:1992cs}
  M.~E.~Luke and A.~V.~Manohar,
  Phys.\ Lett.\  B {\bf 286}, 348 (1992)
  [arXiv:hep-ph/9205228].
   
\bibitem{Brambilla:2003nt}
  N.~Brambilla, D.~Gromes and A.~Vairo,
  Phys.\ Lett.\  B {\bf 576}, 314 (2003)
  [arXiv:hep-ph/0306107].
 
\bibitem{Manohar:1997qy}
  A.~V.~Manohar,
  Phys.\ Rev.\  D {\bf 56}, 230 (1997).
 
\bibitem{Bauer:1997gs}
  C.~W.~Bauer and A.~V.~Manohar,
  Phys.\ Rev.\  D {\bf 57}, 337 (1998)
  [arXiv:hep-ph/9708306].
 
\bibitem{Pineda:2000sz}
  A.~Pineda and A.~Vairo,
  Phys.\ Rev.\  D {\bf 63}, 054007 (2001)
  [Erratum-ibid.\  D {\bf 64}, 039902 (2001)]
  [arXiv:hep-ph/0009145].

\bibitem{Gremm:1997dq}
  M.~Gremm and A.~Kapustin,
  Phys.\ Lett.\  B {\bf 407}, 323 (1997)
  [arXiv:hep-ph/9701353].

\bibitem{Pineda:1998kj}
  A.~Pineda and J.~Soto,
  Phys.\ Rev.\  D {\bf 58}, 114011 (1998).

\bibitem{Beneke:1997zp}
  M.~Beneke and V.~A.~Smirnov,
  Nucl.\ Phys.\  B {\bf 522}, 321 (1998)
  [arXiv:hep-ph/9711391].

\bibitem{Pokorski:1987ed}
  S.~Pokorski,
{\it  Cambridge, Uk: Univ. Pr. ( 1987) 394 P. ( Cambridge Monographs On Mathematical Physics)}

\bibitem{Eichten:1990vp}
  E.~Eichten, B.~R.~Hill,
  Phys.\ Lett.\  {\bf B243}, 427-431 (1990).

\bibitem{KalSar}
G. K{\"a}llen and A. Sarby,
K.\ Dan.\ Vidensk.\ Selsk.\ Mat.-Fis.\ Medd.\ {29},  N17 (1955) 1.

\bibitem{HarBro} I. Harris and L.M. Brown, Phys.\ Rev.\ {105} (1957) 1656.
   
\bibitem{Hoang:1997ui}
  A.~H.~Hoang,
  Phys.\ Rev.\  D {\bf 56}, 5851 (1997).

\bibitem{Czarnecki:1997vz}
  A.~Czarnecki and K.~Melnikov,
  Phys.\ Rev.\ Lett.\  {\bf 80}, 2531 (1998).

\bibitem{Beneke:1997jm}
  M.~Beneke, A.~Signer and V.~A.~Smirnov,
  Phys.\ Rev.\ Lett.\  {\bf 80}, 2535 (1998).

\bibitem{Marquard:2006qi}
  P.~Marquard, J.~H.~Piclum, D.~Seidel and M.~Steinhauser,
  Nucl.\ Phys.\  B {\bf 758}, 144 (2006).

\bibitem{CzaMel2} A. Czarnecki and K. Melnikov,
Phys.\ Rev.\ D {65} (2002) 051501.

\bibitem{Pineda:2001ra}
  A.~Pineda,
  Phys.\ Rev.\  {\bf D65}, 074007 (2002).
  [hep-ph/0109117].

\bibitem{Lepage:1992tx}
  G.~P.~Lepage, L.~Magnea, C.~Nakhleh, U.~Magnea and K.~Hornbostel,
  Phys.\ Rev.\  D {\bf 46}, 4052 (1992)
  [arXiv:hep-lat/9205007].

\bibitem{Labelle:1996en}
  P.~Labelle,
  Phys.\ Rev.\  D {\bf 58} (1998) 093013
  [arXiv:hep-ph/9608491].
 
\bibitem{Luke:1996hj}
  M.~E.~Luke and A.~V.~Manohar,
  Phys.\ Rev.\  D {\bf 55}, 4129 (1997)
  [arXiv:hep-ph/9610534].
  
\bibitem{Grinstein:1997gv}
  B.~Grinstein and I.~Z.~Rothstein,
  Phys.\ Rev.\  D {\bf 57}, 78 (1998)
  [arXiv:hep-ph/9703298].

\bibitem{Luke:1997ys}
  M.~E.~Luke and M.~J.~Savage,
  Phys.\ Rev.\  D {\bf 57}, 413 (1998)
  [arXiv:hep-ph/9707313].
 
\bibitem{Lepage:1997cs}
  G.~P.~Lepage,
  arXiv:nucl-th/9706029.

\bibitem{Kaplan:1996xu}
  D.~B.~Kaplan, M.~J.~Savage and M.~B.~Wise,
  Nucl.\ Phys.\  B {\bf 478}, 629 (1996)
  [arXiv:nucl-th/9605002].
  
\bibitem{Pascual:1984zb}
  P.~Pascual and R.~Tarrach,
  Lect.\ Notes Phys.\  {\bf 194} (1984) 1.

\bibitem{Pineda:1997ie}
  A.~Pineda and J.~Soto,
  Phys.\ Lett.\  B {\bf 420}, 391 (1998).

\bibitem{Kniehl:2001ju}
  B.~A.~Kniehl, A.~A.~Penin, M.~Steinhauser and V.~A.~Smirnov,
  Phys.\ Rev.\  D {\bf 65} (2002) 091503
  [arXiv:hep-ph/0106135].

\bibitem{Kniehl:2002br}
  B.~A.~Kniehl, A.~A.~Penin, V.~A.~Smirnov and M.~Steinhauser,
  Nucl.\ Phys.\  B {\bf 635}, 357 (2002)
  [arXiv:hep-ph/0203166].

\bibitem{Gupta:1981pd}
  S.~N.~Gupta and S.~F.~Radford,
  Phys.\ Rev.\  D {\bf 24} (1981) 2309.

\bibitem{Gupta:1982qc}
  S.~N.~Gupta and S.~F.~Radford,
  Phys.\ Rev.\  D {\bf 25} (1982) 3430.

\bibitem{Buchmuller:1981aj}
  W.~Buchmuller, Y.~J.~Ng and S.~H.~H.~Tye,
  Phys.\ Rev.\  D {\bf 24}, 3003 (1981).

\bibitem{Pantaleone:1985uf}
  J.~T.~Pantaleone, S.~H.~H.~Tye and Y.~J.~Ng,
  Phys.\ Rev.\  D {\bf 33}, 777 (1986).

\bibitem{Titard:1993nn}
  S.~Titard and F.~J.~Yndurain,
  Phys.\ Rev.\  D {\bf 49}, 6007 (1994)
  [arXiv:hep-ph/9310236].

\bibitem{Brambilla:1999xj}
  N.~Brambilla, A.~Pineda, J.~Soto and A.~Vairo,
  Phys.\ Lett.\  B {\bf 470}, 215 (1999).

\bibitem{Manohar:2000hj}
  A.~V.~Manohar and I.~W.~Stewart,
  Phys.\ Rev.\  D {\bf 62}, 074015 (2000)
  [arXiv:hep-ph/0003032].

\bibitem{Fischler:1977yf}
  W.~Fischler,
  Nucl.\ Phys.\  B {\bf 129}, 157 (1977).

\bibitem{Schroder:1998vy}
  Y.~Schroder,
  Phys.\ Lett.\  {\bf B447}, 321-326 (1999).
  [arXiv:hep-ph/9812205 [hep-ph]].

\bibitem{Brambilla:1999qa}
  N.~Brambilla, A.~Pineda, J.~Soto and A.~Vairo,
  Phys.\ Rev.\  D {\bf 60}, 091502 (1999).

\bibitem{Kniehl:1999ud}
  B.~A.~Kniehl and A.~A.~Penin,
  Nucl.\ Phys.\  B {\bf 563}, 200 (1999).

\bibitem{Smirnov:2008pn}
  A.~V.~Smirnov, V.~A.~Smirnov and M.~Steinhauser,
  Phys.\ Lett.\  B {\bf 668}, 293 (2008).

\bibitem{Anzai:2009tm}
  C.~Anzai, Y.~Kiyo, Y.~Sumino,
  Phys.\ Rev.\ Lett.\  {\bf 104}, 112003 (2010).
  [arXiv:0911.4335 [hep-ph]].

\bibitem{Smirnov:2009fh}
  A.~V.~Smirnov, V.~A.~Smirnov, M.~Steinhauser,
  Phys.\ Rev.\ Lett.\  {\bf 104}, 112002 (2010).
  [arXiv:0911.4742 [hep-ph]].

\bibitem{Eichten:1980mw}
  E.~Eichten and F.~Feinberg,
  Phys.\ Rev.\  D {\bf 23}, 2724 (1981).

\bibitem{Brambilla:2006wp}
  N.~Brambilla, X.~Garcia i Tormo, J.~Soto and A.~Vairo,
  Phys.\ Lett.\  B {\bf 647}, 185 (2007).

\bibitem{Pineda:2000gza}
  A.~Pineda and J.~Soto,
  Phys.\ Lett.\  B {\bf 495}, 323 (2000).
 
\bibitem{Eidemuller:1997bb}
  M.~Eidemuller and M.~Jamin,
  Phys.\ Lett.\  B {\bf 416}, 415 (1998)
  [arXiv:hep-ph/9709419].

\bibitem{Pineda:2011db}
  A.~Pineda, M.~Stahlhofen,
  Phys.\ Rev.\  {\bf D84}, 034016 (2011).
  [arXiv:1105.4356 [hep-ph]].
 
\bibitem{Pineda:2011aw}
  A.~Pineda,
  Phys.\ Rev.\  {\bf D84}, 014012 (2011).
  [arXiv:1101.3269 [hep-ph]].

\bibitem{Schroder:1999sg}
  Y.~Schroder,
  ``The static potential in QCD'', DESY-THESIS-1999-021.

\bibitem{Brambilla:2009bi}
  N.~Brambilla, A.~Vairo, X.~Garcia i Tormo and J.~Soto,
  Phys.\ Rev.\  D {\bf 80}, 034016 (2009)
  [arXiv:0906.1390 [hep-ph]].

\bibitem{Luke:1999kz}
  M.~E.~Luke, A.~V.~Manohar and I.~Z.~Rothstein,
  Phys.\ Rev.\  D {\bf 61}, 074025 (2000)
  [arXiv:hep-ph/9910209].

\bibitem{Hoang:2002yy}
  A.~H.~Hoang and I.~W.~Stewart,
  Phys.\ Rev.\  D {\bf 67}, 114020 (2003)
  [arXiv:hep-ph/0209340].

\bibitem{Hoang:2006ht}
  A.~H.~Hoang and M.~Stahlhofen,
  Phys.\ Rev.\  D {\bf 75}, 054025 (2007)
  [arXiv:hep-ph/0611292].
 
\bibitem{Hoang:2011gy}
  A.~H.~Hoang and M.~Stahlhofen,
  JHEP {\bf 1106}, 088 (2011)
  [arXiv:1102.0269 [hep-ph]].

\bibitem{Czarnecki:1999mw}
  A.~Czarnecki, K.~Melnikov and A.~Yelkhovsky,
  Phys.\ Rev.\  A {\bf 59}, 4316 (1999).

\bibitem{Kniehl:2003ap}
  B.~A.~Kniehl, A.~A.~Penin, A.~Pineda, V.~A.~Smirnov and M.~Steinhauser,
  Phys.\ Rev.\ Lett.\  {\bf 92}, 242001 (2004)
  [Erratum-ibid.\  {\bf 104}, 199901 (2010)]
  [arXiv:hep-ph/0312086].

\bibitem{Penin:2004xi}
  A.~A.~Penin, A.~Pineda, V.~A.~Smirnov and M.~Steinhauser,
  Phys.\ Lett.\  B {\bf 593}, 124 (2004)
  [Erratum-ibid.\  {\bf 677}, 343 (2009)]
  [Erratum-ibid.\  {\bf 683}, 358 (2010)]
  [arXiv:hep-ph/0403080].

\bibitem{Yelkhovsky:2001tx}
  A.~Yelkhovsky,
  Phys.\ Rev.\  A {\bf 64}, 062104 (2001)
  [arXiv:hep-ph/0103241].

\bibitem{Pineda:2001et}
  A.~Pineda,
  Phys.\ Rev.\  {\bf D66}, 054022 (2002).
  [hep-ph/0110216].

\bibitem{Kniehl:1999mx}
  B.~A.~Kniehl and A.~A.~Penin,
  Nucl.\ Phys.\  B {\bf 577}, 197 (2000).

\bibitem{Penin:2004ay}
  A.~A.~Penin, A.~Pineda, V.~A.~Smirnov and M.~Steinhauser,
  Nucl.\ Phys.\  B {\bf 699}, 183 (2004).
  [arXiv:hep-ph/0406175].

\bibitem{Pineda:2006ri}
  A.~Pineda, A.~Signer,
  Nucl.\ Phys.\  {\bf B762}, 67-94 (2007).
  [hep-ph/0607239].
 
\bibitem{Hoang:2001mm}
  A.~H.~Hoang, A.~V.~Manohar, I.~W.~Stewart, T.~Teubner,
  Phys.\ Rev.\  {\bf D65}, 014014 (2002).
  [hep-ph/0107144].

\bibitem{Hoang:2003ns}
  A.~H.~Hoang,
  Phys.\ Rev.\  D {\bf 69}, 034009 (2004).
 
\bibitem{Kniehl:2002yv}
  B.~A.~Kniehl, A.~A.~Penin, M.~Steinhauser and V.~A.~Smirnov,
  Phys.\ Rev.\ Lett.\  {\bf 90}, 212001 (2003).

\bibitem{Pineda:1997hz}
  A.~Pineda and F.~J.~Yndurain,
  Phys.\ Rev.\  D {\bf 58}, 094022 (1998)
  [arXiv:hep-ph/9711287].

\bibitem{Penin:2002zv}
  A.~A.~Penin and M.~Steinhauser,
  Phys.\ Lett.\  B {\bf 538}, 335 (2002)
  [arXiv:hep-ph/0204290].

\bibitem{Leutwyler:1980tn}
  H.~Leutwyler,
  Phys.\ Lett.\  B {\bf 98}, 447 (1981).

\bibitem{Voloshin:1979uv}
  M.~B.~Voloshin,
  Sov.\ J.\ Nucl.\ Phys.\  {\bf 36}, 143 (1982)
  [Yad.\ Fiz.\  {\bf 36}, 247 (1982)].

\bibitem{Pineda:1996uk}
  A.~Pineda,
  Nucl.\ Phys.\  B {\bf 494}, 213 (1997)
  [arXiv:hep-ph/9611388].

\bibitem{Kiselev:1999sc}
  V.~V.~Kiselev, A.~K.~Likhoded and A.~I.~Onishchenko,
  Nucl.\ Phys.\  B {\bf 569}, 473 (2000)
  [arXiv:hep-ph/9905359].
  
\bibitem{Fadin:1988fn}
  V.~S.~Fadin and V.~A.~Khoze,
  Sov.\ J.\ Nucl.\ Phys.\  {\bf 48}, 309 (1988)
  [Yad.\ Fiz.\  {\bf 48}, 487 (1988)].
  
\bibitem{Strassler:1990nw}
  M.~J.~Strassler, M.~E.~Peskin,
  Phys.\ Rev.\  {\bf D43}, 1500-1514 (1991).
 
\bibitem{Hoang:2000yr}
  A.~H.~Hoang {\it et al.},
  Eur.\ Phys.\ J.\ direct C {\bf 2}, 1 (2000)
  [arXiv:hep-ph/0001286].

\bibitem{Penin:2005eu}
  A.~A.~Penin, V.~A.~Smirnov and M.~Steinhauser,
  Nucl.\ Phys.\  B {\bf 716}, 303 (2005).
 
\bibitem{Beneke:2005hg}
  M.~Beneke, Y.~Kiyo and K.~Schuller,
  Nucl.\ Phys.\  B {\bf 714}, 67 (2005).

\bibitem{Beneke:2007gj}
  M.~Beneke, Y.~Kiyo and K.~Schuller,
  Phys.\ Lett.\  B {\bf 658}, 222 (2008).

\bibitem{Beneke:2007pj}
  M.~Beneke, Y.~Kiyo and A.~A.~Penin,
  Phys.\ Lett.\  B {\bf 653}, 53 (2007).

\bibitem{Kuhn:1998uy}
  J.~H.~Kuhn, A.~A.~Penin, A.~A.~Pivovarov,
  Nucl.\ Phys.\  {\bf B534}, 356-370 (1998).
  [hep-ph/9801356].

\bibitem{Penin:1998zh}
  A.~A.~Penin, A.~A.~Pivovarov,
  Phys.\ Lett.\  {\bf B435}, 413-419 (1998).
  [hep-ph/9803363].

\bibitem{Hoang:1998uv}
  A.~H.~Hoang,
  Phys.\ Rev.\  {\bf D59}, 014039 (1999).
  [hep-ph/9803454].

\bibitem{Melnikov:1998ug}
  K.~Melnikov, A.~Yelkhovsky,
  Phys.\ Rev.\  {\bf D59}, 114009 (1999).
  [hep-ph/9805270].

\bibitem{Beneke:1999fe}
  M.~Beneke and A.~Signer,
  Phys.\ Lett.\  B {\bf 471}, 233 (1999).

\bibitem{Penin} 
A.~A.~Penin and A.~A.~Pivovarov,
Nucl.\ Phys.\ B {\bf 549}, 217 (1999)
[arXiv:hep-ph/9807421].

\bibitem{Brambilla:2004wf}
  N.~Brambilla {\it et al.}  [Quarkonium Working Group],
  arXiv:hep-ph/0412158.
  
\bibitem{Asner:2008nq}
  D.~M.~Asner {\it et al.},
  arXiv:0809.1869 [hep-ex].

\bibitem{Brambilla:2010cs}
  N.~Brambilla {\it et al.},
  Eur.\ Phys.\ J.\  C {\bf 71}, 1534 (2011)
  [arXiv:1010.5827 [hep-ph]].

\bibitem{Recksiegel:2001xq}
  S.~Recksiegel and Y.~Sumino,
  Phys.\ Rev.\  D {\bf 65}, 054018 (2002).

\bibitem{Pineda:2002se}
  A.~Pineda,
  J.\ Phys.\ G {\bf 29}, 371 (2003).

\bibitem{Lee:2002sn}
  T.~Lee,
  Phys.\ Rev.\  D {\bf 67}, 014020 (2003).

\bibitem{Brambilla:2010pp}
  N.~Brambilla, X.~Garcia i Tormo, J.~Soto and A.~Vairo,
  Phys.\ Rev.\ Lett.\  {\bf 105}, 212001 (2010).

\bibitem{Pineda:2001zq}
  A.~Pineda,
  JHEP {\bf 0106}, 022 (2001).

\bibitem{Necco:2001xg}
  S.~Necco and R.~Sommer,
  Nucl.\ Phys.\  B {\bf 622}, 328 (2002).

\bibitem{Brambilla:2001fw}
  N.~Brambilla, Y.~Sumino and A.~Vairo,
  Phys.\ Lett.\  B {\bf 513}, 381 (2001).

\bibitem{Brambilla:2001qk}
  N.~Brambilla, Y.~Sumino and A.~Vairo,
  Phys.\ Rev.\  D {\bf 65}, 034001 (2002).

\bibitem{Lee:2003hh}
  T.~Lee,
  JHEP {\bf 0310}, 044 (2003).

\bibitem{Recksiegel:2003fm}
  S.~Recksiegel and Y.~Sumino,
  Phys.\ Lett.\  B {\bf 578}, 369 (2004).
  
\bibitem{:2008vj}
  B.~Aubert {\it et al.}  [BABAR Collaboration],
  Phys.\ Rev.\ Lett.\  {\bf 101}, 071801 (2008)
  [E-ibid.\  {\bf 102}, 029901 (2009)].

\bibitem{Bonvicini:2009hs}
  G.~Bonvicini {\it et al.}  [The CLEO Collaboration],
  arXiv:0909.5474 [hep-ex].

\bibitem{QWG11}
R.~Mizuk, talk on behalf the [Belle Collaboration] at the QWG2011 - 8th International Workshop on Heavy Quarkonium.
  
\bibitem{Brambilla:2000db}
  N.~Brambilla and A.~Vairo,
  Phys.\ Rev.\  D {\bf 62}, 094019 (2000).
  
\bibitem{Aaltonen:2007gv}
  T.~Aaltonen {\it et al.}  [CDF Collaboration],
  Phys.\ Rev.\ Lett.\  {\bf 100}, 182002 (2008).

\bibitem{Abazov:2008kv}
  V.~M.~Abazov {\it et al.}  [D0 Collaboration],
  Phys.\ Rev.\ Lett.\  {\bf 101}, 012001 (2008).

\bibitem{GarciaiTormo:2005bs}
  X.~Garcia i Tormo and J.~Soto,
  Phys.\ Rev.\ Lett.\  {\bf 96}, 111801 (2006).
  
\bibitem{DomenechGarret:2008vk}
  J.~L.~Domenech-Garret and M.~A.~Sanchis-Lozano,
  Phys.\ Lett.\  B {\bf 669}, 52 (2008).
  
\bibitem{Recksiegel:2002za}
  S.~Recksiegel and Y.~Sumino,
  Phys.\ Rev.\  D {\bf 67}, 014004 (2003).
  
\bibitem{Aubert:2003pt}
  B.~Aubert {\it et al.}  [BABAR Collaboration],
  Phys.\ Rev.\ Lett.\  {\bf 92}, 142002 (2004).

\bibitem{Asner:2003wv}
  D.~M.~Asner {\it et al.} [CLEO Collaboration],
  Phys.\ Rev.\ Lett.\  {\bf 92}, 142001 (2004).

\bibitem{Choi:2002na}
  S.~K.~Choi {\it et al.}  [BELLE collaboration],
  Phys.\ Rev.\ Lett.\  {\bf 89}, 102001 (2002)
  [E-ibid.\  {\bf 89}, 129901 (2002)].

\bibitem{Edwards:1981mq}
  C.~Edwards {\it et al.},
  Phys.\ Rev.\ Lett.\  {\bf 48}, 70 (1982).

\bibitem{Hoang:1998ng}
  A.~H.~Hoang, Z.~Ligeti and A.~V.~Manohar,
  Phys.\ Rev.\ Lett.\  {\bf 82}, 277 (1999).
 
\bibitem{Brambilla:2005zw}
  N.~Brambilla, Y.~Jia and A.~Vairo,
  Phys.\ Rev.\  D {\bf 73}, 054005 (2006)
  [arXiv:hep-ph/0512369].

\bibitem{PDG2010}
K. Nakamura et al. (Particle Data Group), J. Phys. G 37, 075021 (2010). 

\bibitem{PDG2004}
S. Eidelman et al. (Particle Data Group), Phys. Lett. B592, 1 (2004).     

\bibitem{Brambilla:2010ey}
  N.~Brambilla, P.~Roig and A.~Vairo,
  arXiv:1012.0773 [hep-ph].

\bibitem{:2008fb}
  R.~E.~Mitchell {\it et al.}  [CLEO Collaboration],
  Phys.\ Rev.\ Lett.\  {\bf 102} (2009) 011801.

\bibitem{Manohar:2003xv}
  A.~V.~Manohar and P.~Ruiz-Femenia,
  Phys.\ Rev.\  D {\bf 69}, 053003 (2004)
  [arXiv:hep-ph/0311002].

\bibitem{Voloshin:2003hh}
  M.~B.~Voloshin,
  Mod.\ Phys.\ Lett.\  A {\bf 19}, 181 (2004)
  [arXiv:hep-ph/0311204].

\bibitem{Djouadi:2007ik}
  G.~Aarons {\it et al.} [ ILC Collaboration ],
  [arXiv:0709.1893 [hep-ph]].

\bibitem{Hoang:2000ib}
  A.~H.~Hoang, A.~V.~Manohar, I.~W.~Stewart and T.~Teubner,
  Phys.\ Rev.\ Lett.\  {\bf 86}, 1951 (2001).

\bibitem{Martinez:2002st}
  M.~Martinez, R.~Miquel,
  Eur.\ Phys.\ J.\  {\bf C27}, 49-55 (2003).
  [hep-ph/0207315].

\bibitem{Beneke:1998rk}
  M.~Beneke,
  Phys.\ Lett.\  B {\bf 434}, 115 (1998)
  [arXiv:hep-ph/9804241].

\bibitem{Pineda:2006gx}
  A.~Pineda and A.~Signer,
  Phys.\ Rev.\  D {\bf 73}, 111501 (2006).

\bibitem{Signer:2008da}
  A.~Signer,
  Phys.\ Lett.\  B {\bf 672}, 333 (2009)
  [arXiv:0810.1152 [hep-ph]].
 
\bibitem{Czarnecki:2001zc}
  A.~Czarnecki and K.~Melnikov,
  Phys.\ Lett.\  B {\bf 519}, 212 (2001).

\bibitem{Kiyo:2010jm}
  Y.~Kiyo, A.~Pineda and A.~Signer,
Nucl.\ Phys.\  B {\bf 841}, 231 (2010).
  [arXiv:1006.2685 [hep-ph]].

\bibitem{GarciaiTormo:2005ch}
  X.~Garcia i Tormo and J.~Soto,
  Phys.\ Rev.\  D {\bf 72}, 054014 (2005).

\bibitem{Brambilla:2007cz}
  N.~Brambilla, X.~Garcia i Tormo, J.~Soto and A.~Vairo,
  Phys.\ Rev.\  D {\bf 75}, 074014 (2007).
 
\bibitem{Kinoshita:1995mt}
  T.~Kinoshita and M.~Nio,
  Phys.\ Rev.\  D {\bf 53}, 4909 (1996)
  [arXiv:hep-ph/9512327].
 
\bibitem{Bodwin:1998mn}
  G.~T.~Bodwin and Y.~Q.~Chen,
  Phys.\ Rev.\  D {\bf 60}, 054008 (1999)
  [arXiv:hep-ph/9807492].

\bibitem{Amoros:1997rx}
  G.~Amoros, M.~Beneke and M.~Neubert,
  Phys.\ Lett.\  B {\bf 401}, 81 (1997)
  [arXiv:hep-ph/9701375].

\bibitem{bb:beth} H.~A.~Bethe and E.~E.~Salpeter,
{\it Quantum Mechanics of One and Two--Electron Atoms},
Plenum, 1977.

\bibitem{bb:gali} A.~Galindo and P.~Pascual,
{\it Quantum Mechanics}, Springer, 1989.



\end{thebibliography}
\end{document}